\documentclass[sn-basic]{sn-jnl}
\usepackage{graphicx}%
\usepackage{multirow}%
\usepackage{amsmath,amssymb,amsfonts}%
\usepackage{amsthm}%
\usepackage{mathrsfs}%
\usepackage{xcolor}%
\usepackage{textcomp}%
\usepackage{manyfoot}%
\usepackage{booktabs}%
\usepackage{algorithm}%
\usepackage{algorithmicx}%
\usepackage{algpseudocode}%
\usepackage{listings}%
\usepackage{subcaption} %

\usepackage{bm}

\newtheorem{proposition}{Proposition}

\newcommand{\bmu}{\mbox{\boldmath $\mu$}}

\newcommand{\bnu}{\mbox{\boldmath $\nu$}}

\newcommand{\bSigma}{\mbox{\boldmath $\Sigma$}}

\newcommand{\bbeta}{\mbox{\boldmath $\beta$}}
\newcommand{\btheta}{\mbox{\boldmath $\theta$}}

\newcommand{\x}{\mathbf{x}}
\newcommand{\w}{\mathbf{w}}

\newcommand{\CN}{\textrm{CN}}
\newcommand{\SMN}{\textrm{SMN}}
\newcommand{\A}{\mathbf{A}}

\newcommand{\bV}{\mathbf{V}}

\newcommand{\bgamma}{\mbox{\boldmath $\gamma$}}

\newcommand{\bC}{\mathbf{C}}

\newcommand{\xp}{\mathbf{x}}
\newcommand{\y}{\mathbf{y}}
\newcommand{\Y}{\mathbf{Y}}

\newcommand{\C}{\mathbf{C}}

\newcommand{\Z}{\mathbf{Z}}

\newcommand{\X}{\mathbf{X}}

\newcommand{\ap}{\mathbf{a}}
\newcommand{\bp}{\mathbf{b}}
\newcommand{\bx}{\mathbf{x}}

\newcommand{\binfty}{\boldsymbol{\infty}}

\newcommand{\EE}{{\rm E}}

\newcommand{\Ha}{\textrm{H}}

\newcommand{\CG}[1]{\textcolor{black}{#1}}

\newcommand{\CB}[1]{\textcolor{black}{#1}}

\theoremstyle{thmstyleone}%

\theoremstyle{thmstyletwo}%

\theoremstyle{thmstylethree}%

\begin{document}

\title[Bayesian analysis of flexible Heckman selection models using Hamiltonian Monte Carlo]{Bayesian analysis of flexible Heckman selection models using Hamiltonian Monte Carlo}


\author[1]{Heeju Lim}
\author[2]{Victor E. Lachos}
\author[1]{Victor H. Lachos}

\affil[1]{Department of Statistics, University of Connecticut, Storrs, CT 06269, U.S.A}
\affil[2]{Department of Applied Mathematics and Statistics, University of S\~ao Paulo, S\~{a}o Carlos, Brazil}

\abstract{The Heckman selection model is widely used in econometric analysis and other social sciences to address sample selection bias in data modeling. A common assumption in Heckman selection models is that the error terms follow an independent bivariate normal distribution. However, real-world data often deviates from this assumption, exhibiting heavy-tailed behavior, which can lead to inconsistent estimates if not properly addressed. In this paper, we propose a Bayesian analysis of Heckman selection models that replace the Gaussian assumption with well-known members of the class of scale mixture of normal distributions, such as the Student's-t and contaminated normal distributions. For these complex structures, Stan’s default No-U-Turn sampler is utilized to obtain posterior simulations. Through extensive simulation studies, we compare the performance of the Heckman selection models with normal, Student's-t and contaminated normal distributions. We also demonstrate the broad applicability of this methodology by applying it to medical care and labor supply data. The proposed algorithms are implemented in the \textsf{R} package \texttt{HeckmanStan}.}


\maketitle

\section{Introduction}
{Sample selection (SL) bias is a significant challenge in addressing missing data in health-related research in various fields, including biostatistics, sociology, and economics. This bias occurs when a variable of interest is only observable within a specific subset of the population, leading to distortions in data interpretation. The Heckman SL model, introduced by \citet{heckman1974shadow}, presents a parametric approach under the assumption of bivariate normality (SLn). However, the classical SLn model often falls short in accurately capturing residuals from real-world data, particularly when the data exhibits heavy tails or skewness. }

{To address this limitation within parametric modeling, \citet{marchenko2012heckman} introduced the Heckman selection-$t$ model (SLt), which replaces the traditional SLn model's error structure with a bivariate Student's-$t$ distribution. This modification enhances flexibility for modeling heavier-tailed data by introducing a single additional parameter referred to as degrees of freedom, which governs the tail behavior of the distribution. Furthermore, the Heckman selection contaminated normal (SLcn) model , introduced by \citet{lim2024heckman}, assumes a bivariate contaminated normal (CN) distribution for the error terms. This novel SLcn model allows for a more nuanced representation of the data, effectively capturing the complexities of real-world scenarios. In the CN distribution, one component accounts for the majority of typical observations, while the other addresses the presence of atypical points, such as outliers and inliers \citep{lim2024heckman}. Thus, this innovative framework not only improves the model's performance but also offers a more flexible approach to understanding the intricacies of sample selection bias and its implications in SL data modeling.}

Parameter estimation in these models is often performed using maximum likelihood estimation (MLE), such as the Expectation–Maximization (EM) algorithm. Detailed descriptions of the EM algorithm under the SLn, SLt, and SLcn models are provided by \citet{zhao2020new}, \citet{lachosHeckman}, and \citet{lim2024heckman}, respectively. These algorithms have been implemented in the \textsf{R} package \texttt{HeckmanEM}, providing an accessible framework for practitioners working with Heckman SL models. Although various methodological approaches have been proposed from a likelihood-based perspective, no work has been published from a Bayesian perspective. Bayesian methods for SL models offer several important advantages, including flexibility, the incorporation of prior information, comprehensive uncertainty quantification, and robust handling of complex data structures, such as missing responses. Nevertheless, Bayesian procedures typically rely on iterative sampling techniques such as Markov Chain Monte Carlo (MCMC), which often involve a larger number of iterations and greater computational load than EM algorithms.

Moreover, in this work, we propose a comprehensive Bayesian framework for these flexible models, seamlessly integrating the Hamiltonian Monte Carlo (HMC) algorithm introduced by \citet{neal2011mcmc}. As discussed in \citet{ThomasTu2021}, HMC provides a substantial improvement over conventional Markov Chain Monte Carlo (MCMC) techniques, such as the Metropolis-Hastings algorithm, by dramatically boosting computational efficiency, particularly when dealing with high-dimensional and complex models. Although HMC requires the computation of the gradient of the log density of the target distribution, the \texttt{Stan} software \citep{carpenter2017stan} excels at this task by using automatic differentiation. It also optimizes tuning parameters using the \texttt{No-U-turn} Sampler (NUTS) \citep{hoffman2014no}. By leveraging \texttt{Stan}, we need only to articulate the Bayesian model in its intuitive modeling language, after which the software promptly provides samples from the target distribution. Furthermore, we aim to assess and compare the performance of the normal, Student's t, and contaminated normal SL models by using popular Bayesian model selection criteria.

The remainder of the paper is organized as follows. In Section~\ref{model}, after introducing some notation, we briefly discuss some key properties related to the multivariate Student's-$t$ and contaminated normal distributions.  In Section~\ref{SLnModel}, we present the SLn model proposed by \citet{heckman1979}, then we present the robust SLt and SLcn models, including the related likelihood functions, which are instrumental for both ML estimation and Bayesian inference implemented through \texttt{Stan} and the \textsf{R} package \texttt{HeckmanStan}.  In Section~\ref{BI}, we outline the priors and posterior of the Bayesian model and its implementation using
\texttt{Stan}, including Bayesian model selection tools. In Sections~\ref{secSim} and \ref{secApp}, numerical examples using both simulated and real data are given to illustrate the performance of the proposed method. Finally, some concluding remarks are presented in Section~\ref{sec:6}.

\section{Background}\label{model}
In this section, we begin our exposition by defining the notation and presenting some basic concepts used throughout our methodology's development. As is usual in probability theory and its applications, we denote a random variable by an upper-case letter and its realization by the corresponding
lower case and use boldface letters for vectors and matrices. Let $\mathbf{I}_p$ represent a $p\times p$ identity matrix, $\A^{\top}$ be the transpose of $\A$. Throughout this paper, ${\cal N}_p(\bmu, \bSigma)$ denotes the $p$-variate normal (MN) distribution with mean vector $\bmu$ and covariance matrix $\bSigma$; 
and $\phi_p\left(\cdot\mid\bmu,\bSigma \right)$ and $\bm{\Phi}_p(\cdot \mid \bm{\mu}, \bm{\Sigma}) $
denote its probability density function ({pdf}) and cumulative density function ({cdf}), respectively. When $p=1$ we drop the index $p$, and in this case, if $\mu=0$ and $\sigma^2=1$, we write $\phi(\cdot)$
for the {pdf} and $\Phi(\cdot)$ for the {cdf}.  

For multiple integrals, we use the shorthand notation $$\int_{\ap}^{\bp}f(\x)d\x=\int_{a_1}^{b_1}\ldots\int_{a_p}^{b_p}f(x_1,\ldots,x_p)\mathrm{d} x_p\ldots \mathrm{d} x_1.$$
where  $\mathbf{a}=(a_1,\ldots,a_p)^\top$  and $\mathbf{b}=(b_1,\ldots,b_p)^\top$. If the Borel set  in $\mathbb{R}^p$ has the form
\begin{equation} \label{hyper1}
\mathbb{A} = \{(x_1,\ldots,x_p)\in \mathbb{R}^p:\,\,\, a_1\leq x_1 \leq b_1,\ldots, a_p\leq x_p \leq b_p \}=\{\mathbf{x}\in\mathbb{R}^p:\mathbf{a}\leq\mathbf{x}\leq\mathbf{b}\}.
\end{equation}
we  use the notation $\{\X \in \mathbb{A}\}=\{\ap\leq \X\leq\bp\}$.

{The Cauchy distribution, denoted by $\text{Cauchy}(\mu, \sigma^2)$, with location parameter $\mu\in {\cal R}$ and scale parameter \(\sigma^2>0\), has pdf  given by:}

$$\text{Cauchy}(y \mid \mu, \sigma^2) = \frac{1}{\pi \sigma \left(1 + \displaystyle\frac{(y - \mu)^2}{\sigma^2}\right)}, \quad y \in \mathbb{R}.$$

For the uniform and Beta distributions, we will use the notation $\text{Uniform}(a,b),\,\, b>a$ and $\rm{Beta}(\alpha,\beta)$, respectively.

\subsection{Scale mixtures of normal distributions}
 The symmetric class of SMN distributions \citep{andrews1974scale, Lange93} is defined as the distribution of the $p$-variate random vector
\begin{equation}\label{stoNI1} \X=\bmu+U^{-1/2}\Z,
\end{equation}
where $\bmu$ is a location vector, $\Z$ is a normal random vector with mean vector $\mathbf{0}$ and covariance matrix $\bSigma$, $U$ is a positive random variable with cumulative distribution function ({\it cdf}) $\Ha(u\mid \bnu)$ and probability density function ({\it pdf}) $h(u\mid \bnu)$, independent of $\Z$, where $\bnu$ is a scalar or parameter vector indexing the distribution of $U$. Given $U=u$, $\X$ follows a multivariate normal distribution with mean vector $\bmu$ and variance-covariance matrix $u^{-1}\bSigma$. Hence, the {\it pdf} of $\X$ is $\psi^{\SMN}_p(\xp\mid \bmu,\bSigma,\bnu)=\int^{\infty}_0{\phi_p(\xp\mid\bmu,u^{-1}\bSigma)}d\Ha(u\mid\bnu). $  By convention, we shall write $\mathbf{X}\sim\SMN_p(\bmu,\bSigma,\bnu)$. The $\SMN$ class constitutes a class of thick-tailed distributions, some of which are the multivariate versions of the Student's-$t$, slash, and the contaminated normal distribution. For further properties of the SMN class, see \citet{lachos2011linear} and \citet{m2024bayesian}. 
\subsubsection{The multivariate Student's-$t$  distribution}
In this case, $U\sim \mathcal{G}(\nu/2,\nu/2)$, where ${\mathcal{G}}(a,b)$ denotes a
gamma distribution with mean $a/b$. A random variable $\X$ having a $p$-variate Student's-$t$ distribution with
location vector $\bmu$, positive-definite scale-covariance
matrix $\bSigma$ and degrees of freedom $\nu$, denoted by $\X \sim
{\cal T}_p(\bmu,\bSigma,\nu)$, has the pdf:
$$
\psi^{T}_p(\xp\mid\bmu,\bSigma,\nu)=
\frac{\Gamma(\frac{p+\nu}{2})}{\Gamma(\frac{\nu}{2})\pi^{p/2}}\nu^{-p/2}|\bSigma|^{-1/2}\left(1+\frac{\delta(\xp)}{\nu}\right)^{-(p+\nu)/2},\label{lsdefAB1}
$$
where $\Gamma{(\cdot)}$ is the standard gamma function and $\delta(\xp) = (\xp-\bmu)^{\top}\bSigma^{-1}(\xp-\bmu)$ is the squared Mahalanobis
distance. It is \CB{known} that  as {$\nu \to \infty$,} $\X$ converges in
distribution to the ${\cal N} _p(\bmu,\bSigma)$.  

Let $\Psi^{T}_p(\ap,\bp\mid\bmu,\bSigma,\nu)$ represent
$$\Psi^{T}_p(\ap,\bp\mid\bmu,\bSigma,\nu)=\int_{\ap}^{\bp}{{\psi}^T_p}(\x\mid\bmu,\bSigma,\bnu)\textrm{d}\bx,$$ where  $\mathbf{a}=(a_1,\ldots,a_p)^\top$  and $\mathbf{b}=(b_1,\ldots,b_p)^\top$. When $\ap=-\binfty$ we will write simply $\Psi^T_p(\bp\mid\bmu,\bSigma,\nu)$ and when $p=1$ we will omit the sub-index $p$.

\begin{figure}[!ht]
\centering
    \begin{subfigure}[b]{0.32\textwidth}
        \includegraphics[width=\textwidth]{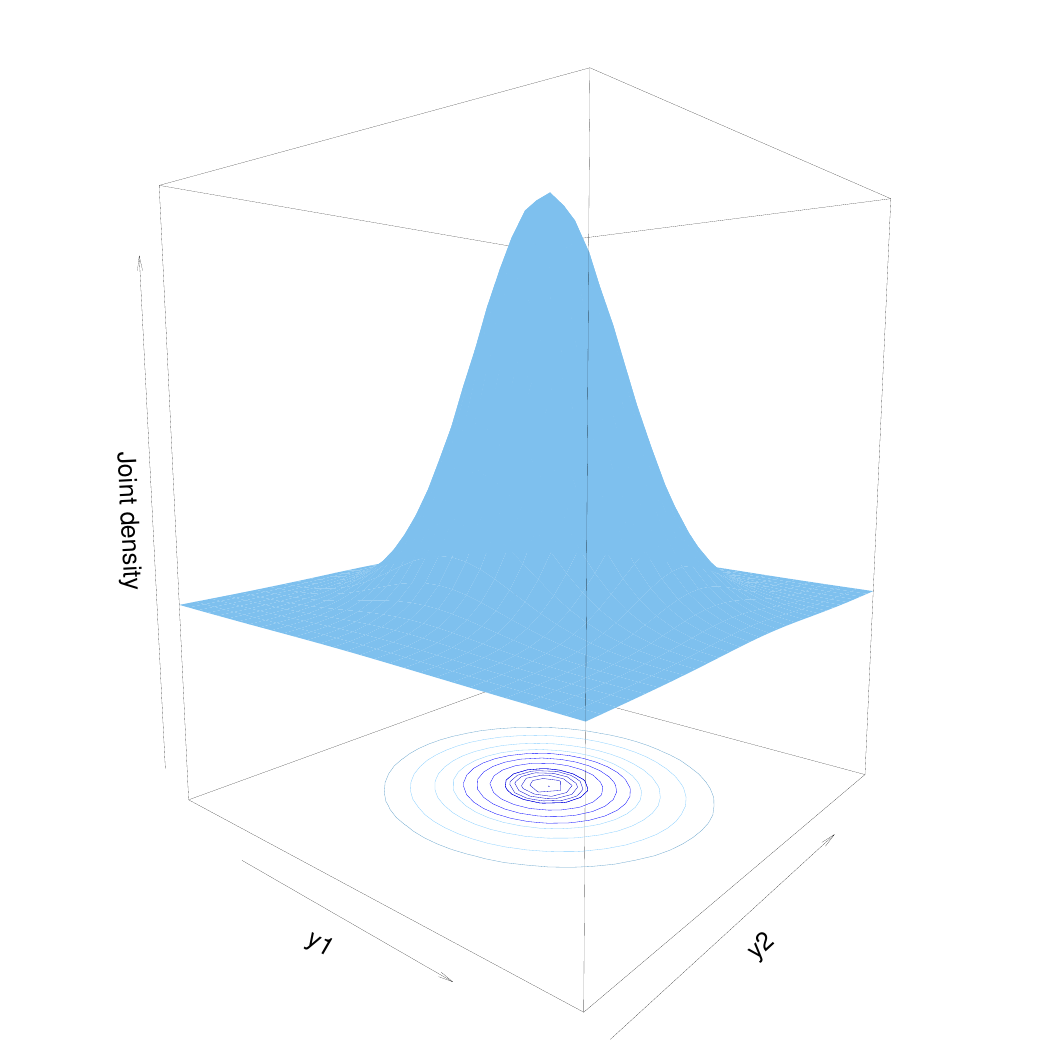}
        \caption{Normal distribution with $\bSigma= \begin{bmatrix} 1 & 0 \\ 0 & 1 \end{bmatrix}$.}
        \label{fig:normalvar1}
    \end{subfigure}
    \begin{subfigure}[b]{0.32\textwidth}
        \centering
        \includegraphics[width=\textwidth]{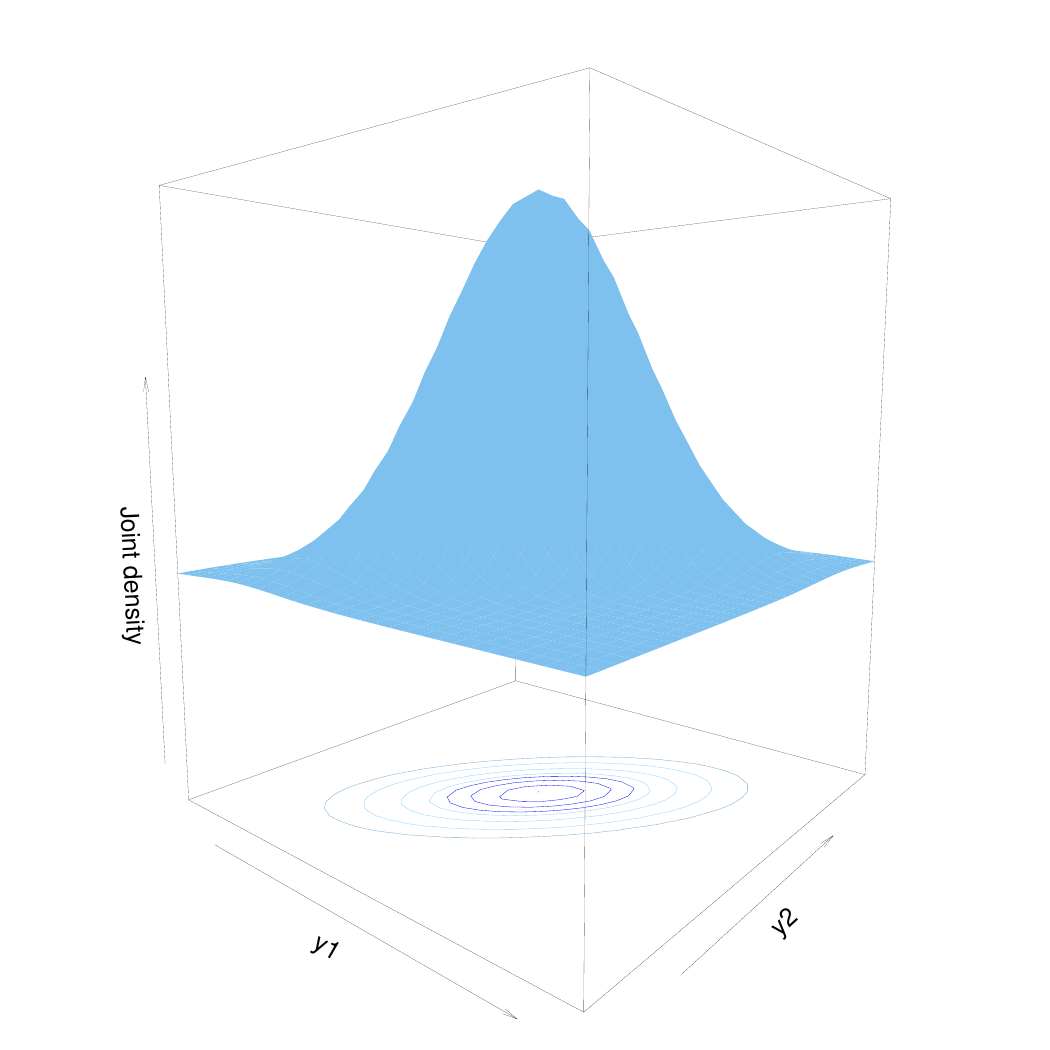}
        \caption{Normal distribution with $\bSigma= \begin{bmatrix} 3 & 1.2 \\ 1.2 & 1 \end{bmatrix}$.}
        \label{fig:normalvar2}
    \end{subfigure}
    \begin{subfigure}[b]{0.32\textwidth}
        \centering
        \includegraphics[width=\textwidth]{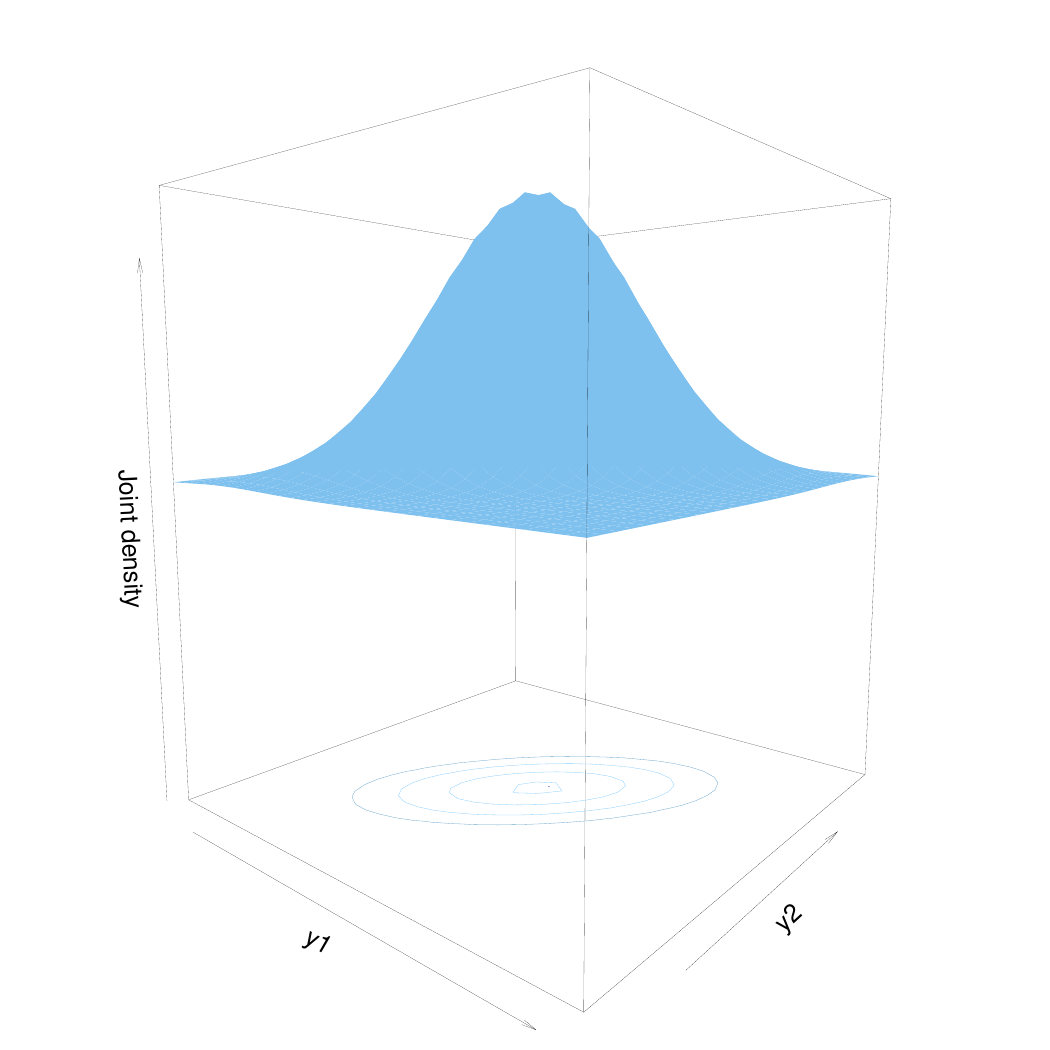}
        \caption{Normal distribution with $\bSigma= \begin{bmatrix} 5 & 1.6 \\ 1.6 & 1 \end{bmatrix}$.}
        \label{fig:normalvar3}
    \end{subfigure}
    \begin{subfigure}[b]{0.32\textwidth}            
        \centering
        \includegraphics[width=\textwidth]{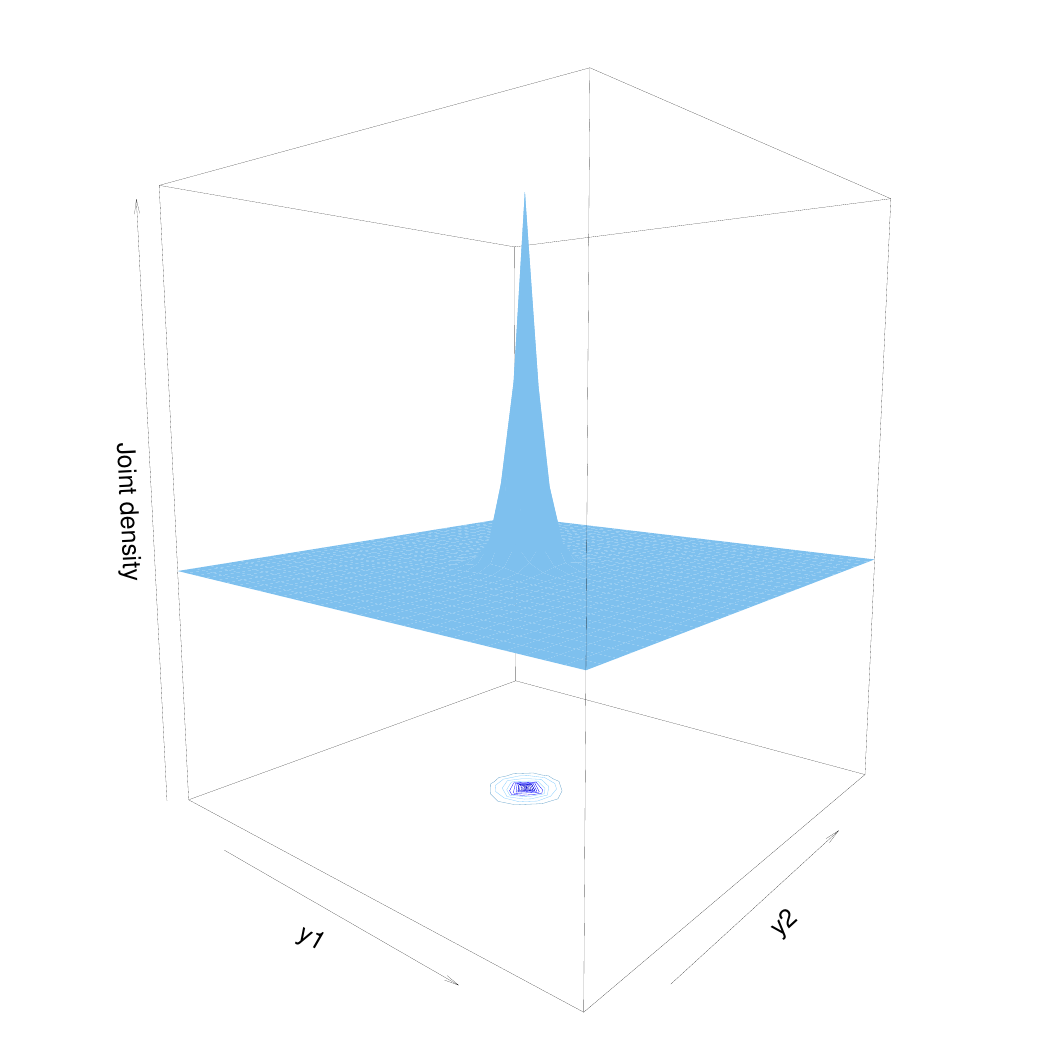}
        \caption{Student’s-t distribution with $\nu=3$.}
        \label{fig:t1}
    \end{subfigure}
    \begin{subfigure}[b]{0.32\textwidth}
        \centering
        \includegraphics[width=\textwidth]{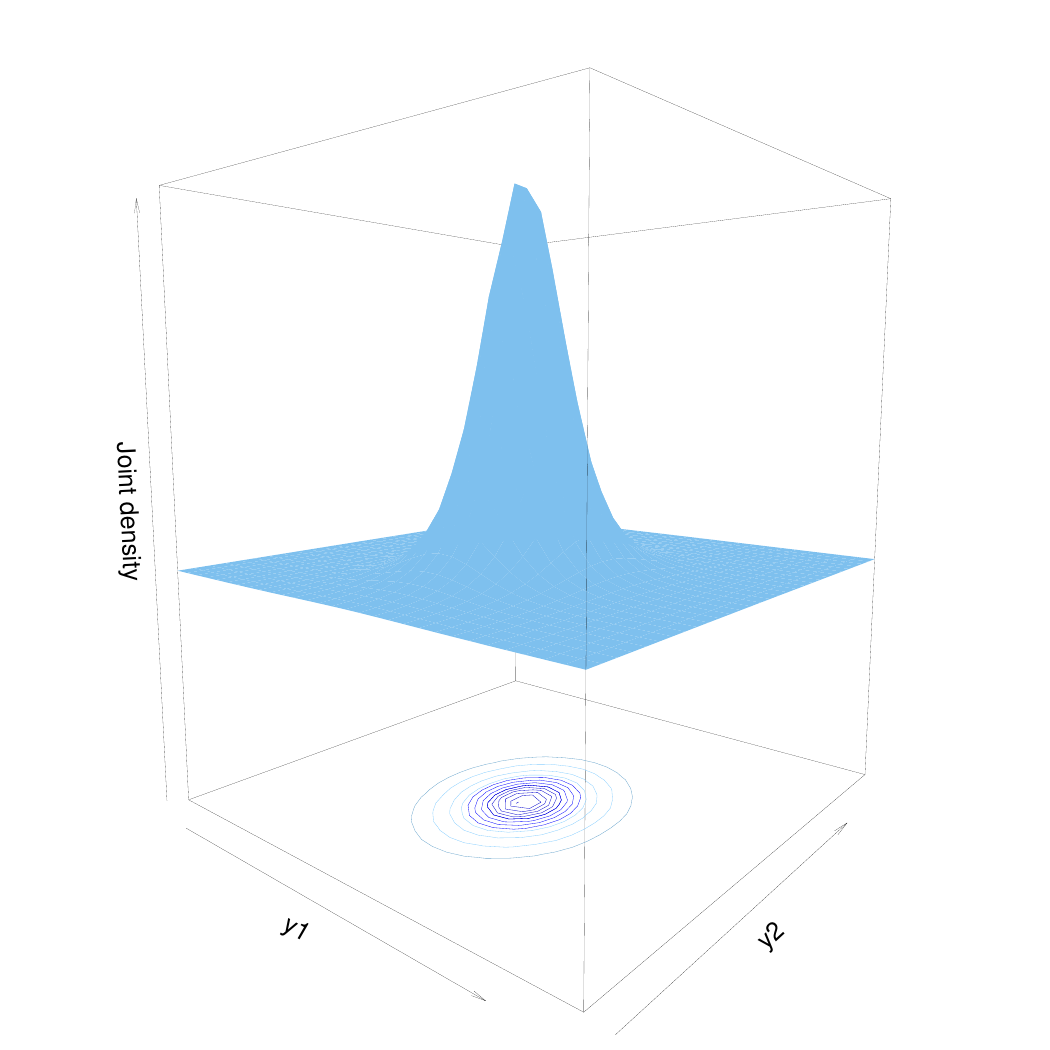}
        \caption{Student’s-t distribution with $\nu=5$.}
        \label{fig:t2}
    \end{subfigure}
    \begin{subfigure}[b]{0.32\textwidth}
        \centering
        \includegraphics[width=\textwidth]{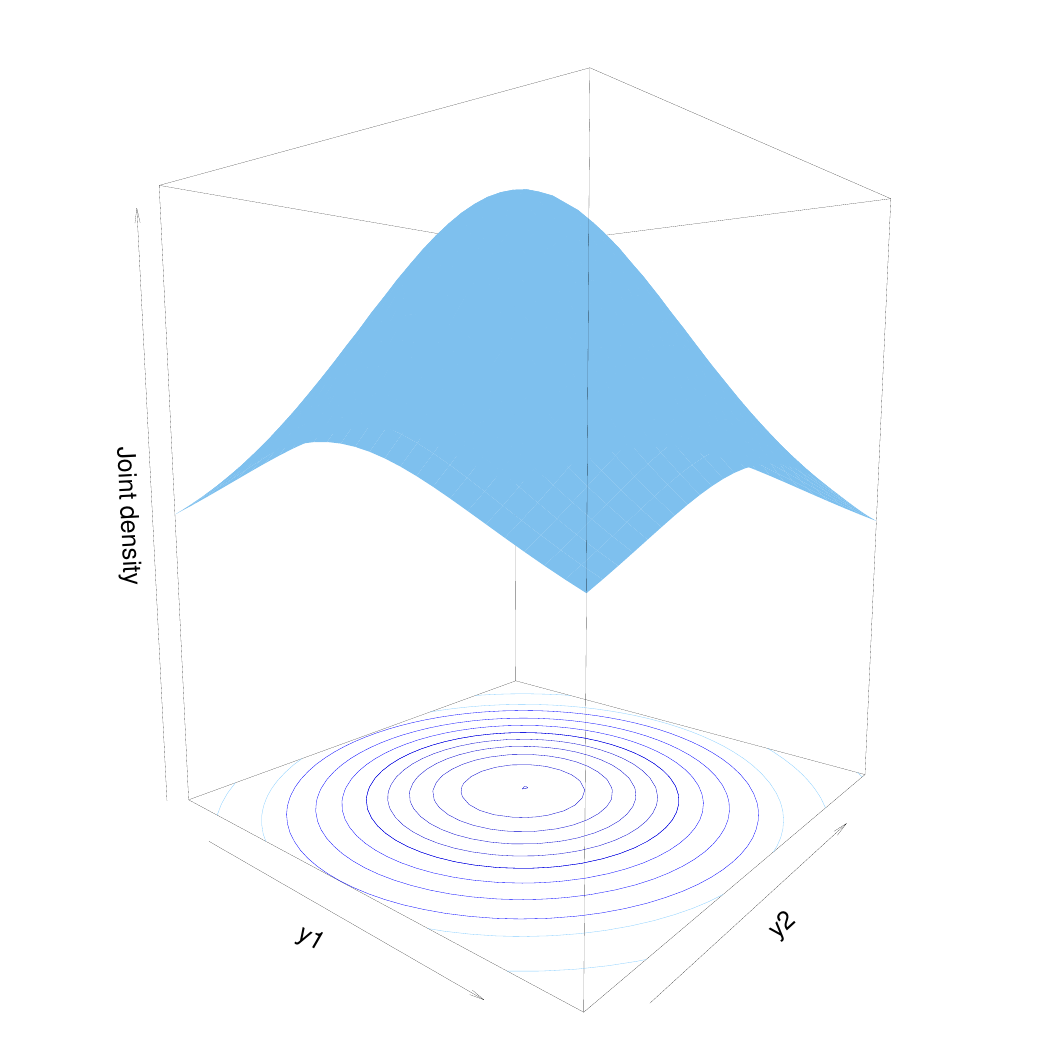}
        \caption{Student’s-t distribution with $\nu=9$.}
        \label{fig:t3}
    \end{subfigure}
    \begin{subfigure}[b]{0.32\textwidth}            
        \centering
        \includegraphics[width=\textwidth]{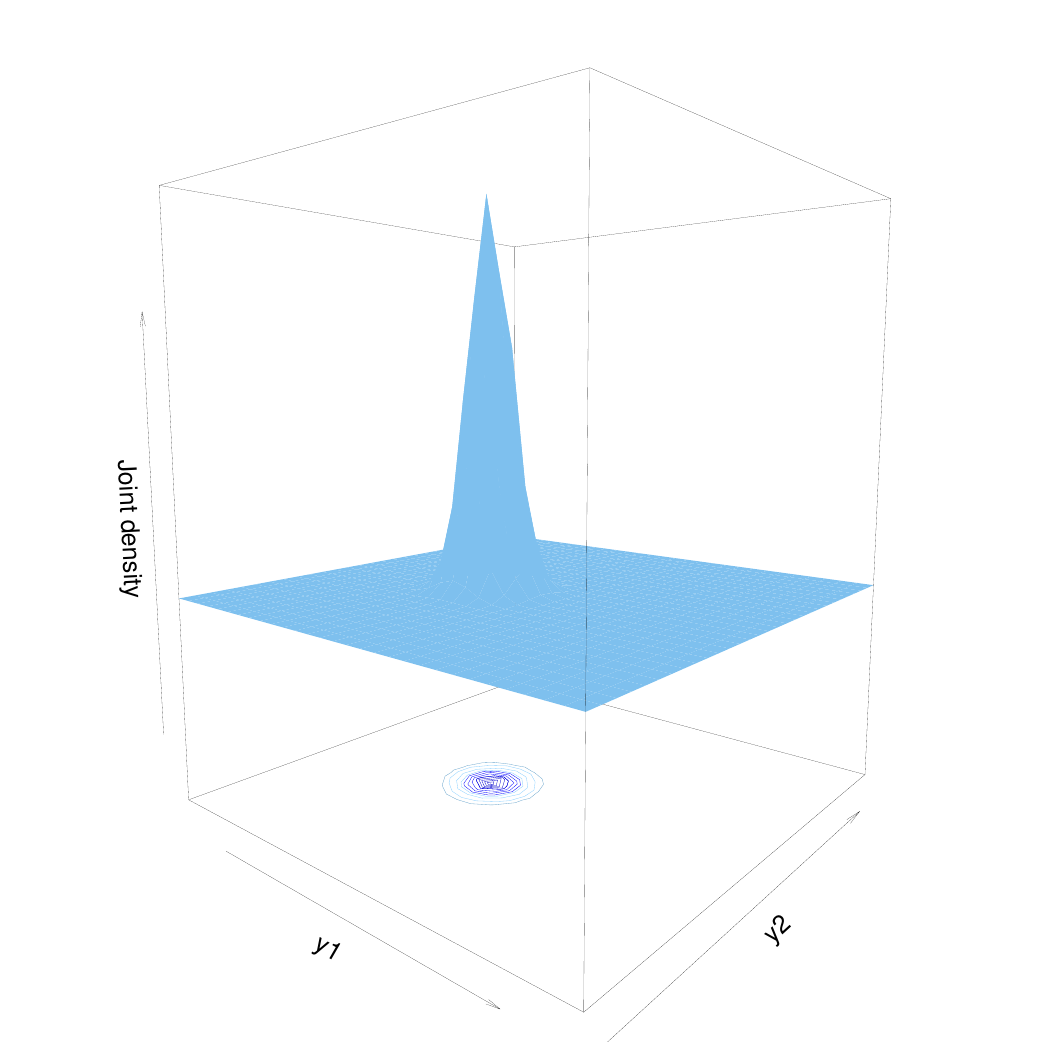}
        \caption{CN distribution with $\nu_1=0.1$ and $\nu_2=0.1$.}
        \label{fig:cn1}
    \end{subfigure}
    \begin{subfigure}[b]{0.32\textwidth}
        \centering
        \includegraphics[width=\textwidth]{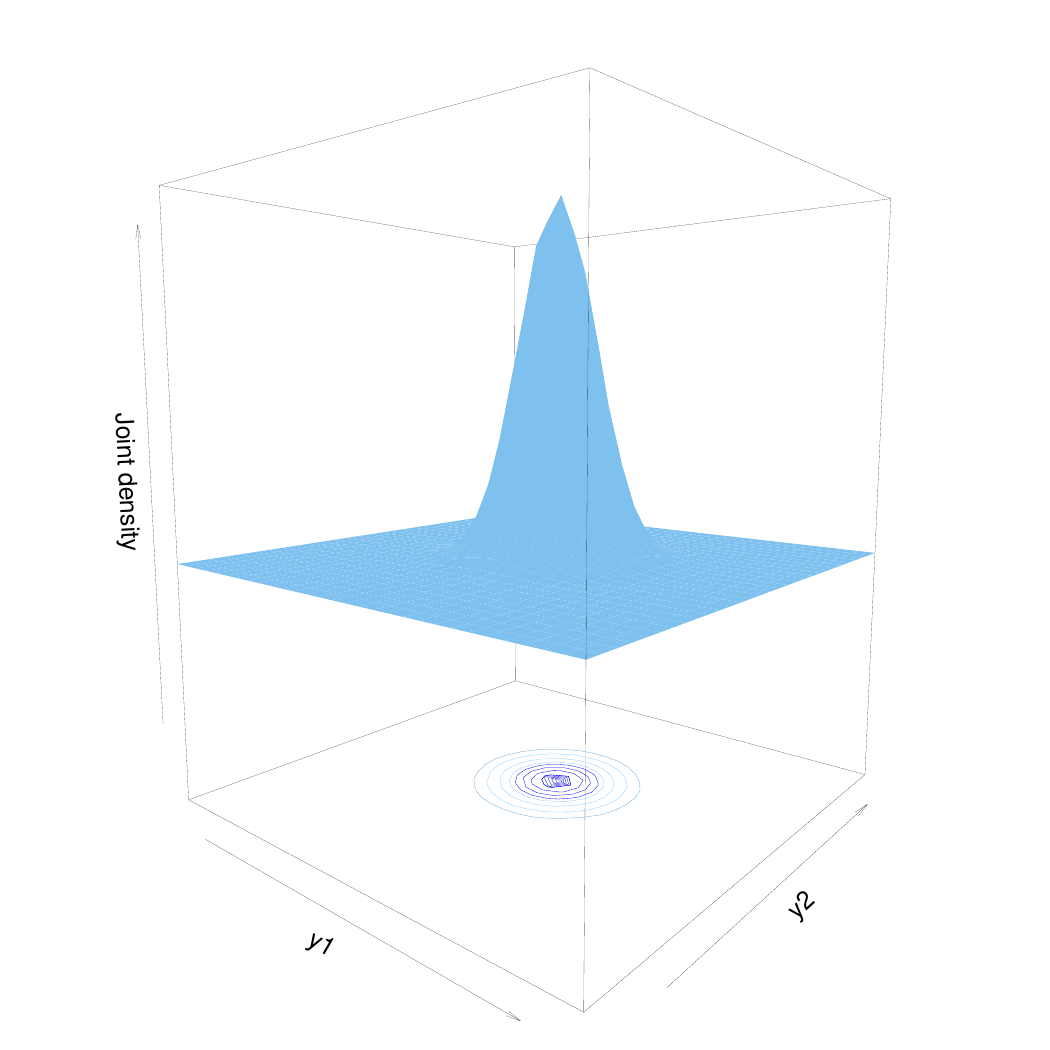}
        \caption{CN distribution with $\nu_1=0.3$ and $\nu_2=0.3$.}
        \label{fig:cn2}
    \end{subfigure}
    \begin{subfigure}[b]{0.32\textwidth}
        \centering
        \includegraphics[width=\textwidth]{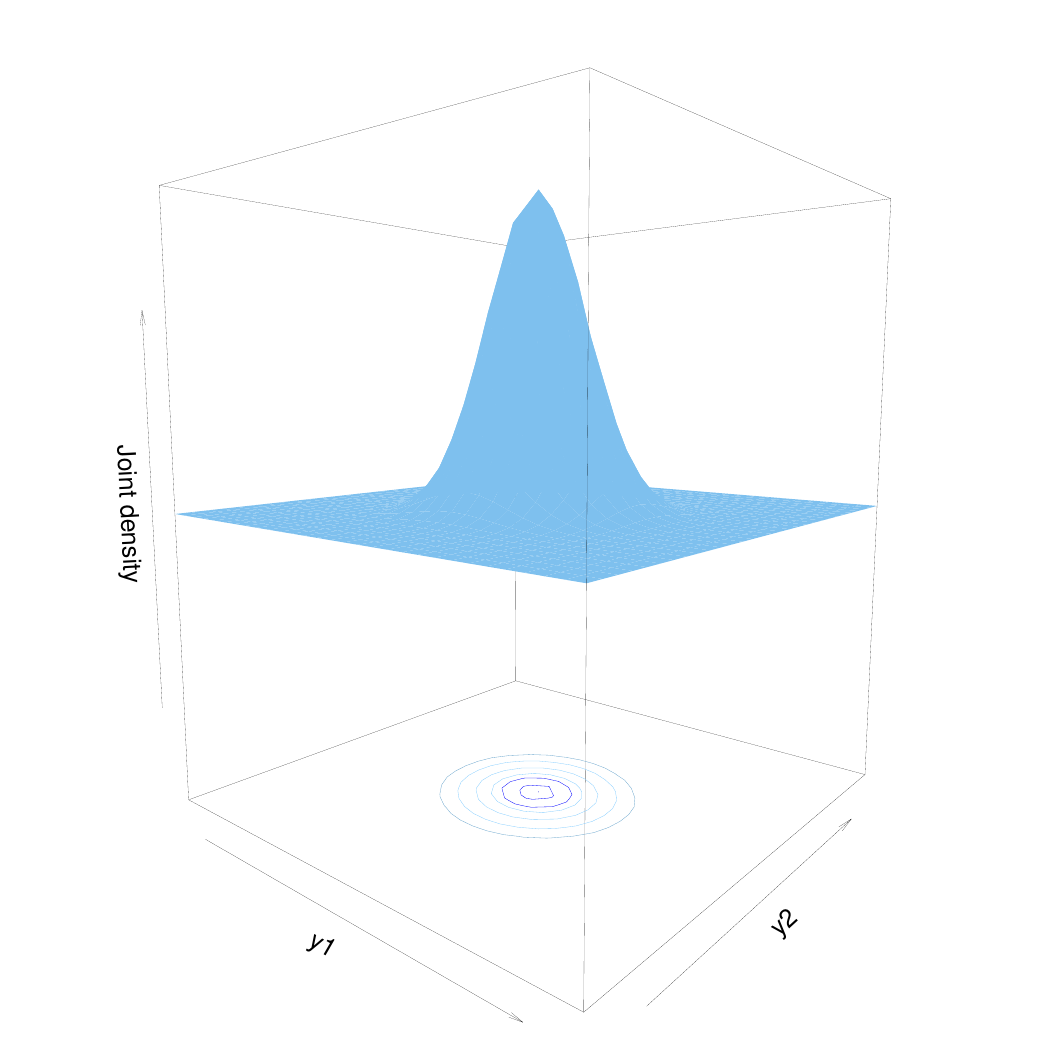}
        \caption{CN distribution with $\nu_1=0.5$ and $\nu_2=0.5$.}
        \label{fig:cn3}
    \end{subfigure}
\caption{Perspective plots overlaid on contour plots for three distributions with different scales. The location parameter is set as $\mu = (0,0)^\top$ for all distributions, and the scale matrix is set as the identity matrix, $\bSigma = \mathbf{I}_2$, for the Student’s-t and contaminated normal (CN) distributions.}
\label{fig:ContourPlot}
\end{figure}

The following properties of the  $p$-variate Student's-$t$ distribution are useful for our theoretical developments. We start with the
marginal-conditional decomposition of a  $p$-variate Student's-$t$ random vector. The proof of the following propositions can be found in
\citet{ArellanoBolfarine95}.

\begin{proposition}\label{prop1}
	Let  $\X\sim {\cal T}_p(\bmu,\bSigma,\nu)$ \CG{partitioned} as
	$\X^{\top}=(\X^{\top}_1,\X^{\top}_2)^{\top}$ \CG{with} $dim(\X_1) =
	p_1$, $dim(\Y_2) = p_2$, \CG{where} $p_1 + p_2 = p$. Let
	$\bmu=(\bmu^{\top}_1,\bmu^{\top}_2)^{\top}$ and
	$\bSigma=\left(\begin{array}{cc}
	\bSigma_{11} & \bSigma_{12} \\
	\bSigma_{21} & \bSigma_{22}
	\end{array}
	\right)$ be the corresponding
	partitions of \CG{$\bmu$ and $\bSigma$}. Then, we have\\
	{\rm (i)} $\X_1\sim {\cal T}_{p_1}(\bmu_1,\bSigma_{11},\nu)$; and \\
	{\rm (ii)} 		$
		\X_2 \mid \CG{(\X_1=\xp_1)}\sim {\cal T}_{p_2}\left(\bmu_{2.1},\widetilde{\bSigma}_{22.1},\nu+p_1
		\right),
		$\\
		where \CG{$\bmu_{2.1}=\bmu_2+\bSigma_{21}\bSigma^{-1}_{11}(\xp_1-\bmu_1)$
			and}
		$\widetilde{\bSigma}_{22.1}=\left(\displaystyle\frac{\nu+\delta_1}{\nu+p_1}\right)\bSigma_{22.1}$,
		\CG{with }$\delta_1=(\xp_1-\bmu_1)^{\top}\bSigma_{11}^{-1}(\xp_1-\bmu_1)$
		\CG{and }
		$\bSigma_{22.1}=\bSigma_{22}-\bSigma_{21}\bSigma^{-1}_{11}\bSigma_{12}$.
\end{proposition}

\subsubsection{The multivariate contaminated normal  distribution}

In this case, $U\sim \nu_1I_{\nu_2}(U)+(1-\nu_1)I_1(U)$, where $I_{\nu_2}(U)$ is an indicator variable with value 1 if $U=\nu_2$ and $0$ otherwise,  and $I_{1}(U)$ is an indicator variable with value 1 if $U=1$ and $0$ otherwise. 
A random variable $\X$ having a $p$-variate contaminated normal (MCN) distribution with
location vector $\bmu$, positive-definite scale-covariance
matrix $\bSigma$, proportion of bad points $\nu_1\in (0,1)$ and degree of contamination $\nu_2\in (0,1)$, denoted by $\X \sim
{\cal CN}_p(\bmu,\bSigma,\nu_1,\nu_2)$, has the following pdf:
$$
\psi^{CN}_p(\xp\mid \bmu,\bSigma,\nu_1,\nu_2)=\nu_1\phi_p(\xp\mid\bmu,\nu_2^{-1}\bSigma)+(1-\nu_1)\phi_p(\xp\mid\bmu,\bSigma),\,\,\xp\in {\mathbb R}^p.\label{lsdefAB1}
$$
Let $\Phi_p(\ap,\bp\mid\bmu,\bSigma)$ represent
$$\Phi_p(\ap,\bp\mid\bmu,\bSigma)=\int_{\ap}^{\bp}{{\phi}_p}(\x\mid\bmu,\bSigma)\textrm{d}\bx,$$ where  $\mathbf{a}$  and $\mathbf{b}$ as denoted in (\ref{hyper1}). Thus, we write
$$
\Psi^{\CN}_p(\ap,\bp\mid \bmu,\bSigma,\nu_1,\nu_2)=\nu_1\Phi_p(\ap,\bp\mid\bmu,\nu_2^{-1}\bSigma)+(1-\nu_1)\Phi_p(\ap,\bp\mid~\bmu,\bSigma).\label{lsdefAB1}
$$
When $\ap=-\binfty$ we will  simply write $\Phi_p(\bp\mid~\bmu,\bSigma)$, and in this case the cdf of $\X$ is
$$
\Psi^{\CN}_p(\bp\mid \bmu,\bSigma,\nu_1,\nu_2)=\nu_1\Phi_p(\bp\mid\bmu,\nu_2^{-1}\bSigma)+(1-\nu_1)\Phi_p(\bp\mid\bmu,\bSigma).\label{lsdefAB1}
$$

It is \CB{known} that  as {$\nu_1 \to 0$} or  {$\nu_2 \to 1$}, the MCN distribution $\X$ converges in
distribution to the ${\cal N}_p(\bmu,\bSigma)$.  From (\ref{stoNI1}), it follows that $\EE[U\mid\X]=\nu_2\mathcal{ P}_{\nu_2}+\mathcal{ P}_1$ and $\EE[U^2\mid\X]=\nu^2_2\mathcal{ P}_{\nu_2}+\mathcal{ P}_1$ , where
\begin{equation}\label{propor}
    \mathcal{ P}_{\nu_2}=P(U=\nu_2~|~\X)=\frac{\nu_1\phi_p(\xp\mid~\bmu,\nu_2^{-1}\bSigma)}{\psi^{CN}_p(\xp\mid ~\bmu,\bSigma,\nu_1,\nu_2)}\,\,\,{\it and}\,\,\, \mathcal{ P}_{1}=P(U=1~|~\X)=1- \mathcal{ P}_{\nu_2}.
\end{equation}

The following properties are related to the marginal-conditional decomposition of a  $p$-variate MCN random vector. The proof of the following propositions can be found in \citet{wang2023multivariate}.

\begin{proposition}\label{prop2}
	Let  $\X\sim {\cal CN}_p(\bmu,\bSigma,\nu_1,\nu_2)$ partitioned as
	$\X^{\top}=(\X^{\top}_1,\X^{\top}_2)^{\top}$ with $dim(\X_1) =
	p_1$, $dim(\Y_2) = p_2$, where $p_1 + p_2 = p$. Let
	$\bmu=(\bmu^{\top}_1,\bmu^{\top}_2)^{\top}$ and
	$\bSigma=\left(\begin{array}{cc}
	\bSigma_{11} & \bSigma_{12} \\
	\bSigma_{21} & \bSigma_{22}
	\end{array}
	\right)$ be the corresponding
	partitions of $\bmu$ and $\bSigma$. Then, we have\\
	
	{\rm (i)} $\X_1\sim {\cal CN}_{p_1}(\bmu_1,\bSigma_{11},\nu_1,\nu_2)$; and \\

	{\rm (ii)}	$\X_2 \mid (\X_1=\xp_1)\sim {\cal CN}_{p_2}\left(\bmu_{2.1},{\bSigma}_{22.1},\omega_{\nu_2},\nu_2
		\right),
		$
  
		where $\bmu_{2.1}=\bmu_2+\bSigma_{21}\bSigma^{-1}_{11}(\xp_1-\bmu_1)$
		and 
		$\bSigma_{22.1}=\bSigma_{22}-\bSigma_{21}\bSigma^{-1}_{11}\bSigma_{12}$, and $\omega_{\nu_2}=\displaystyle\frac{\nu_1\phi_{p_1}(\xp_1\mid~\bmu_1,\nu_2^{-1}\bSigma_{11})}{\psi^{CN}_{p_1}(\xp_1\mid ~\bmu_1,\bSigma_{11},\nu_1,\nu_2)}.$

\end{proposition}

{Fig.\ref{fig:ContourPlot} presents the perspective plots overlaid on the contour plots for three bivariate distributions: normal, Student's-t, and contaminated normal (CN) distributions. The location parameter is set as $\bmu = (0,0)^\top$ for all distributions. The first row illustrates different variances for the normal distribution, the second row shows varying degrees of freedom ($\nu$) for the Student's-t distribution, and the third row presents CN distribution with different $\nu_1$ and $\nu_2$. The scale matrix is set as the identity matrix, $\bSigma = \mathbf{I}_2$, for the Student’s-t and CN distributions. These visualizations help us to understand how different distributional assumptions affect joint density, particularly in terms of tail behavior and sensitivity to outliers.}

\section{Heckman selection models}\label{SLnModel}

The Heckman SL model, or missing data, is common in applied research.  The Heckman SL model consists of a linear
equation for the outcome and a probit equation for the sample selection mechanism. The outcome equation is
\begin{equation}\label{1HS}
Y_{1i}=\x^{\top}_i\bbeta+\epsilon_{1i},
\end{equation}
and the sample selection mechanism is characterized by the following latent linear equation, for $i \in \{1, \ldots, n\}$,
\begin{equation}\label{2HS}
Y_{2i}=\w^{\top}_i\bgamma+\epsilon_{2i}.
\end{equation}
The vectors $\bbeta\in {\mathbb R}^p$ and 
$\bgamma \in {\mathbb R}^q$ are unknown regression parameters. $\x^{\top}_i=(x_{i1},\ldots,x_{ip})$ and  $\w^{\top}_i=(w_{i1},\ldots,w_{iq})$ are known characteristics. The covariates in $\x_i$ and $\w_i$ may overlap with each other, and the exclusion restriction holds when at least
one of the elements of $\w_i$ is not in $\x_i$. 

Thus, the observed data for the
$i$th subject is given by $(\bV_i, C_i),$ where $\bV_i=(V_{1i},V_{2i})$ represents
the vector of censored readings  and $C_i=I_{\{Y_{2i}>0\}}$ is the censoring indicators. In other words,

\begin{equation}
Y_{1i}=
\begin{cases}
 V_{1i} & \mbox{if $C_{i}=1$,} \\
 V_{2i} = \texttt{NA} & \mbox{if $ C_{i}=0$,}
\end{cases}
\label{CensL1}
\end{equation}\\

for all $i \in \{ 1, \ldots, n\}$.  
Note that $V_{2i}=\texttt{NA}$ is equivalent to write $-\infty<V_{2i}< \infty$. 

\subsection{The classical Heckman sample selection model}
\citet{heckman1979} assumes independent 
bivariate normal distribution for the error terms (SLn), as follows:
		\begin{eqnarray}\label{nerror}
		\left(\begin{array}{c}
			\epsilon_{1i}  \\
			\epsilon_{2i} \\
		\end{array}
		\right) \sim {\mathcal N}_{2}\left( \bm{0},
		\bSigma
		\right), \; \bSigma=\left(\begin{array}{cc}
			\sigma^2 & \rho\sigma \\
			\rho\sigma & 1 \\
		\end{array}\right),
	\end{eqnarray}
where the second diagonal element equals 1 in order to achieve full identifiability.  The SLn model (\ref{1HS})-(\ref{nerror}) is known as "Type 2 tobit model" in the econometrics literature  and is sometimes also referred to as the “Heckman model.” Absence of selection effect ($\rho=0$) implies that the outcomes are missing at random, and the observed outcomes are representative for inference of the population given the observed covariates.

To obtain the likelihood function of the SLn model, first note that if $C_i=1$, then $Y_{1i}\sim \mathcal{N}(\x^{\top}_i\bbeta,\sigma^2)$ and $Y_{2i}|Y_{1i}=V_{1i}\sim \mathcal{N}(\mu_{c},\sigma^2_c)$, where

$$\mu_c=\w^{\top}_i\bgamma+\frac{\rho}{\sigma}(V_{1i}-\x^{\top}_i\bbeta),\,\,\sigma^2_c=(1-\rho^2).$$
Thus, the contribution to the likelihood is
\begin{eqnarray*}
f(Y_{1i}\mid\btheta)P(Y_{2i}>0\mid Y_{1i}=V_{1i})&=&\phi\left(V_{1i}\mid\x^{\top}_i\bbeta,\sigma^2\right)\Phi\left(\displaystyle\frac{\mu_{c}}{\sigma_c}\right)\\
&=&\phi(V_{1i}\mid\x^{\top}_i\bbeta,\sigma^2)\Phi\left(\displaystyle\frac{\w^{\top}_i\bgamma+\displaystyle\frac{\rho}{\sigma}(V_{1i}-\x^{\top}_i\bbeta)}{\sqrt{1-\rho^2}}\right).
\end{eqnarray*}
If $C_i=0$, then the contribution in the likelihood is
$$P(Y_{2i}\leq 0)=\Phi(-\w^{\top}_i\bgamma).$$

Therefore, the likelihood function of $\btheta=(\bbeta^{\top},\bgamma^{\top},\sigma^2,\rho)^{\top}$ is
\begin{align}
L_{\text{SLn}}(\btheta\mid\bV,\bC)&=\prod_{i=1}^{n}\left\{\phi(V_{1i}\mid\x^{\top}_i\bbeta,\sigma^2)\Phi\left(\displaystyle\frac{\w^{\top}_i\bgamma+\displaystyle\frac{\rho}{\sigma}(V_{1i}-\x^{\top}_i\bbeta)}{\sqrt{1-\rho^2}}\right)\right\}^{C_i}\left\{\Phi(-\w^{\top}_i\bgamma)\right\}^{1-C_i},\label{equ8.1}
\end{align}
where   $\bV=(\bV_1,\ldots,\bV_n)$
and $\bC=(C_1,\ldots,C_n)$.  The log-likelihood function for the observed data is given by $\ell_{\text{SLn}}(\btheta)=\ell_{\text{SLn}}(\btheta\mid\bV,\bC)=\ln L_{\text{SLn}}(\btheta\mid\bV,\bC)$,

The SLn model presented in this section has faced criticism in the literature due to its sensitivity to the assumption of normality. In the subsequent sections, we will establish a new connection between the SL model and heavy-tailed distributions, such as the Student's-t distribution (SLt) introduced by \citet{marchenko2012heckman} and the contaminated normal (SLcn) introduced by \citet{lim2024heckman}.

\subsection{The Heckman selection-t model} \label{SLtModel}

In order to accommodate for heavy-tailedness, \citet{marchenko2012heckman} proposed the SLt model, replacing the normal assumption of error terms in (\ref{nerror}) by an independent bivariate Student'-t distribution with an unknown number of degrees of freedom $\nu$:
	\begin{eqnarray}\label{modeleqt}
	\left(\begin{array}{c}
	\epsilon_{1i}  \\
	\epsilon_{2i} \\
	\end{array}
	\right)\sim {\cal T}_{2}\left( \bm{0},
	\bSigma,\nu
	\right), \; i \in \{0, \ldots, n\},
	\end{eqnarray}
where $\bSigma$ as defined in (\ref{nerror}).
To obtain the likelihood function of the SLt model, first  from Proposition \ref{prop1}  note that if $C_i=1$, then $Y_{1i}\sim {\cal T}(\x^{\top}_i\bbeta,\sigma^2,\nu)$ and $Y_{2i}|Y_{1i}=V_{1i}\sim {\cal T}(\mu_{ti},\sigma^2_{ti},\nu+1)$, where
$$\mu_{ti}=\w^{\top}_i\bgamma+\frac{\rho}{\sigma}(V_{1i}-\x^{\top}_i\bbeta),\,\,\sigma^2_{ti}=\frac{\nu+\delta(V_{1i})}{\nu+1}(1-\rho^2),$$
where $\delta(V_i)={(V_{1i}-\x^{\top}_i\bbeta)^2}/{\sigma^2}$. Thus, the contribution in the likelihood function of $\btheta=(\bbeta^{\top},\bgamma^{\top},\sigma^2,\rho,\nu)^{\top}$, given the observed sample $(\bV, \C)$, is
$$f(Y_{1i}\mid\btheta)P(Y_{2i}>0\mid Y_{1i}=V_i)=\psi^T(V_{1i}\mid\x^{\top}_i\bbeta,\sigma^2,\nu){\Psi^T}(-\infty,0\mid-\mu_{ti},\sigma^2_{ti},\nu+1).$$

If $C_i=0$, then the contribution to the likelihood function is
$$P(Y_{2i}\leq 0)=\Psi^T(-\infty,0\mid\w^{\top}_i\bgamma,1,\nu).$$

Therefore, the likelihood function of $\btheta=(\bbeta^{\top},\bgamma^{\top},\sigma^2,\rho,\nu)^{\top}$,  given the observed sample $(\bV, \C)$, is
\begin{align}
L_{\text{SLt}}(\btheta\mid\bV,\bC)&=\prod_{i=1}^{n}\left\{\psi^T(V_{1i}\mid\x^{\top}_i\bbeta,\sigma^2,\nu)\Psi^T(-\infty,0\mid -\mu_{ti},\sigma^2_{ti},\nu+1)\right\}^{C_i}\nonumber \\
& ~~~~~~~~\times \left\{\Psi^T(-\infty,0\mid\w^{\top}_i\bgamma,1,\nu)\right\}^{1-C_i},\label{equ8.2}
\end{align}
where   $\bV=(\bV_1,\ldots,\bV_n)$
and $\bC=(C_1,\ldots,C_n)$.  The log-likelihood function for the observed data is given by $\ell_{\text{SLt}}(\btheta)=\ell_{\text{SLt}}(\btheta\mid\bV,\bC)=\ln L_{\text{SLt}}(\btheta\mid\bV,\bC)$, that is,
\begin{align}
\ell_{\text{SLt}}(\btheta)&=\sum_{i=1}^{n}\left\{C_i\ln \psi^T(V_{1i}\mid\x^{\top}_i\bbeta,\sigma^2,\nu)
 C_i\ln \Psi^T(-\infty,0\mid-\mu_{ti},\sigma^2_{ti},\nu+1)\right\} \nonumber \\
 &~~~~~~ + \sum_{i=1}^{n}(1-C_i)\ln\left\{ \Psi^T(-\infty,0\mid\w^{\top}_i\bgamma,1,\nu)\right\}. \label{equ8.3}
\end{align}

\subsection{The Heckman selection-contaminated normal model} \label{SLcnModel}

To protect the reference SLn model from the occurrence of atypical points and heavier-than-normal error tails, \citet{lim2024heckman} have proposed a novel Heckman selection-contaminated normal model, replacing the normal assumption of error terms in \eqref{nerror} by a bivariate CN distribution as follows:

	\begin{eqnarray}\label{modeleqt2}
	\left(\begin{array}{c}
	\epsilon_{1i}  \\
	\epsilon_{2i} \\
	\end{array}
	\right)\sim {{\cal CN}_{2}}\left( \bm{0},
	\bSigma,\nu_1,\nu_2
	\right), \; i \in \{0, \ldots, n\},
	\end{eqnarray}
where $\bSigma$ as defined in (\ref{nerror}). 

To obtain the likelihood function of the SLcn model, first  from Proposition \ref{prop2}  note that if $C_i=1$, then $Y_{1i}\sim {\cal CN}(\x^{\top}_i\bbeta,\sigma^2,\nu_1,\nu_2)$ and $Y_{2i}\mid Y_{1i}=V_{1i}\sim {\cal CN}(\mu_{ti},\sigma^2_{t},\omega_{\nu_{2i}},\nu_2)$, where
\begin{equation}\label{condiy2}
\mu_{ti}=\w^{\top}_i\bgamma+\frac{\rho}{\sigma}(V_{1i}-\x^{\top}_i\bbeta),\,\,\sigma^2_{t}=(1-\rho^2),\,\, \omega_{\nu_{2i}}=\frac{\nu_1\phi(V_{1i}\mid\x^{\top}_i\bbeta,\nu_2^{-1}\sigma^2)}{\psi^{CN}(V_{1i}\mid\x^{\top}_i\bbeta,\sigma^2,\nu_1,\nu_2)}.
\end{equation}
Thus, the contribution in the likelihood function of $\btheta=(\bbeta^{\top},\bgamma^{\top},\sigma^2,\rho,\nu_1,\nu_2)^{\top}$, given the observed sample $(\bV, \C)$, is
$$f(Y_{1i}|\btheta)P(Y_{2i}>0|Y_{1i}=V_i)=\psi^{CN}(V_{1i}\mid\x^{\top}_i\bbeta,\sigma^2,\nu_1,\nu_2)\Psi^{CN}(-\infty,0\mid-\mu_{ti},\sigma^2_{t},\omega_{\nu_{2i}},\nu_2).$$

If $C_i=0$, then the contribution to the likelihood function is
$$P(Y_{2i}\leq 0)=\Psi^{CN}(-\infty,0\mid\w^{\top}_i\bgamma,1,\nu_1,\nu_2).$$

Therefore, the likelihood function of $\btheta=(\bbeta^{\top},\bgamma^{\top},\sigma^2,\rho,\nu_1,\nu_2)^{\top}$,  given the observed sample $(\bV, \C)$, is
\begin{align}
L_{\text{SLcn}}(\btheta\mid\bV,\bC)&=\prod_{i=1}^{n}\left\{\psi^{CN}(V_{1i}\mid\x^{\top}_i\bbeta,\sigma^2,\nu_1,\nu_2)\Psi^{CN}(-\infty,0\mid -\mu_{ti},\sigma^2_{t},\omega_{\nu_{2i}},\nu_2)\right\}^{C_i} \nonumber\\
&~~~~~~\times
`\left\{\Psi^{CN}(-\infty,0\mid\w^{\top}_i\bgamma,1,\nu_1,\nu_2)\right\}^{1-C_i},\label{equ8.4}
\end{align}
where   $\bV=(\bV_1,\ldots,\bV_n)$
and $\bC=(C_1,\ldots,C_n)$.  The log-likelihood function for the observed data is given by $\ell_{\text{SLcn}}(\btheta)=\ell_{\text{SLcn}}(\btheta\mid\bV,\bC)=\ln L_{\text{SLcn}}(\btheta\mid\bV,\bC)$, that is,
\begin{eqnarray}
    \ell_{\text{SLcn}}(\btheta)&=&\sum_{i=1}^{n}\left\{C_i\ln \psi^{CN}(V_{1i}\mid\x^{\top}_i\bbeta,\sigma^2,\nu_1,\nu_2)+C_i\ln \Psi^{CN}(-\infty,0\mid -\mu_{ti},\sigma^2_{t},\omega_{\nu_{2i}},\nu_2)\right\}\nonumber\\
    &&+\sum_{i=1}^{n}(1-C_i)\ln\left\{\Psi^{CN}(-\infty,0\mid\w^{\top}_i\bgamma,1,\nu_1,\nu_2)\right\}. \label{equ8.5}
\end{eqnarray}

\subsubsection{Automatic inlier/outlier detection}
\label{sec:Automatic mild outlier detection}

An outlier is defined as an observation with a low posterior probability of being generated by the linear model, while an inlier is an observation that aligns well with the model and has a relatively high posterior probability of generation. Following the work of \citet{lim2024heckman}, let ${\varepsilon}_i$ represent the posterior probability, as described in Equation (\ref{propor}), that the $i$th unit belongs to the contaminated normal component of the MCN distribution. If ${\varepsilon}_i > 0.5$, then unit $i$ is considered an outlier; otherwise, it is classified as an inlier.

In practice, the term ${\varepsilon}_i$ can be calculated using the posterior probability for every unit in the sample, regardless of whether $Y_{1i}$ is observed ($C_i=1$) or not ($C_i=0$). For further exploration of detecting inliers and outliers, readers may refer to the studies by  \citet{falkenhagen2019likelihood}, \citet{mazza2020mixtures}, and \citet{punzo2017robust}. This feature is unique to the SLcn model, as it is not possible to establish an automatic inlier/outlier detection rule in the SLn and SLt models.

\section{Bayesian inference}\label{BI}

This section outlines the Bayesian estimation approach for the SL model, incorporating both the likelihood function and the prior distributions of the parameters. The selection of prior distributions can vary depending on the available knowledge and assumptions. When combined with the likelihood, these priors yield the posterior distributions, which provide parameter estimates and help quantify uncertainty through credible intervals. The following implementation will focus on the SLn, SLt, and SLcn models.

\subsection{Commonly used priors}\label{priors}

Prior knowledge plays a crucial role in incorporating existing knowledge or expert opinions into the model, which is particularly important for heavy-tailed distributions. In such cases,  well-specified priors enhance the convergence properties of MCMC algorithms. We introduce prior distributions for all unknown parameters $\btheta=(\bbeta,\bgamma,\sigma^2, \rho, \bnu)$.
Specifically, we define the weakly informative prior as {$\bbeta \sim {\cal N}_p(0, 10^2\mathbf{I}_p)$, $\bgamma \sim {\cal N}_q (0, 10^2\mathbf{I}_q)$}, $\rho \sim \text{Uniform}(-1, 1)$, $\sigma^2 \sim \text{Cauchy}(0, 4^2)$. 
Also, the prior on $\nu$ for the specific models in subsections \ref{SLtModel} and \ref{SLcnModel} was chosen as follows: $\nu \sim {\cal T}(0, 5, 4)$ for the SLt model, and $\nu_1 \sim \text{Beta}(2, 6)\text{ and } \nu_2 \sim \text{Beta}(2, 12)$ for SLcn model.
Therefore, we consider an independent structure where the following joint prior densities for each model are specified.

\begin{enumerate}[(i)]
    \item For SLn model, $p_{\text{SLn}}(\bbeta,\bgamma, \sigma^2, \rho) \propto p(\bbeta) p(\bgamma) p(\sigma^2) p(\rho)$.
    \item For SLt model, \(p_{\text{SLt}}(\bbeta,\bgamma, \sigma^2, \rho, \nu) \propto p(\bbeta) p(\bgamma) p(\sigma^2) p(\rho) p(\nu)\).
    \item For SLcn model, \(p_{\text{SLcn}}(\bbeta,\bgamma, \sigma^2, \rho, \nu_1, \nu_2) \propto p(\bbeta) p(\bgamma) p(\sigma^2) p(\rho) p(\nu_1) p(\nu_2)\).
\end{enumerate}

\subsection{Posterior analysis}\label{BI2}
Assuming the components of the parameter vector are independent, the posterior distributions are derived by combining the likelihood from equations (\ref{equ8.1}), (\ref{equ8.3}), and (\ref{equ8.4}) with the joint prior densities provided in the previous Subsection. Thus, 

\begin{equation}\label{post1}
\pi_{\text{SLn}}(\btheta|\y) \propto  p_{\text{SLn}}(\bbeta,\bgamma, \sigma^2,\rho)  L_{\text{SLn}}(\btheta\mid\bV,\bC),
\end{equation}

\begin{equation}\label{post2}
\pi_{\text{SLt}}(\btheta|\y) \propto    p_{SLt}(\bbeta,\bgamma, \sigma^2,\rho, \nu)  L_{\text{SLt}}(\btheta\mid\bV,\bC),
\end{equation}

\begin{equation}\label{post3}
\pi_{\text{SLcn}}(\btheta|\y) \propto   p_{\text{SLcn}}(\bbeta,\bgamma, \sigma^2,\rho, \nu_1, \nu_2)  L_{\text{SLcn}}(\btheta\mid\bV,\bC),
\end{equation}
where the left side, $\pi_{\text{model}}(\btheta|\y)$, denotes the posterior distribution, the right side corresponds to the joint prior densities, and $ L_{\text{model}}(\btheta\mid\bV,\bC)$ denotes the likelihood of the specific model foe each model.

\subsection{{Implementation in \texttt{Stan}}}\label{BI3}

\texttt{Stan} is a probabilistic programming language and compiler for Bayesian inference, and \texttt{rstan} provides an R interface to Stan, enabling users to specify models in \texttt{Stan}, execute posterior sampling, and analyze MCMC outputs directly within the R environment. For posterior computation, we employ Hamiltonian Monte Carlo (HMC) \citep{hoffman2014no}, which utilizes Hamiltonian dynamics to efficiently explore high-dimensional parameter spaces.

The Bayesian framework combines the priors specified in Section~\ref{priors} with the likelihood functions (Equations~\ref{post1}--\ref{post3}) corresponding to the SLn, SLt, and SLcn models. While these likelihoods differ due to their error distributions, they yield distinct posterior densities. Posterior inference from these posteriors is consistently carried out using the HMC algorithm, which efficiently samples from the resulting high-dimensional distributions. Model-specific pseudo-code templates in the Supplementary Material provide details of prior definitions. \texttt{Stan} requires only the specification of the joint log-posterior; all gradient calculations are performed automatically using reverse-mode autodifferentiation. Thus, no manually derived full conditional distributions are needed. Posterior diagnostics were monitored using standard \texttt{Stan} outputs, confirming convergence across all simulation scenarios and empirical applications: $\hat{R}$ values were consistently near 1, effective sample sizes exceeded 2000, no divergent transitions were observed, and acceptance rates averaged between 0.88 and 0.92. Finally, we developed the \textsf{R} package \texttt{Heckmanstan}, which leverages \texttt{rstan} to provide a practical framework for Bayesian analysis of SL models.\\
To assess the robustness of inference under alternative prior specifications for the degrees of freedom parameter $\nu$ for the SLt model and contamination parameter $\nu_1$ for the SLcn model, we conducted a prior sensitivity analysis. Results in the Supplementary materials indicate that posterior estimates remained stable across diverse prior choices, supporting the reliability of our conclusions.

\subsection{Bayesian model selection criteria}\label{sec modsel diag}

To compare model performance in the context of Bayesian estimation, we employ three widely used criteria\citep{Ando.2010}: the Leave-One-Out Information Criterion (LOOIC), the Watanabe-Akaike Information Criterion (WAIC), and the Conditional Predictive Ordinate (CPO). These methods are particularly suitable for Bayesian models estimated via MCMC, as they provide fully Bayesian assessments of predictive accuracy while accommodating complex posterior geometries and non-conjugate structures. We do not consider the Deviance Information Criterion (DIC) due to its known limitations in non-Gaussian and hierarchical models, as discussed in \citet{vehtari2017practical, watanabe2010asymptotic, celeux2006deviance}. WAIC and LOOIC are computed using the \texttt{waic} and \texttt{loo} functions from the \texttt{loo} package, respectively. CPO is calculated by extracting the log inverse likelihood from \texttt{Rstan}. These three criteria are employed throughout our simulations and real-data applications, as they offer sample-wise assessments of predictive performance.

LOOIC \citep{vehtari2017practical} is designed to assess the predictive performance of a model by excluding one data point at a time and calculating how well the model predicts the excluded point. It is based on leave-one-out cross-validation, and smaller values indicate a better model. Let $\y=\{y_1,\ldots,y_n \}$ be an
observed random sample from $p(\cdot|\btheta)$, with the likelihood function expressed as $L(\btheta\mid\bV,\bC)$ in Section~\ref{SLnModel}, and the posterior distribution $\pi(\btheta \mid \y)$. The \textrm{LOOIC} is defined as

\begin{align} \label{eqn LOOIC formula}
\textrm{LOOIC} &= \sum_{i=1}^{n} \left( \log p(y_i \mid \y) - \log p(y_i \mid y_{(-i)}) \right) \notag \\
&=\sum_{i=1}^{n} \left( \int \log p(y_i \mid \btheta) \pi(\btheta \mid \y) \, d\btheta - \int \log p(y_i \mid \btheta) \pi(\btheta \mid y_{(-i)}) \, d\btheta \right).
\end{align}

The notation \( y_{(-i)} \) signifies that the \( i \)th data point where $i=1,2,\ldots,n$  has been omitted from the dataset $\y$. As the number of model parameters increases, the in-sample predictive performance, denoted by \( \log p(y_i \mid y) \), improves more rapidly than the out-of-sample predictive performance, represented by \( \log p(y_i \mid y_{(-i)}) \). This results in a growing sum of their pointwise differences. This behavior provides insight into why LOOIC reflects model simplicity. Its interpretation as an effective parameter count stems from this key observation: when priors are broad or flat, LOOIC aligns closely with the actual number of parameters, but it decreases when prior information is introduced \citep{vehtari2017practical}.

\citet{watanabe2010} proposed the \textit{Watanabe-Akaike information criterion} (WAIC), a model selection criterion balancing goodness of fit and complexity. WAIC, asymptotically equivalent to Bayes cross-validation loss, evaluates predictive performance by incorporating model complexity through the effective number of parameters, mitigating overfitting. Estimates the expected log-pointwise predictive density and is defined as:

\begin{equation} \label{eqn WAIC formula}
\textrm{WAIC} = \sum_{i=1}^{n} \mathrm{Var}_{p\left(\btheta \mid y\right)} [\log p(y_i \mid \btheta)]
\end{equation}

As the number of parameters increases, the epistemic uncertainty in the posterior also grows, causing a rise in the variance of posterior predictive quantities like \( \log p(y_n \mid \btheta) \). This is particularly relevant for models with hierarchical priors, where the extent of hierarchical shrinkage (i.e., the influence of the priors) depends on the data itself \citep{gelman2014understanding}. 

Conditional predictive ordinate
($\textrm{CPO}$) \citep{carlin2008bayesian} is based on the cross-validation criterion to compare the models.  For the $i$th observation, $i=1,2,\ldots,n$ the $\textrm{CPO}_i$ is  written as:
\begin{equation} \label{eqn CPO formula}
 \textrm{CPO}_i = \int p(y_i |\btheta)\pi(\btheta| \y_{(-i)})d\btheta = \left( \int \frac{\pi(\btheta| \y_{(-i)})}{p(y_i|\btheta)}d\btheta \right)^{-1},
\end{equation}
where $\y_{(-i)}$ is the sample  without the $i$th observation.
In the proposed model, the CPO does not have a closed-form expression. However, as shown in Equation (\ref{eqn CPO formula}), a Monte Carlo approximation can be obtained using an MCMC sample ${\btheta_1, \dots, \btheta_Q}$ drawn from the posterior distribution $\pi(\btheta|\y)$ after burn-in and thinning steps. The approximation is given by \citet{dipak1997} as: 
$\widehat{\textrm{CPO}_i} = \left( \frac{1}{Q} \sum\limits_{q=1}^Q  \frac{1}{p(y_i|\btheta_{q})} \right)^{-1}.$ A summary measure of the CPO values is the logarithm of the pseudo marginal likelihood (LPML), defined as: $\textrm{LPML} = \sum_{i=1}^n \log(\widehat{\textrm{CPO}_i}).$ Higher LPML values indicate a better model fit.

\begin{table}[!ht]
\caption{Simulation study 1. Mean estimates (ME), mean standard deviations (SD), and mean high posterior density (HPD) 95\% interval of the $100$ Monte Carlo replicates for the data generated from the normal distribution. }
	\label{tab:sim_normal}
\centering
\setlength{\tabcolsep}{3pt}
\begin{tabular}{cccccccccccccc}
  \toprule
 & & \multicolumn{4}{c}{SLn} & \multicolumn{4}{c}{SLt} & \multicolumn{4}{c}{SLcn} \\
 n & TRUE & ME & SD & \multicolumn{2}{c}{HPD (95\%)} & ME & SD & \multicolumn{2}{c}{HPD (95\%)}& ME & SD & \multicolumn{2}{c}{HPD (95\%)} \\ 
  \midrule
200 & $\beta_1=1$  & 1.132 & 0.207 & 0.739 & 1.538 & 1.135 & 0.200 & 0.757 & 1.529 & 1.137 & 0.199 & 0.758 & 1.528 \\
	& $\beta_2=0.5$  & 0.447 & 0.154 & 0.152 & 0.746 & 0.444 & 0.153 & 0.149 & 0.742 & 0.445 & 0.153 & 0.149 & 0.742 \\  
	& $\gamma_1=1$   & 1.033 & 0.124 & 0.797 & 1.272 & 1.136 & 0.159 & 0.843 & 1.445 & 1.284 & 0.339 & 0.817 & 1.911 \\
	& $\gamma_2=0.3$ & 0.297 & 0.122 & 0.062 & 0.532 & 0.330 & 0.139 & 0.064 & 0.600 & 0.373 & 0.185 & 0.046 & 0.727 \\   
	& $\gamma_3=-.5$ & -0.547 & 0.125 & -0.790 & -0.308 & -0.614 & 0.149 & -0.904 & -0.332 & -0.694 & 0.239 & -1.149 & -0.315 \\  
	& $\sigma^2=3$   &2.905 & 0.441 & 2.147 & 3.735 & 2.480 & 0.441 & 1.714 & 3.305 & 2.187 & 0.639 & 0.950 & 3.334 \\   
	& $\rho=0.7$     &0.460 & 0.270 & -0.078 & 0.877 & 0.463 & 0.271 & -0.078 & 0.881 & 0.464 & 0.271 & -0.070 & 0.884 \\    
 	& $\nu$    & & & & &11.615 & 7.161 & 3.135 & 24.269 &&&\\ 
   	& $\nu_1$  & & & & & & & & & 0.218 & 0.176 & 0.012 & 0.582  \\
        & $\nu_2$  & & & & & & & & & 0.251 & 0.103 & 0.055 & 0.445 \\ 
  & LOOIC& \multicolumn{4}{c}{780.512}& \multicolumn{4}{c}{783.056 }& \multicolumn{4}{c}{783.618}\\
  & WAIC & \multicolumn{4}{c}{780.390}& \multicolumn{4}{c}{782.968 }& \multicolumn{4}{c}{783.531}\\
  & CPO & \multicolumn{4}{c}{-390.256}& \multicolumn{4}{c}{-391.522}& \multicolumn{4}{c}{-391.802}\\
400 & $\beta_1=1$   & 1.055 & 0.141 & 0.789 & 1.328 & 1.061 & 0.137 & 0.802 & 1.326 & 1.059 & 0.137 & 0.802 & 1.324 \\  
	& $\beta_2=0.5$  & 0.486 & 0.104 & 0.287 & 0.685 & 0.483 & 0.104 & 0.284 & 0.681 & 0.483 & 0.104 & 0.285 & 0.683 \\  
	& $\gamma_1=1$   & 1.000 & 0.087 & 0.837 & 1.165 & 1.071 & 0.104 & 0.879 & 1.269 & 1.144 & 0.199 & 0.860 & 1.502 \\  
	& $\gamma_2=0.3$ & 0.305 & 0.086 & 0.140 & 0.468 & 0.333 & 0.095 & 0.152 & 0.515 & 0.355 & 0.118 & 0.144 & 0.575 \\  
	& $\gamma_3=-.5$ & -0.511 & 0.081 & -0.664 & -0.361 & -0.558 & 0.091 & -0.734 & -0.388 & -0.595 & 0.134 & -0.842 & -0.373 \\
	& $\sigma^2=3$   & 2.969 & 0.344 & 2.382 & 3.587 & 2.606 & 0.351 & 2.014 & 3.231 & 2.414 & 0.501 & 1.392 & 3.328 \\
	& $\rho=0.7$     & 0.589 & 0.175 & 0.239 & 0.864 & 0.586 & 0.175 & 0.238 & 0.867 & 0.590 & 0.174 & 0.242 & 0.868 \\  
 	& $\nu$    & & & & &14.807 & 8.446 & 4.687 & 29.926 \\ 
   	& $\nu_1$  & & & & & & & & & 0.177 & 0.161 & 0.011 & 0.519 \\ 
        & $\nu_2$  & & & & & & & & & 0.292 & 0.107 & 0.084 & 0.493 \\ 
  & LOOIC& \multicolumn{4}{c}{1562.162}& \multicolumn{4}{c}{1564.731}& \multicolumn{4}{c}{1565.061}\\
  & WAIC & \multicolumn{4}{c}{1562.104}& \multicolumn{4}{c}{1564.689}& \multicolumn{4}{c}{1565.017 }\\
  & CPO & \multicolumn{4}{c}{-781.073}& \multicolumn{4}{c}{-782.356}& \multicolumn{4}{c}{-782.520}\\
800 & $\beta_1=1$   & 1.024 & 0.099 & 0.841 & 1.211 & 1.030 & 0.096 & 0.852 & 1.213 & 1.029 & 0.097 & 0.849 & 1.213 \\  
	& $\beta_2=0.5$  & 0.486 & 0.071 & 0.352 & 0.619 & 0.483 & 0.070 & 0.350 & 0.616 & 0.484 & 0.071 & 0.351 & 0.616 \\  
	& $\gamma_1=1$   & 1.009 & 0.063 & 0.893 & 1.126 & 1.059 & 0.076 & 0.927 & 1.193 & 1.111 & 0.148 & 0.907 & 1.386 \\  
	& $\gamma_2=0.3$ & 0.300 & 0.056 & 0.194 & 0.405 & 0.319 & 0.062 & 0.206 & 0.434 & 0.333 & 0.077 & 0.200 & 0.480 \\  
	& $\gamma_3=-.5$ & -0.509 & 0.059 & -0.619 & -0.400 & -0.543 & 0.067 & -0.665 & -0.423 & -0.567 & 0.098 & -0.751 & -0.409 \\
	& $\sigma^2=3$   &2.981 & 0.279 & 2.542 & 3.434 & 2.692 & 0.390 & 2.240 & 3.162 & 2.547 & 0.423 & 1.688 & 3.270 \\
	& $\rho=0.7$     & 0.641 & 0.111 & 0.421 & 0.827 & 0.634 & 0.112 & 0.411 & 0.825 & 0.636 & 0.112 & 0.417 & 0.829 \\   
 	& $\nu$    & & & & &19.876 & 11.556 & 6.788 & 39.784 \\   
   	& $\nu_1$  & & & & & & & & & 0.149 & 0.151 & 0.010 & 0.471 \\ 
        & $\nu_2$  & & & & & & & & & 0.329 & 0.112 & 0.112 & 0.539  \\ 
  & LOOIC& \multicolumn{4}{c}{3106.898 }& \multicolumn{4}{c}{3110.026}& \multicolumn{4}{c}{3110.040}\\
  & WAIC & \multicolumn{4}{c}{3106.867}& \multicolumn{4}{c}{3110.000 }& \multicolumn{4}{c}{3110.014 }\\
  & CPO & \multicolumn{4}{c}{-1553.439}& \multicolumn{4}{c}{-1555.003}& \multicolumn{4}{c}{-1555.009}\\
   \bottomrule
\end{tabular}
\end{table}

\section{Simulation study} 
\label{secSim}

In this section, we present various simulation studies to investigate the performance of the Bayesian estimation considering different sample sizes: $n=200$, $400$, and $800$. We generate 100 artificial Monte Carlo (MC) samples of size $n$ to ensure the robustness of the results, and then we calculate the mean estimate (ME), standard deviation (SD), and highest posterior density (HPD) interval across all 100 samples. {The HPD interval is computed by identifying the shortest interval containing 95\% of the posterior probability distribution with the highest density.} For the model selection, we use LOOIC, WAIC, and CPO defined in Section \ref{sec modsel diag}. We compare the SLn, SLt, and SLcn models under different configurations. After discarding the initial 1,000 burn-in samples, we used an additional 20,000 samples (with a thinning of $5$) from a single chain that started with initial values obtained via the two-step approach \citep{heckman1979} to compute posterior summaries. Comprehensive details on the prior and posterior distributions are provided in Section~\ref{BI}. The analysis is performed using the R package \texttt{HeckmanStan}, available on GitHub at https://github.com/heeju-lim/HeckmanStan.

\begin{figure}[!ht]
	\begin{center}
		\includegraphics[ scale=0.35]{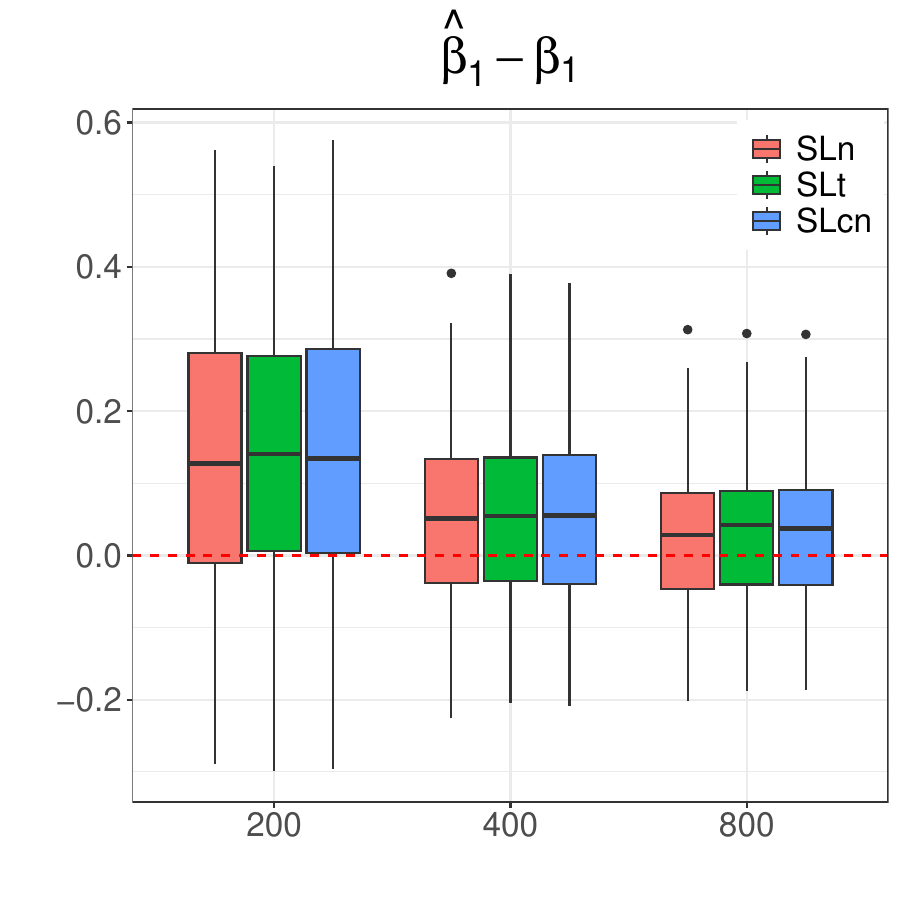}
  	\includegraphics[ scale=0.35]{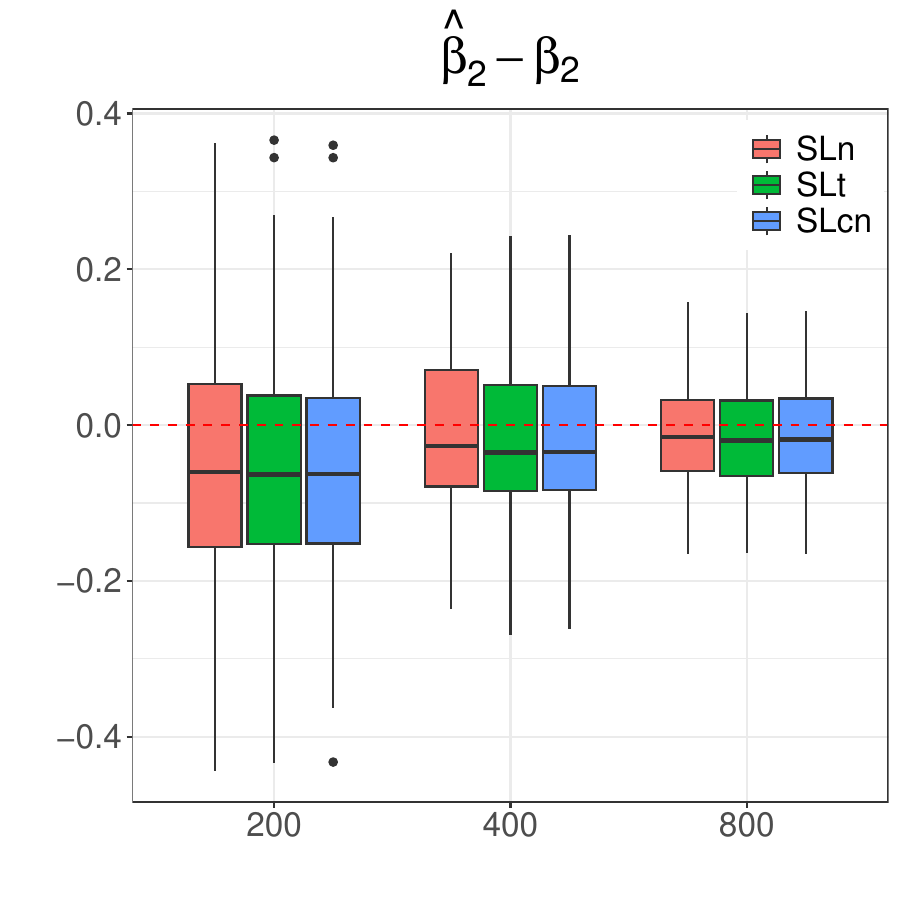}\\
   	\includegraphics[ scale=0.35]{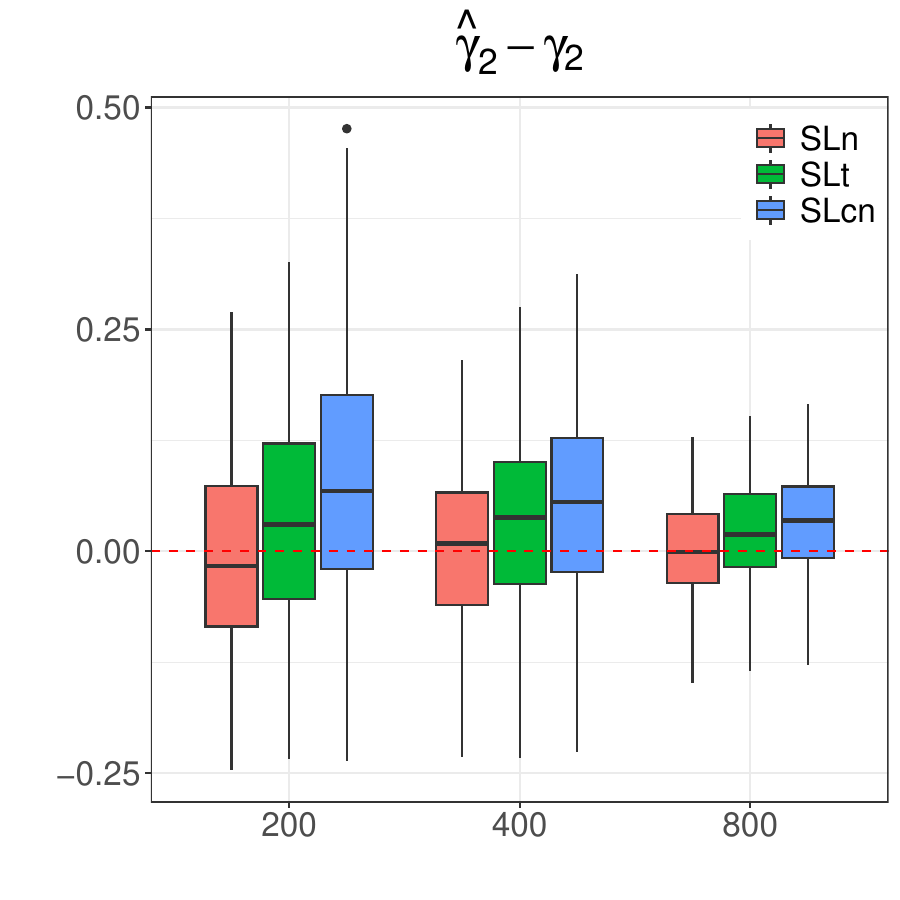}
		\includegraphics[ scale=0.35]{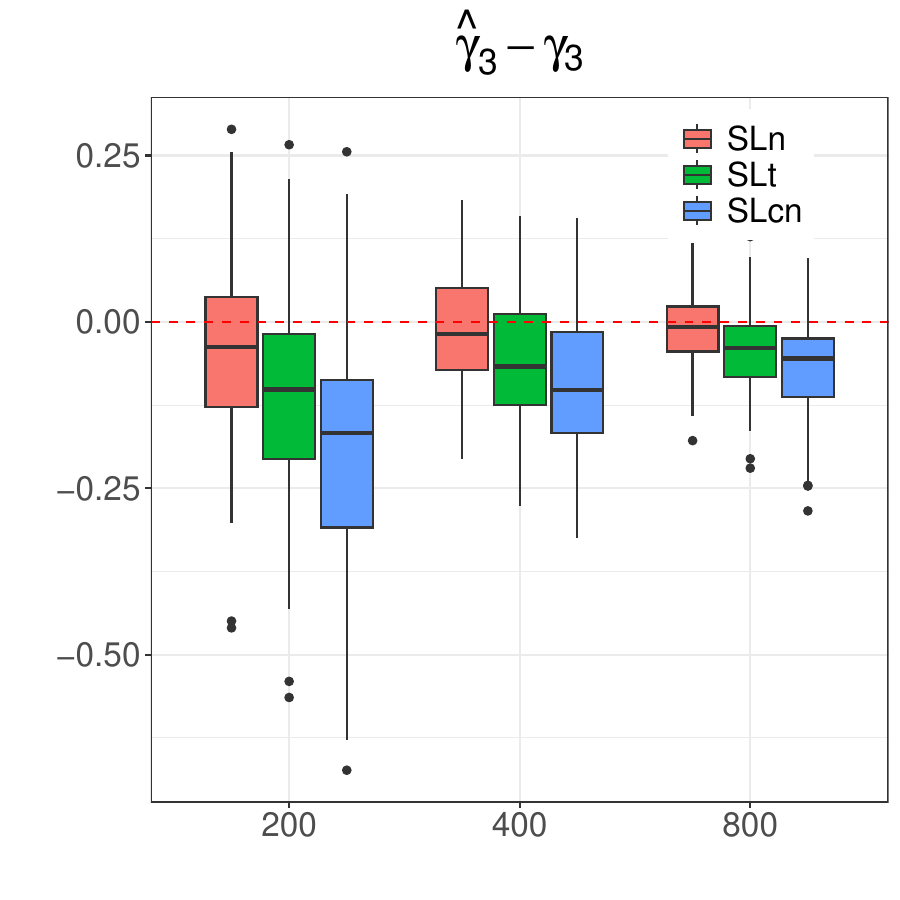}
        \caption{Simulation study. Boxplot of the 100 Monte Carlo estimates for $\beta_1$, $\beta_2$,  $\gamma_2$ and $\gamma_3$ from the normal distribution with n=200, 400 and 800, respectively}
		\label{pic_normal}
	\end{center}
\end{figure}

\begin{table}[!ht]
\caption{Simulation study. Mean estimates (ME), mean standard deviations (SD), and mean highest posterior density (HPD) 95\% interval of the $100$ Monte Carlo replicates for the data generated from the Student's-t distribution with $\nu=3$ degrees of freedom.}
	\label{tab:sim_t}
\centering
\setlength{\tabcolsep}{3pt}
\begin{tabular}{cccccccccccccc}
  \toprule
 & & \multicolumn{4}{c}{SLn} & \multicolumn{4}{c}{SLt} & \multicolumn{4}{c}{SLcn} \\
 $n$ & TRUE & ME & SD & \multicolumn{2}{c}{HPD (95\%)} & ME & SD & \multicolumn{2}{c}{HPD (95\%)}& ME & SD & \multicolumn{2}{c}{HPD (95\%)} \\ 
  \midrule
200 & $\beta_1=1$   & 1.056 & 0.320 & 0.453 & 1.688 & 1.128 & 0.223 & 0.699 & 1.569 & 1.137 & 0.226 & 0.707 & 1.580 \\
	& $\beta_2=0.5$  & 0.529 & 0.240 & 0.062 & 0.997 & 0.464 & 0.183 & 0.111 & 0.821 & 0.468 & 0.183 & 0.111 & 0.824 \\ 
	& $\gamma_1=1$   & 0.813 & 0.111 & 0.600 & 1.025 & 1.059 & 0.176 & 0.735 & 1.405 & 1.199 & 0.275 & 0.747 & 1.738 \\
	& $\gamma_2=0.3$ & 0.229 & 0.112 & 0.014 & 0.445 & 0.314 & 0.152 & 0.022 & 0.610 & 0.359 & 0.187 & 0.014 & 0.727 \\  
	& $\gamma_3=-.5$ & -0.368 & 0.107 & -0.575 & -0.166 & -0.550 & 0.160 & -0.868 & -0.250 & -0.616 & 0.209 & -1.029 & -0.247 \\  
	& $\sigma^2=3$  & 7.583 & 1.147 & 5.590 & 9.752 & 3.092 & 0.698 & 1.924 & 4.390 & 2.740 & 0.722 & 1.446 & 4.154 \\
	& $\rho=0.7$    & 0.580 & 0.217 & 0.155 & 0.886 & 0.523 & 0.222 & 0.082 & 0.880 & 0.527 & 0.223 & 0.084 & 0.881 \\
 	& $\nu$    & & & & &3.920 & 1.460 & 2.036 & 6.622&&&\\ 
   	& $\nu_1$ & & & & & & & & &0.258 & 0.099 & 0.085 & 0.450\\
        & $\nu_2$ & & & & & & & & &0.119 & 0.043 & 0.045 & 0.203\\ 
  & LOOIC& \multicolumn{4}{c}{908.600}& \multicolumn{4}{c}{870.407 }& \multicolumn{4}{c}{872.610}\\
  & WAIC & \multicolumn{4}{c}{908.496}& \multicolumn{4}{c}{870.294}& \multicolumn{4}{c}{872.224}\\
  & CPO & \multicolumn{4}{c}{-454.367}& \multicolumn{4}{c}{-435.198}& \multicolumn{4}{c}{-436.330}\\
400 & $\beta_1=1$   & 0.901 & 0.201 & 0.519 & 1.292 & 1.052 & 0.150 & 0.763 & 1.342 & 1.049 & 0.156 & 0.750 & 1.352 \\
	& $\beta_2=0.5$  & 0.540 & 0.163 & 0.226 & 0.856 & 0.490 & 0.123 & 0.254 & 0.726 & 0.496 & 0.126 & 0.256 & 0.740 \\
	& $\gamma_1=1$   & 0.787 & 0.078 & 0.642 & 0.933 & 1.021 & 0.126 & 0.795 & 1.257 & 1.067 & 0.155 & 0.790 & 1.371 \\ 
	& $\gamma_2=0.3$ & 0.214 & 0.078 & 0.067 & 0.361 & 0.311 & 0.107 & 0.109 & 0.516 & 0.320 & 0.115 & 0.104 & 0.545 \\
	& $\gamma_3=-.5$ & -0.344 & 0.068 & -0.470 & -0.218 & -0.517 & 0.103 & -0.715 & -0.327 & -0.529 & 0.116 & -0.759 & -0.315 \\   
	& $\sigma^2=3$  & 8.222 & 0.942 & 6.668 & 9.852 & 3.112 & 0.537 & 2.220 & 4.063 & 3.087 & 0.583 & 2.022 & 4.207 \\ 
	& $\rho=0.7$    & 0.724 & 0.108 & 0.518 & 0.881 & 0.621 & 0.141 & 0.338 & 0.856 & 0.633 & 0.142 & 0.351 & 0.866 \\  
 	& $\nu $    & & & & &3.430 & 0.813 & 2.137 & 4.990   &&&\\ 
   	& $\nu_1$  & & & & & & & & &0.212 & 0.072 & 0.084 & 0.352\\
        & $\nu_2$  & & & & & & & & &0.112 & 0.031 & 0.057 & 0.172\\ 
  & LOOIC& \multicolumn{4}{c}{1825.719}& \multicolumn{4}{c}{1745.875}& \multicolumn{4}{c}{1751.549}\\
  & WAIC & \multicolumn{4}{c}{1826.506}& \multicolumn{4}{c}{1745.806}& \multicolumn{4}{c}{1751.118}\\
  & CPO & \multicolumn{4}{c}{-912.928}& \multicolumn{4}{c}{-872.936}& \multicolumn{4}{c}{-875.843}\\
800 & $\beta_1=1$ & 0.883 & 0.141 & 0.618 & 1.152 & 1.027 & 0.105 & 0.829 & 1.228 & 1.022 & 0.109 & 0.815 & 1.230 \\ 
	& $\beta_2=0.5$  & 0.541 & 0.112 & 0.330 & 0.756 & 0.482 & 0.083 & 0.325 & 0.640 & 0.488 & 0.086 & 0.327 & 0.651 \\  
	& $\gamma_1=1$ & 0.782 & 0.057 & 0.679 & 0.886 & 1.014 & 0.088 & 0.854 & 1.181 & 1.029 & 0.107 & 0.842 & 1.228 \\
	& $\gamma_2=0.3$ & 0.213 & 0.052 & 0.118 & 0.309 & 0.303 & 0.070 & 0.173 & 0.435 & 0.300 & 0.073 & 0.165 & 0.438 \\ 
	& $\gamma_3=-.5$ & -0.340 & 0.050 & -0.433 & -0.248 & -0.510 & 0.074 & -0.649 & -0.372 & -0.505 & 0.081 & -0.661 & -0.357 \\
	& $\sigma^2=3$  &  8.210 & 0.651 & 7.090 & 9.365 & 3.008 & 0.373 & 2.373 & 3.682 & 3.115 & 0.452 & 2.326 & 3.945 \\ 
	& $\rho=0.7$    & 0.742 & 0.066 & 0.619 & 0.841 & 0.657 & 0.094 & 0.471 & 0.820 & 0.668 & 0.093 & 0.486 & 0.829 \\ 
 	& $\nu $    & & & & &3.185 & 0.506 & 2.284 & 4.177  &&&\\ 
   	& $\nu_1$  & & & & & & & & &0.193 & 0.051 & 0.101 & 0.290\\
        & $\nu_2$  & & & & & & & & &0.101 & 0.022 & 0.064 & 0.141 \\ 
  & LOOIC& \multicolumn{4}{c}{3641.109}& \multicolumn{4}{c}{3459.870}& \multicolumn{4}{c}{3473.647}\\
  & WAIC & \multicolumn{4}{c}{3644.534}& \multicolumn{4}{c}{3459.838}& \multicolumn{4}{c}{3473.309}\\
  & CPO & \multicolumn{4}{c}{-1820.615}& \multicolumn{4}{c}{-1729.924}& \multicolumn{4}{c}{-1736.903}\\
   \bottomrule
\end{tabular}
\end{table}
We follow \citet{lachosHeckman} for the configuration of the simulation study. The elements of $\w_i^\top = (1, w_{i1}, w_{i2})$ are generated from a ${\cal N}(0,1)$ distribution. We set $\bgamma = (1, 0.3, -0.5)$ and $\bbeta = (1, 0.5)$. The variance parameter is given by $\sigma^2 = 3$, and the correlation parameter at $\rho = 0.7$. In addition to generating data from the normal and Students'-t ($\nu=3$) distributions, we also consider the slash distribution with $\nu=1.43$ \citep{rogers1972understanding, Wang_Genton_2006}, which is often used as a challenging distribution for data with heavier tails than the normal one. The average missing rates are 0.20 for the Normal distribution, 0.23 for the Student's t-distribution (df = 3), and 0.26 for the slash distribution (df = 1.43).

\begin{figure}[!htb]
    \centering
    \begin{subfigure}[b]{\textwidth}
        \centering
        \includegraphics[scale=0.28]{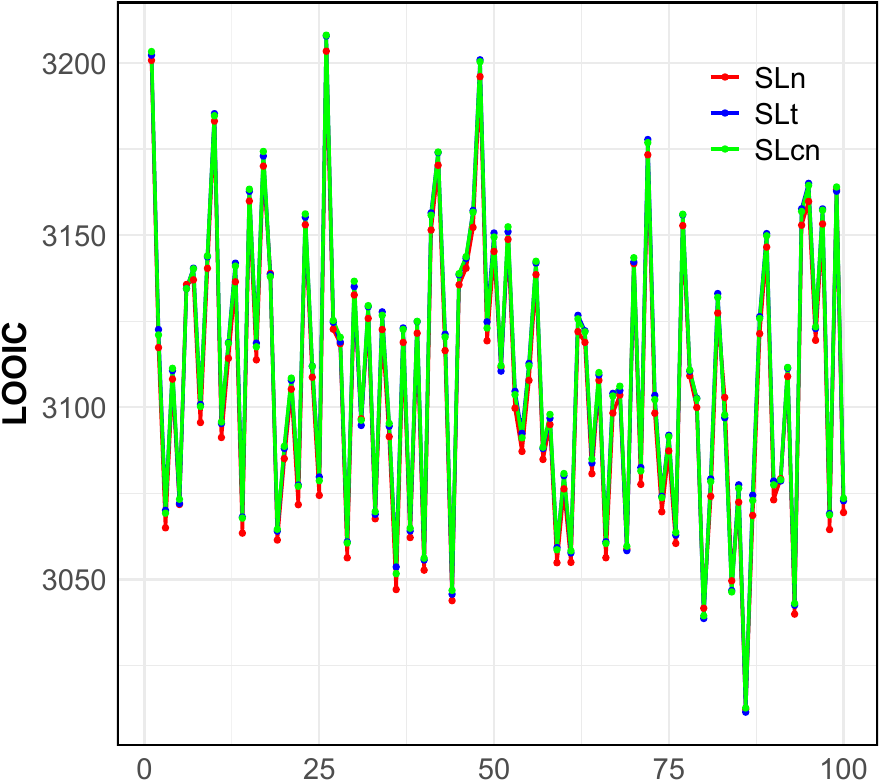}
        \includegraphics[scale=0.28]{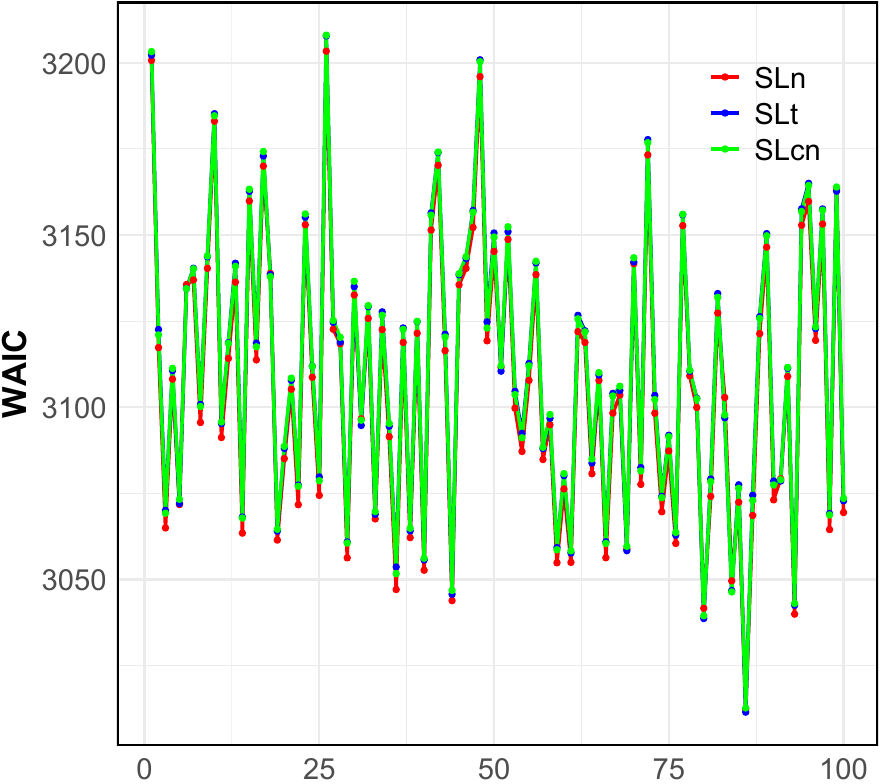}
        \includegraphics[scale=0.28]{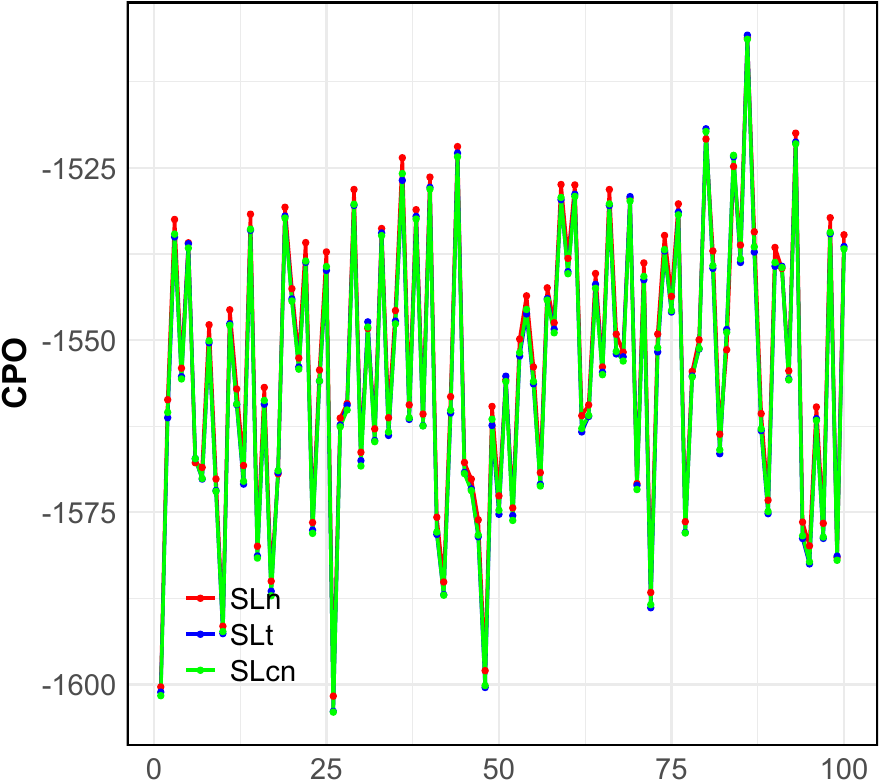}
        \caption{Normal distribution}
        \label{fig:normal}
    \end{subfigure}
    \begin{subfigure}[b]{\textwidth}
        \centering
        \includegraphics[scale=0.28]{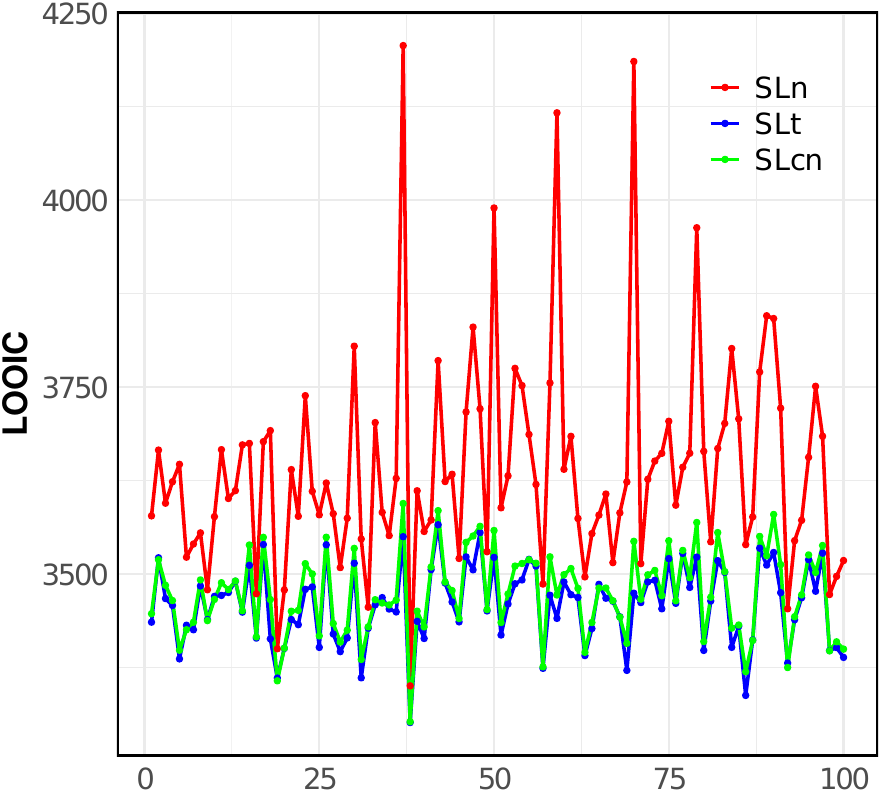}
        \includegraphics[scale=0.28]{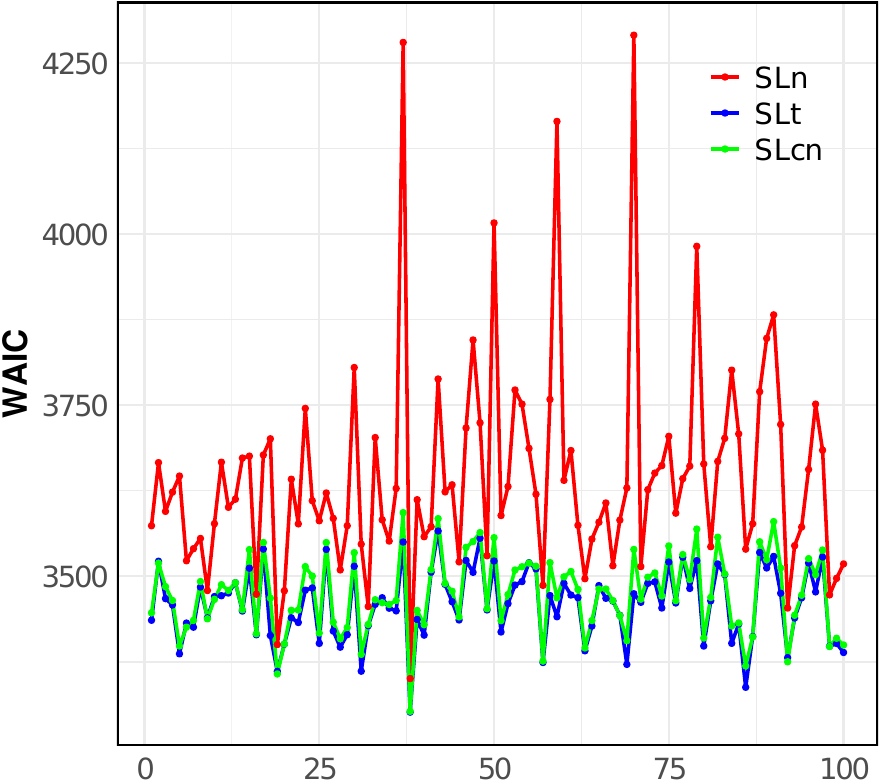}
        \includegraphics[scale=0.28]{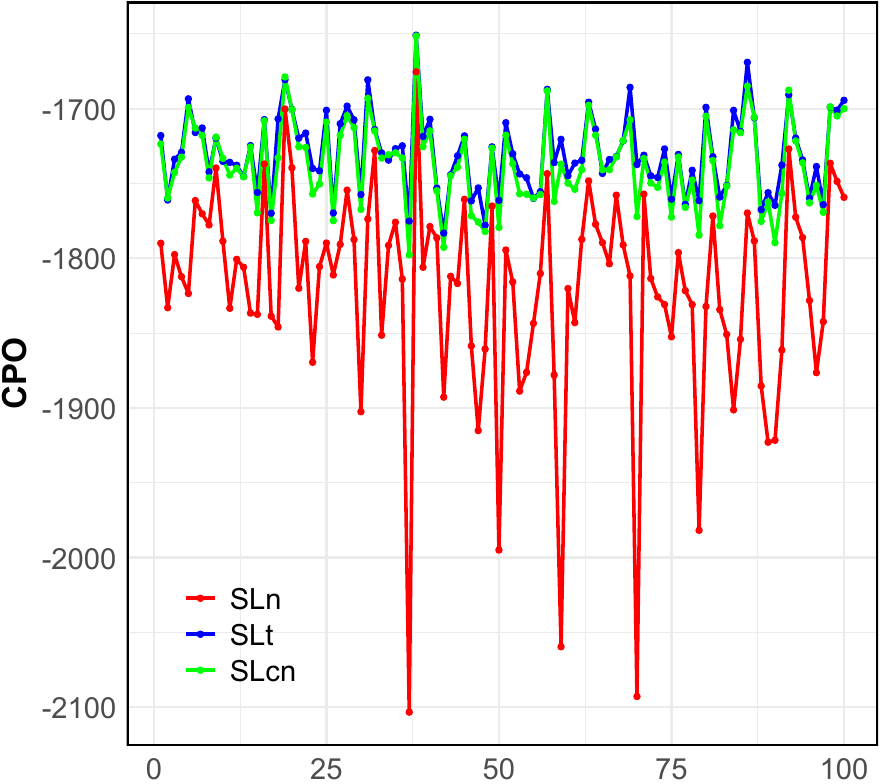}
        \caption{Student's-t distribution}
        \label{fig:t-dist}
    \end{subfigure}
    \begin{subfigure}[b]{\textwidth}
        \centering
        \includegraphics[scale=0.28]{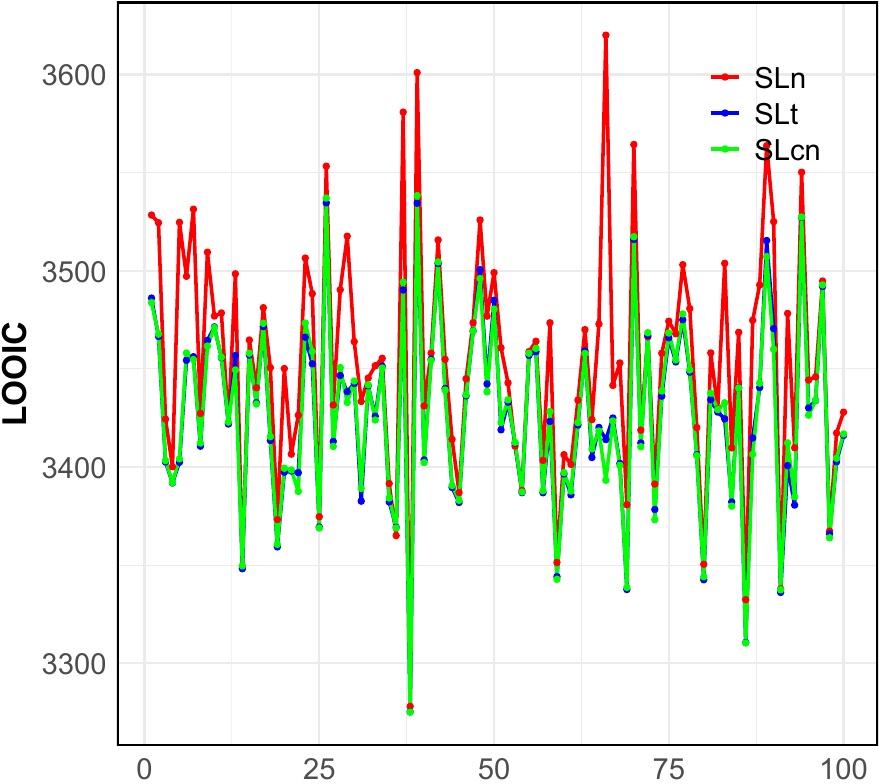}
        \includegraphics[scale=0.28]{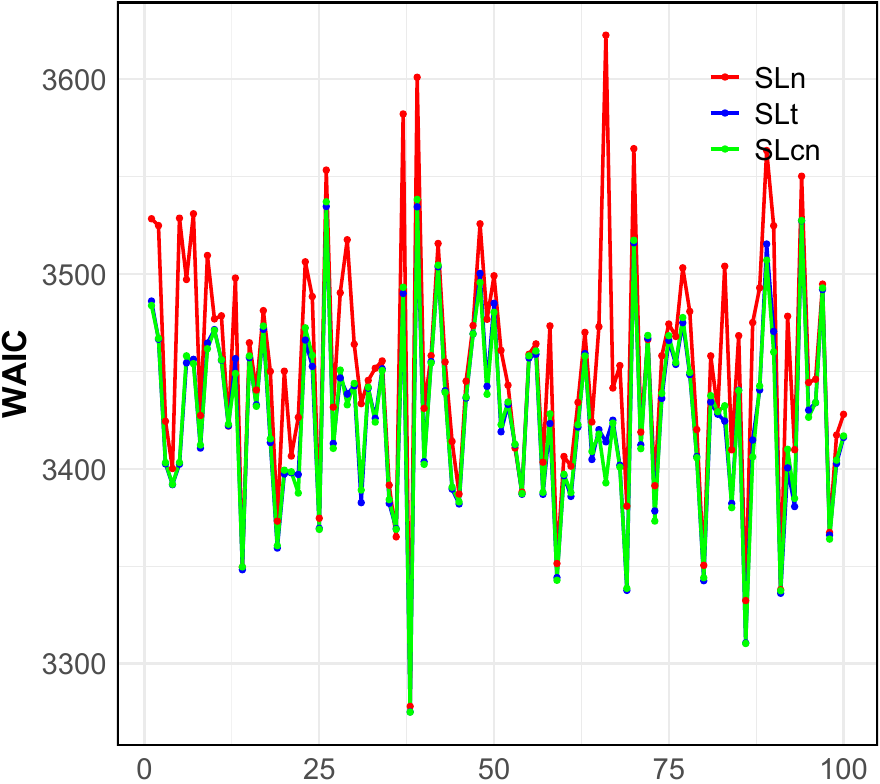}
        \includegraphics[scale=0.28]{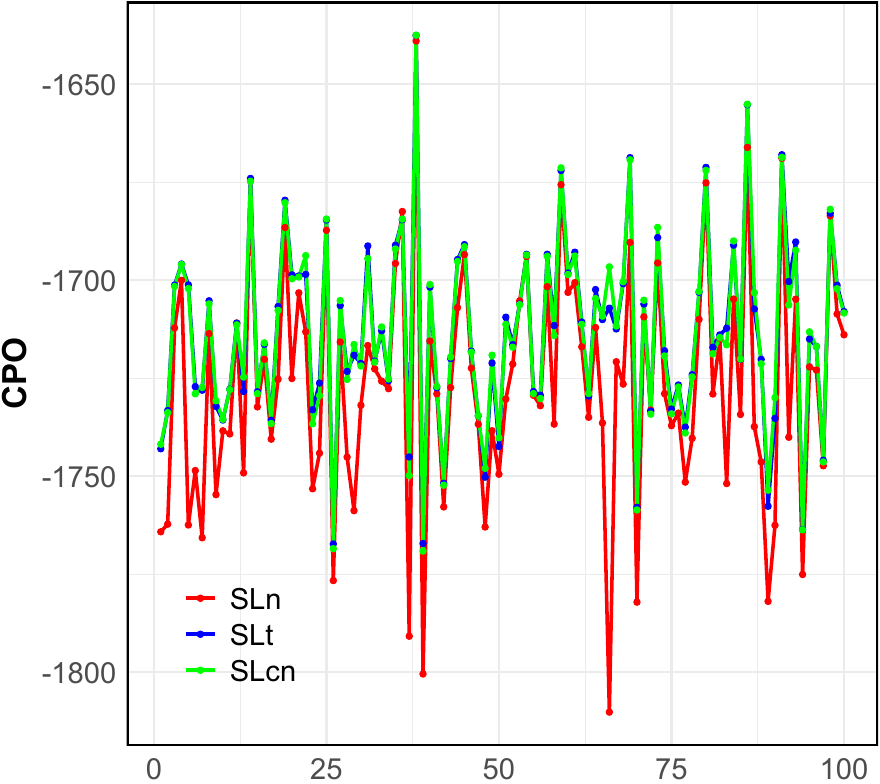}
        \caption{Slash distribution}
        \label{fig:slash}
    \end{subfigure}
    \caption{Distribution of 100 Monte Carlo samples for model selection criteria: LOOIC, WAIC, and CPO across different data-generating distributions.}
    \label{modelselection}
\end{figure}

First, the \tablename~\ref{tab:sim_normal} presents the results for ME, SD, and HPD intervals under the assumption that the data are generated from a normal distribution. Among the models, the SLn model consistently exhibits the lowest LOOIC and WAIC values and the highest CPO values across all sample sizes, indicating the best model fit. Additionally, as depicted in the \figurename~\ref{pic_normal}, the SLn model is represented by the red boxplot, the SLt model by the green boxplot, and the SLcn model by the blue boxplot. The red dashed line indicates the true parameter. Across all parameters, the SLn model demonstrates reduced variation as the sample size increases when it comes to the estimation of $\bgamma$ parameters, effectively satisfying the asymptotic consistency property. However, in the case of the $\bbeta$ parameters, there is no significant difference among the three models. Additionally, when we examine the model selection criteria of 100 Monte Carlo samples in \figurename~\ref{modelselection}, the differences among the three models are not substantial. Thus, when the data are generated from a normal distribution, all three models appear to perform effectively.

\begin{figure}[htbp]
    \centering
    \begin{minipage}{\textwidth}
        \centering
        \includegraphics[ scale=0.35]{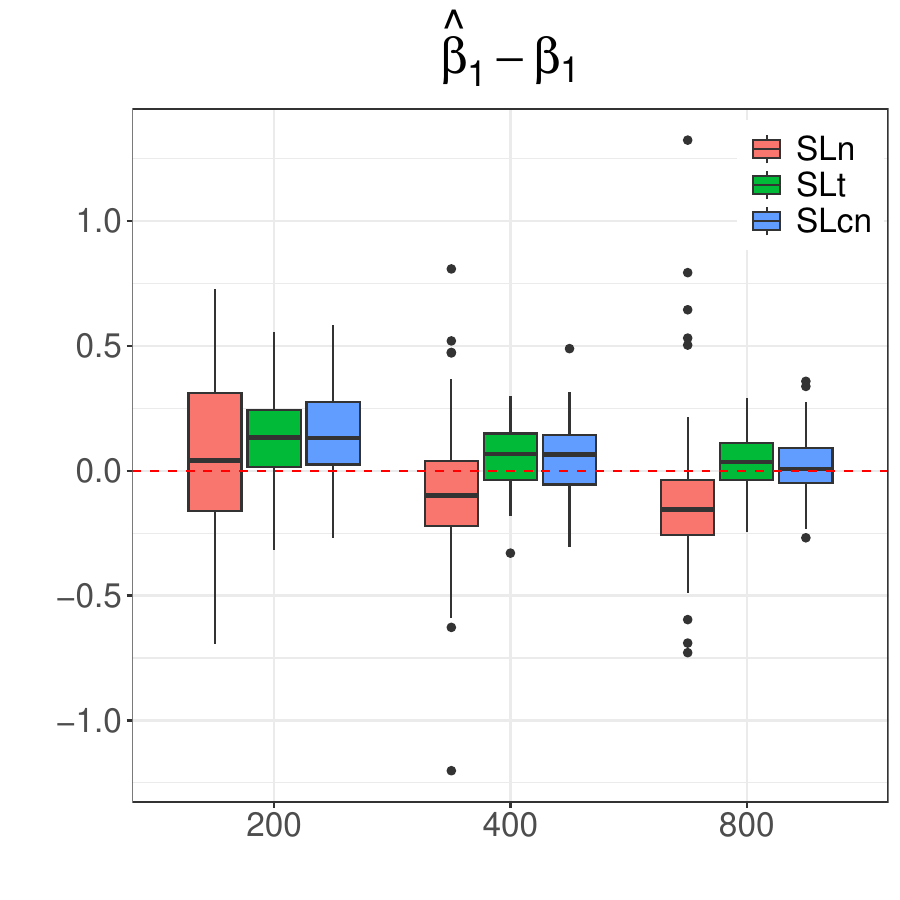} 
        \includegraphics[ scale=0.35]{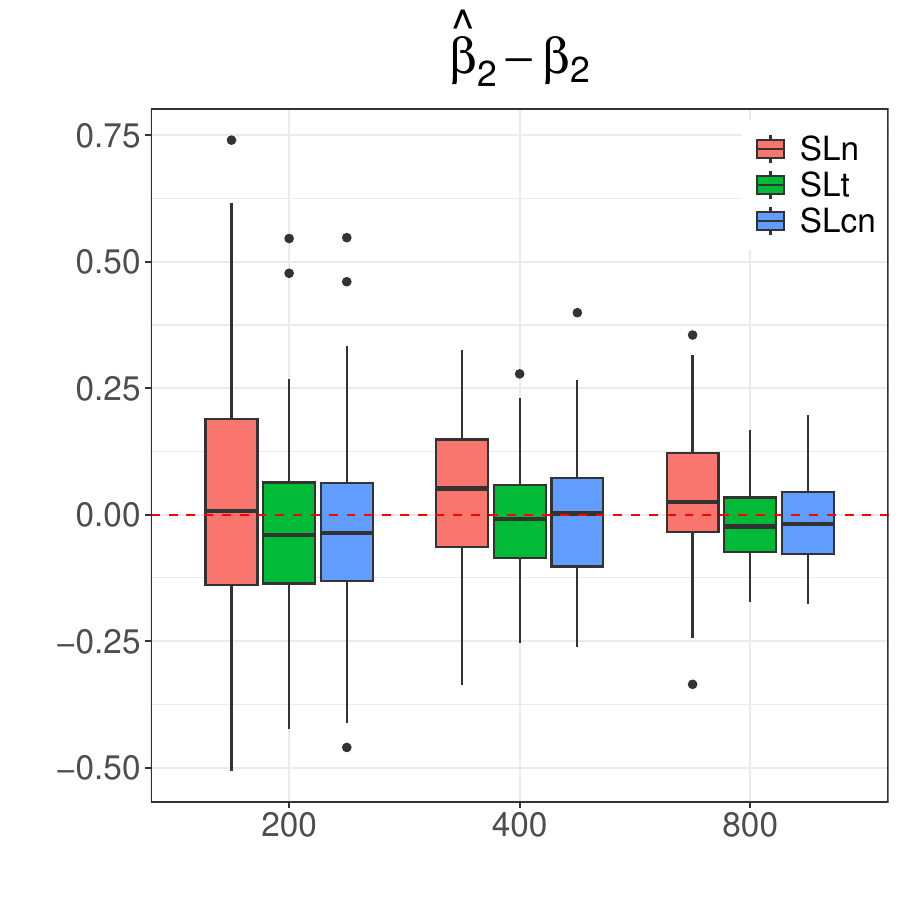} \\
        \includegraphics[ scale=0.35]{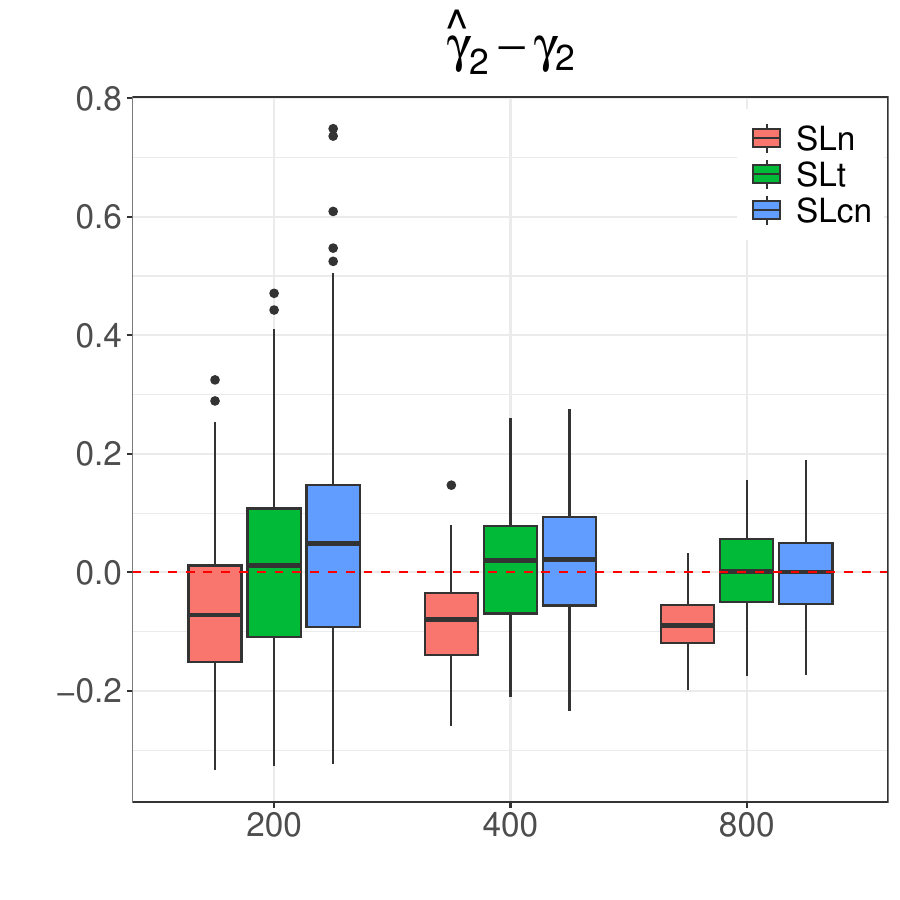} 
        \includegraphics[ scale=0.35]{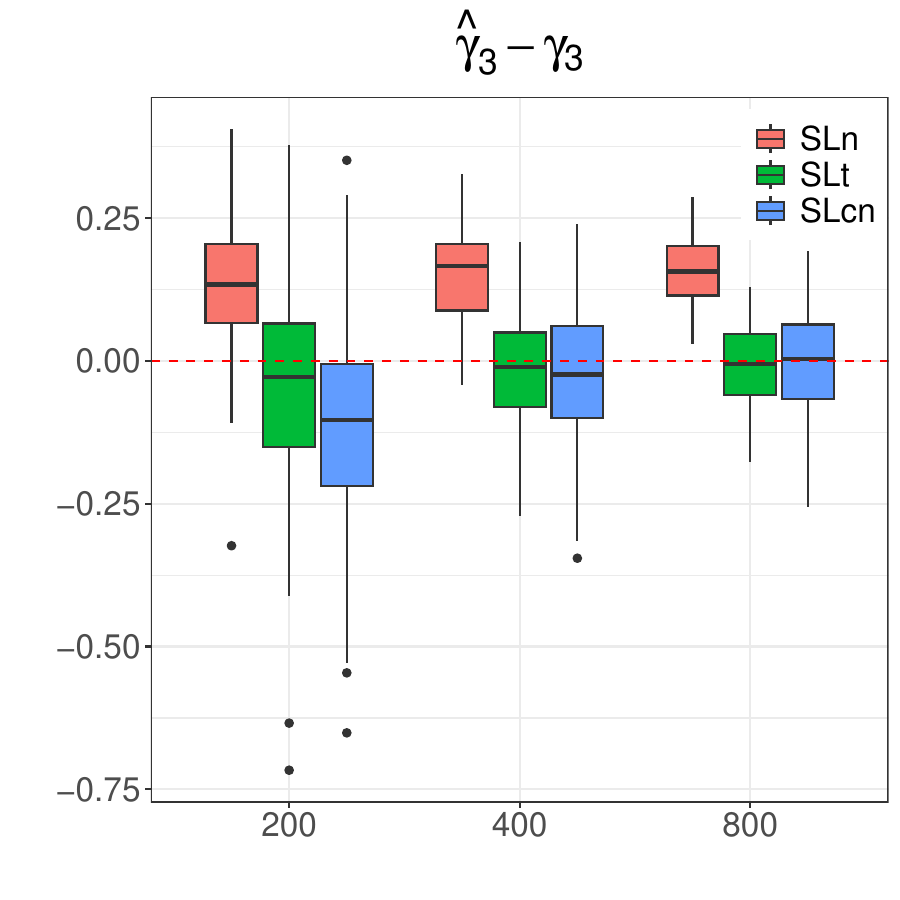}
        \caption{Simulation study. Boxplot of the 100 Monte Carlo estimates for $\beta_1$, $\beta_2$, $\gamma_2$ and $\gamma_3$ from the Student's-t distribution with $n=200$, $400$ and $800$, respectively}
        \label{pic_t}
    \end{minipage}
    \begin{minipage}{\textwidth}
        \centering
        \includegraphics[ scale=0.35]{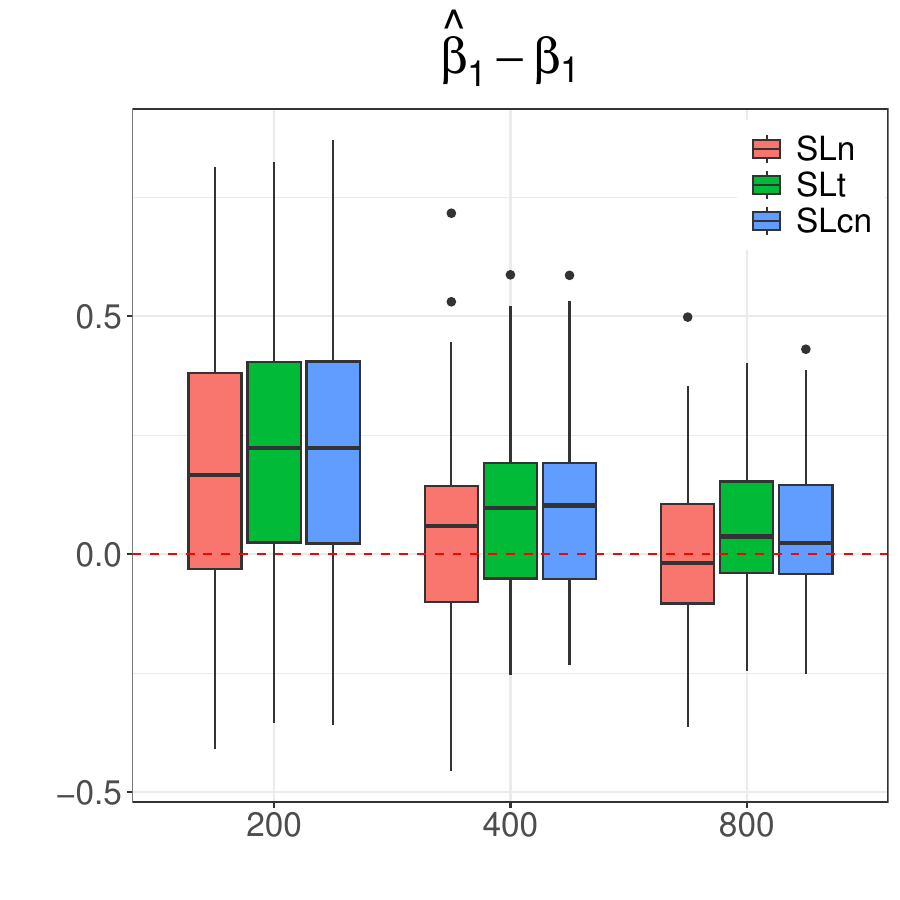} 
        \includegraphics[ scale=0.35]{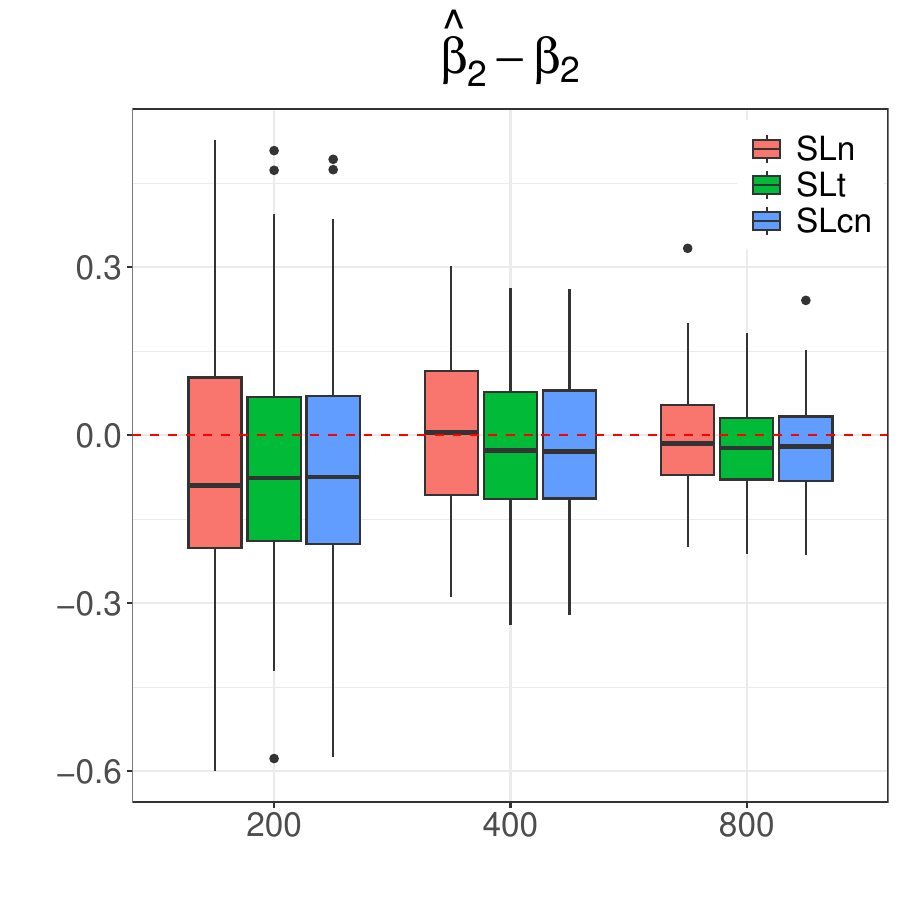} \\
        \includegraphics[ scale=0.35]{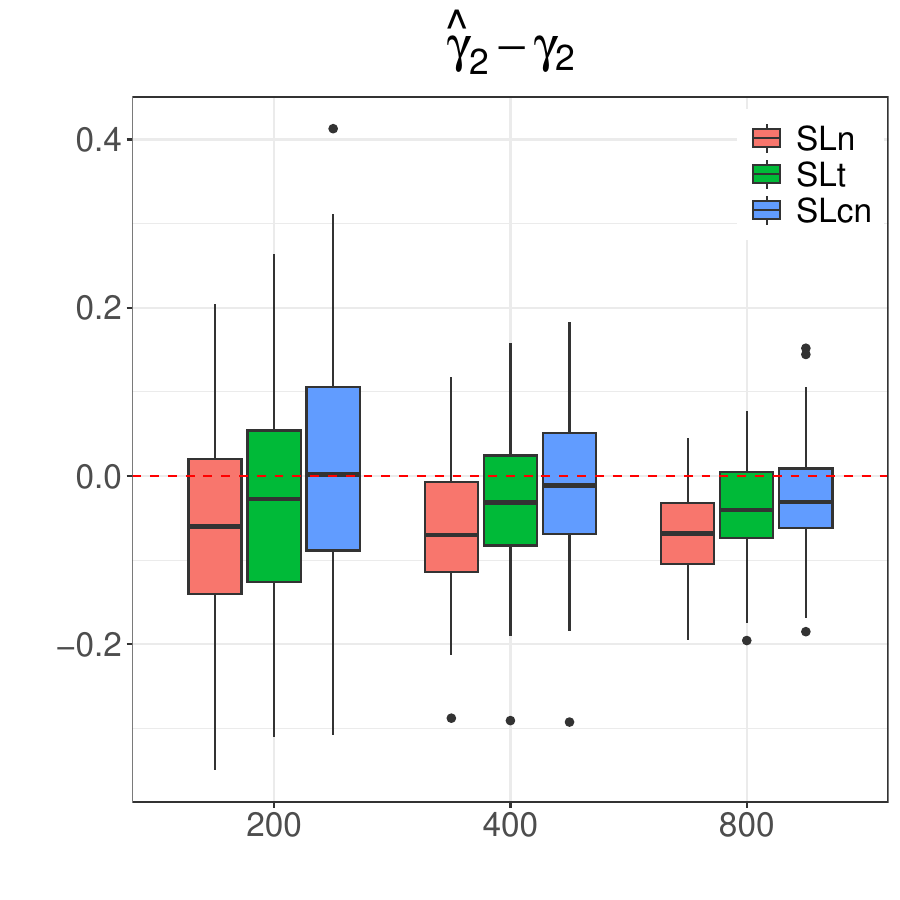} 
        \includegraphics[ scale=0.35]{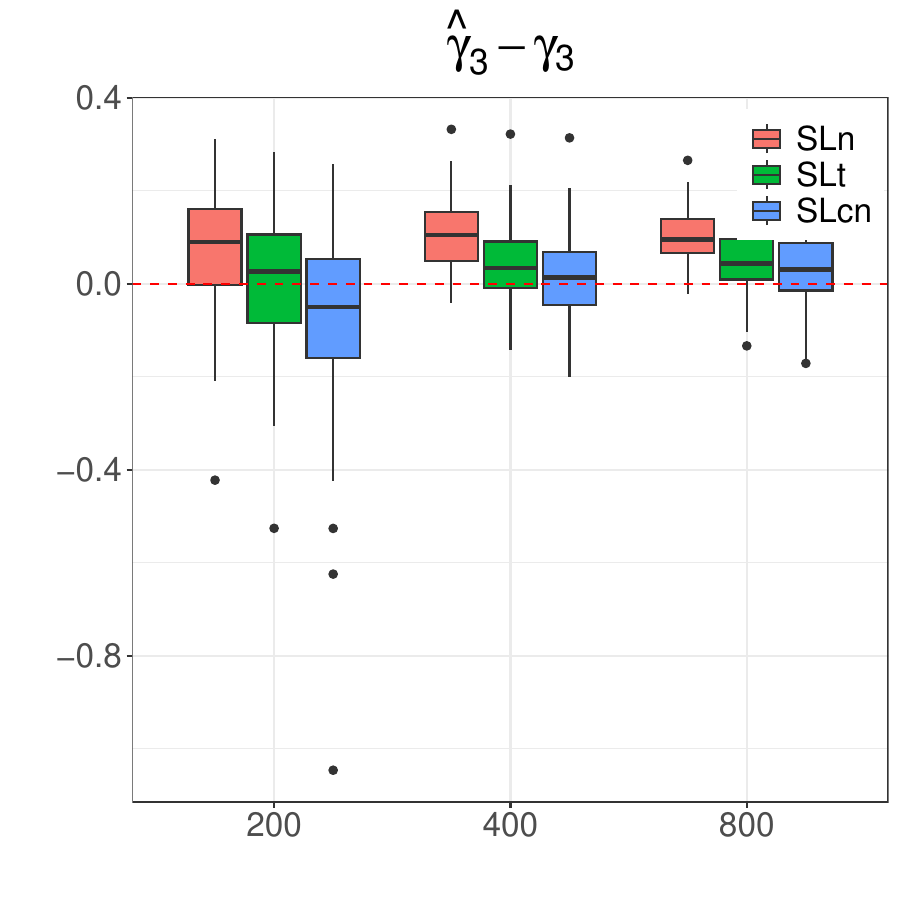}
        \caption{Simulation study. Boxplot of the 100 Monte Carlo estimates for $\beta_1$, $\beta_2$, $\gamma_2$ and $\gamma_3$ from the slash distribution with $n=200$, $400$ and $800$, respectively. }        
        \label{pic_slash}
    \end{minipage}
\end{figure}

Secondly, \tablename~\ref{tab:sim_t} presents the results from the Student's-t distribution with $\nu=3$ degrees of freedom. The SLt model consistently achieves the smallest LOOIC and WAIC values while displaying the highest CPO values across all sample sizes. In this case, the SLcn model also demonstrates a relatively good fit, following the SLt model. However, the SLn model shows signs of bias and exhibits the largest variation when the sample size decreases, as expected. Moreover, the green boxplot in Fig. \ref{pic_t} illustrates that the parameter estimates of the SLt model asymptotically approach the true parameter. The blue boxplot, representing the SLcn model, follows as the next best-performing model. In \figurename~\ref{modelselection}, it is evident that the SLn model performs significantly worse compared to the SLt and SLcn models. Consequently, while the SLt model shows slightly better performance when the data is generated from a Student's-t distribution, the SLcn model also performs comparably well.

\begin{table}[!ht]
\caption{Simulation study. Mean estimates (ME), mean standard deviations (SD) and mean highest posterior density (HPD) 95\% interval of the $100$ Monte Carlo replicates for the data generated from the slash distribution with $2$ degrees of freedom. }
	\label{tab:sim_slash}
\centering
\setlength{\tabcolsep}{3pt}
\begin{tabular}{cccccccccccccc}
  \toprule
 & & \multicolumn{4}{c}{SLn} & \multicolumn{4}{c}{SLt} & \multicolumn{4}{c}{SLcn} \\
 $n$ & TRUE & ME & SD & \multicolumn{2}{c}{HPD (95\%)} & ME & SD & \multicolumn{2}{c}{HPD (95\%)}& ME & SD & \multicolumn{2}{c}{HPD (95\%)} \\ 
  \midrule
200 & $\beta_1=1$ & 1.160 & 0.321 & 0.556 & 1.795 & 1.204 & 0.279 & 0.672 & 1.757 & 1.208 & 0.279 & 0.676 & 1.757 \\
	& $\beta_2=0.5$  &  0.461 & 0.215 & 0.044 & 0.881 & 0.447 & 0.199 & 0.061 & 0.835 & 0.446 & 0.199 & 0.061 & 0.833 \\ 
	& $\gamma_1=1$   & 0.793 & 0.109 & 0.586 & 1.005 & 0.902 & 0.141 & 0.639 & 1.179 & 1.019 & 0.250 & 0.640 & 1.490 \\  
	& $\gamma_2=0.3$ &0.228 & 0.112 & 0.013 & 0.444 & 0.269 & 0.132 & 0.015 & 0.525 & 0.301 & 0.163 & 0.003 & 0.618 \\ 
	& $\gamma_3=-.5$ &-0.408 & 0.108 & -0.617 & -0.203 & -0.486 & 0.134 & -0.749 & -0.233 & -0.546 & 0.189 & -0.910 & -0.226 \\   
	& $\sigma^2=3$  & 5.701 & 0.905 & 4.089 & 7.472 & 3.908 & 1.075 & 2.515 & 5.433 & 3.372 & 0.948 & 1.600 & 5.216 \\ 
	& $\rho=0.7$    &  0.503 & 0.260 & -0.013 & 0.881 & 0.465 & 0.259 & -0.048 & 0.870 & 0.464 & 0.259 & -0.050 & 0.870 \\  
 	& $\nu $    & & & & & 7.501 & 4.007 & 2.433 & 14.794 &&& \\ 
   	& $\nu_1$  & & & & & & & & & 0.238 & 0.142 & 0.030 & 0.518  \\
        & $\nu_2$  & & & & & & & & & 0.200 & 0.076 & 0.059 & 0.349 \\ 
  & LOOIC& \multicolumn{4}{c}{868.880}& \multicolumn{4}{c}{862.217}& \multicolumn{4}{c}{862.648}\\
  & WAIC & \multicolumn{4}{c}{868.635}& \multicolumn{4}{c}{862.104}& \multicolumn{4}{c}{862.472}\\
  & CPO & \multicolumn{4}{c}{-434.491}& \multicolumn{4}{c}{-431.103}& \multicolumn{4}{c}{-431.326}\\
400 & $\beta_1=1$ & 1.031 & 0.216 & 0.627 & 1.460 & 1.098 & 0.193 & 0.733 & 1.479 & 1.097 & 0.195 & 0.728 & 1.479 \\  
	& $\beta_2=0.5$ & 0.501 & 0.145 & 0.223 & 0.780 & 0.477 & 0.135 & 0.217 & 0.739 & 0.481 & 0.136 & 0.220 & 0.742 \\ 
	& $\gamma_1=1$ &0.772 & 0.077 & 0.628 & 0.916 & 0.857 & 0.094 & 0.681 & 1.038 & 0.927 & 0.151 & 0.681 & 1.209 \\ 
	& $\gamma_2=0.3$ & 0.231 & 0.079 & 0.084 & 0.381 & 0.269 & 0.090 & 0.097 & 0.442 & 0.290 & 0.106 & 0.095 & 0.494 \\ 
	& $\gamma_3=-.5$ & -0.379 & 0.070 & -0.511 & -0.250 & -0.445 & 0.083 & -0.606 & -0.288 & -0.478 & 0.112 & -0.688 & -0.286 \\ 
	& $\sigma^2=3$  &5.743 & 0.693 & 4.505 & 7.052 & 4.079 & 0.754 & 2.977 & 5.245 & 3.667 & 0.803 & 2.150 & 5.173 \\ 
	& $\rho=0.7$  &0.630 & 0.164 & 0.303 & 0.871 & 0.575 & 0.171 & 0.236 & 0.852 & 0.577 & 0.171 & 0.235 & 0.853 \\  
    & $\nu $    & & & & & 8.118 & 3.554 & 3.335 & 14.724  &&& \\ 
    & $\nu_1$  & & & & & & & & & 0.211 & 0.125 & 0.029 & 0.459 \\
    & $\nu_2$  & & & & & & & & & 0.221 & 0.071 & 0.088 & 0.360 \\ 
  & LOOIC& \multicolumn{4}{c}{1736.702}& \multicolumn{4}{c}{1724.316}& \multicolumn{4}{c}{1724.727}\\
  & WAIC & \multicolumn{4}{c}{1736.575}& \multicolumn{4}{c}{1724.254}& \multicolumn{4}{c}{1724.594}\\
  & CPO & \multicolumn{4}{c}{-868.368}& \multicolumn{4}{c}{-862.149}& \multicolumn{4}{c}{-862.362}\\
800 & $\beta_1=1$ & 0.970 & 0.146 & 0.700 & 1.256 & 1.058 & 0.136 & 0.800 & 1.327 & 1.053 & 0.140 & 0.789 & 1.323 \\  
	& $\beta_2=0.5$ & 0.509 & 0.098 & 0.322 & 0.695 & 0.476 & 0.091 & 0.302 & 0.652 & 0.479 & 0.093 & 0.305 & 0.654 \\  
	& $\gamma_1=1$ & 0.775 & 0.056 & 0.674 & 0.878 & 0.852 & 0.067 & 0.730 & 0.978 & 0.891 & 0.106 & 0.726 & 1.074 \\ 
	& $\gamma_2=0.3$ & 0.224 & 0.052 & 0.128 & 0.320 & 0.256 & 0.059 & 0.146 & 0.366 & 0.264 & 0.068 & 0.145 & 0.388 \\ 
	& $\gamma_3=-.5$ & -0.373 & 0.052 & -0.470 & -0.279 & -0.436 & 0.061 & -0.551 & -0.324 & -0.451 & 0.078 & -0.592 & -0.320 \\
	& $\sigma^2=3$  & 5.829 & 0.581 & 4.893 & 6.793 & 4.116 & 0.480 & 3.280 & 5.001 & 3.910 & 0.624 & 2.737 & 5.084 \\  
	& $\rho=0.7$    &  0.697 & 0.098 & 0.506 & 0.852 & 0.625 & 0.115 & 0.397 & 0.820 & 0.630 & 0.115 & 0.403 & 0.826 \\
 	& $\nu $    & & & & & 8.376 & 2.805 & 4.248 & 13.711  &&& \\ 
   	& $\nu_1$  & & & & & & & & & 0.173 & 0.098 & 0.029 & 0.362\\
        & $\nu_2$  & & & & & & & & & 0.225 & 0.062 & 0.108 & 0.345  \\ 
    & LOOIC& \multicolumn{4}{c}{3453.776}& \multicolumn{4}{c}{3426.712}& \multicolumn{4}{c}{3426.839}\\
  & WAIC & \multicolumn{4}{c}{3453.779}& \multicolumn{4}{c}{3426.657}& \multicolumn{4}{c}{3426.669}\\
  & CPO & \multicolumn{4}{c}{-1726.900}& \multicolumn{4}{c}{-1713.348}& \multicolumn{4}{c}{-1713.445}\\
   \bottomrule
\end{tabular}
\end{table}

Finally, \tablename~\ref{tab:sim_slash} presents the results when the data are generated from the slash distribution. In this data generation scenario, the SLt, SLcn, and SLn models demonstrate good performance in sequence, although the differences among the three models are relatively minor, as reflected in their lower LOOIC and WAIC values and higher CPO values.

In conclusion, \tablename~\ref{tab:sim_select} shows the percentage of times each model is selected among the SL models across the 100 MC samples, based on having the lowest LOOIC and WAIC or the highest CPO. The results indicate that the SLn model is well-suited for data generated from a normal distribution, while the SLt model exhibits the best performance when the distribution follows a Student’s-t distribution or a slash distribution. Additionally, the SLcn model demonstrated performance comparable to that of the SLt model. Fig. \ref{modelselection} presents a graph of 100 MC samples with a sample size of $n=800$, where the y-axis represents LOOIC, WAIC, and CPO values, and the x-axis corresponds to the sample index. The red line represents the SLn model, the blue line represents the SLt model, and the green line represents the SLcn model. Given that lower LOOIC and WAIC values and higher CPO values indicate better model performance, we can see that the SLn model does not perform well for heavy-tailed distributions when data is generated from the Student’s-t and slash distributions. In contrast, the SLt and SLcn models exhibit similar performance. If we examine more rigorously, the SLt model achieves slightly better indicators. However, since the SLcn model produces nearly comparable values, both models can be considered suitable for data with heavier tails. It is also worth noting that slight estimation bias remains in some parameters, even with larger sample sizes, particularly under non-Gaussian settings. This reflects the inherent complexity of selection models with heavy-tailed errors, and similar behavior has been reported in frequentist estimators such as MLE using the \texttt{R} package \texttt{HeckmanEM}.\\
The computations were run in R 4.4.1 on UConn’s Storrs HPC cluster (400 nodes and $>$20,000 CPU cores across Intel Skylake and AMD EPYC; 10/25 Gb Ethernet between nodes; 100–200 Gb/s InfiniBand; 3.65 PB parallel storage with 220 TB scratch; 135 NVIDIA GPUs). Using simulated data with $n=400$ and averaging over 100 replications, times are reported as user time in seconds. For the SLt example, the EM estimator (HeckmanEM; tol $10^{-6}$ , max 1000 iters) took 44.0 seconds, whereas the Bayesian estimator (HeckmanStan; 10,000 iterations with 1,000 warm-up, thinning 5, one chain) took 236.3 seconds.\\
All R-hat values are close to 1.00, confirming that the Markov chains have converged well. The effective sample sizes (ESS) exceed 3,000 for all parameters, ensuring that Monte Carlo error is negligible. The average acceptance rate also falls within the ideal range (0.65–0.90), indicating stable HMC performance. Further diagnostic details are reported in the Supplementary Material.

\begin{table}[!ht]
\caption{Simulation study. Percentage of times each model is selected among the SL models over 100 Monte Carlo replicates}
	\label{tab:sim_select}
\centering
\renewcommand{\arraystretch}{0.85}
\begin{tabular}{ccccccccccc}
  \toprule
 & & \multicolumn{3}{c}{LOOIC} & \multicolumn{3}{c} {WAIC}  & \multicolumn{3}{c} {CPO} \\
 Distribution & Sample Size & SLn & SLt & SLcn & SLn & SLt & SLcn &  SLn & SLt & SLcn\\ 
 \hline
          & 200 &  94\% & 4\% & 2\% & 94\% & 4\% & 2\%& 95\% & 5\% & 2\%\\\ 
Normal    & 400 &   91\% & 8\% & 1\% & 91\% & 8\% & 1\%& 91\% & 8\% & 1 \%\\\ 
          & 800 &  89\% & 8\% & 3\% & 89\% & 8\% & 3\%& 89\% & 8\% & 3 \%\ \\
          \hline
          & 200 & 0\% & 69\% & 31\% & 0\% & 68\% & 32\%& 0\% & 70\% & 30\% \\ 
Student's-t & 400 & 0\% & 78\% & 22\%& 0\% & 76\% & 24\% & 0\%& 79\% & 21\% \ \\ 
          & 800 & 0\% & 88\% & 12\% & 0\% & 87\% & 13\%& 0\% & 88\% & 12\%\ \\ 
                    \hline
          & 200 &  25\% & 51\% & 24\% & 25\% & 50\% & 25\%& 25\% & 51\% & 24\% \\
Slash     & 400 &  12\% & 57\% & 31\% & 12\% & 58\% & 30\%& 12\% & 57\% & 31\% \\
          & 800&  4\% & 59\% & 37\% & 4\% & 59\% & 37\%& 4\% & 59\% & 37\% \\
   \bottomrule
\end{tabular}
\end{table}

\section{Applications}\label{secApp}
We illustrate the proposed algorithms with the analysis of two real data sets. The analysis also utilized the \textsf{R} package \texttt{HeckmanStan} available on Github (\url{https://github.com/heeju-lim/HeckmanStan}). 

\subsection{Ambulatory expenditures}

The first application concerns a study of ambulatory expenditures taken from \citet{cameron2010microeconometrics}. This dataset was re-analyzed by \citet{marchenko2012heckman} using the ML estimation procedure in \textsf{Stata}, and using the EM algorithm by \citet{lachosHeckman}. 

In our analysis, we choose the same set of covariates as \citet{marchenko2012heckman}. We use the $\ln$ of ambulatory expenditures ($ambexp$) as the outcome variable. The covariates in the outcome equation are $\xp = (1, age, female, educ, blhisp,totchr,ins)$, including age, gender, education status, ethnicity, number of chronic diseases, and insurance status, respectively. The exclusion restriction assumption holds by including the income variable in the selection equation, i.e., $\w = (\xp, \text{income})$. The dataset contains 3,328 observations, with 526 missing values for $ambexp$. More details about the data can be found in Chapter 16 of \citet{cameron2010microeconometrics}.

\begin{figure}[!ht]
    \centering
    \includegraphics[scale=0.24]{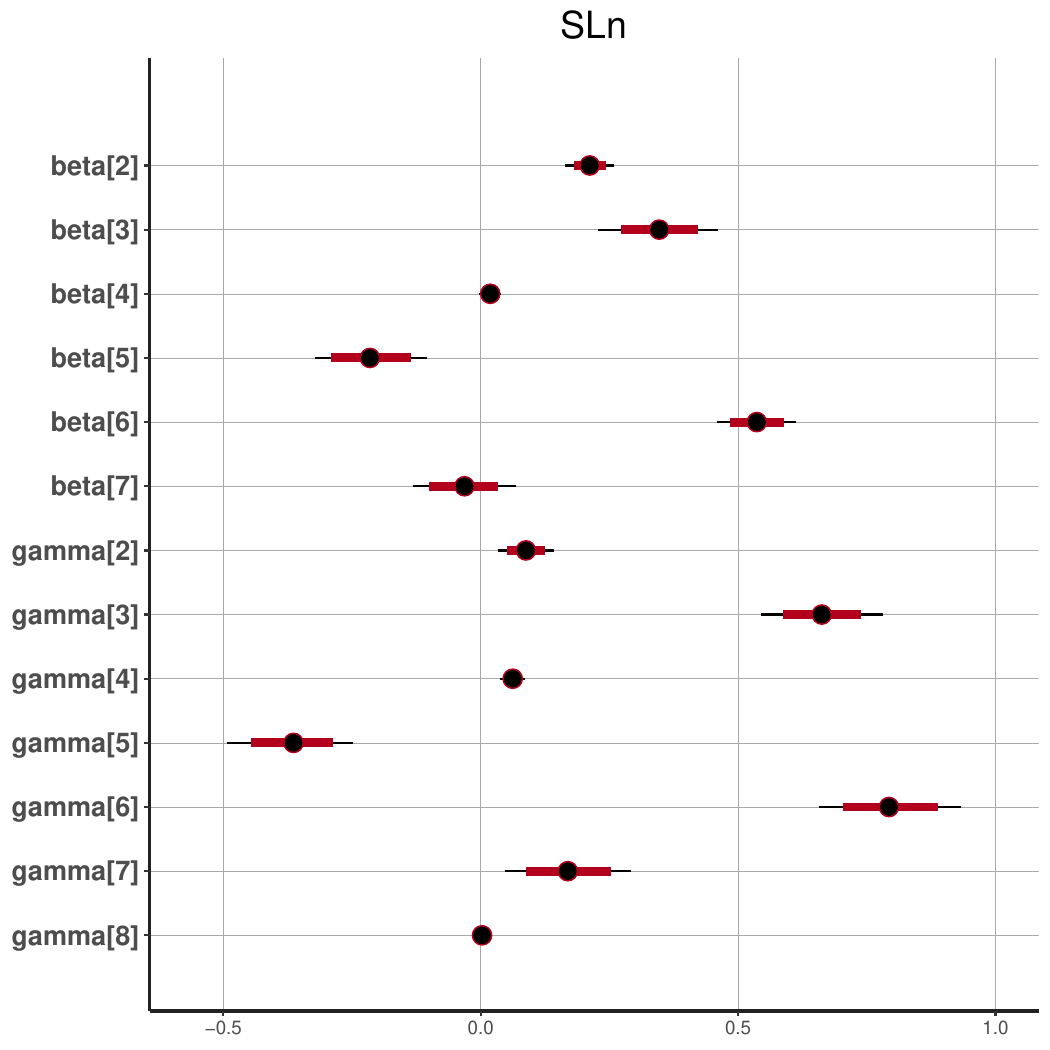} 
     \includegraphics[scale=0.24]{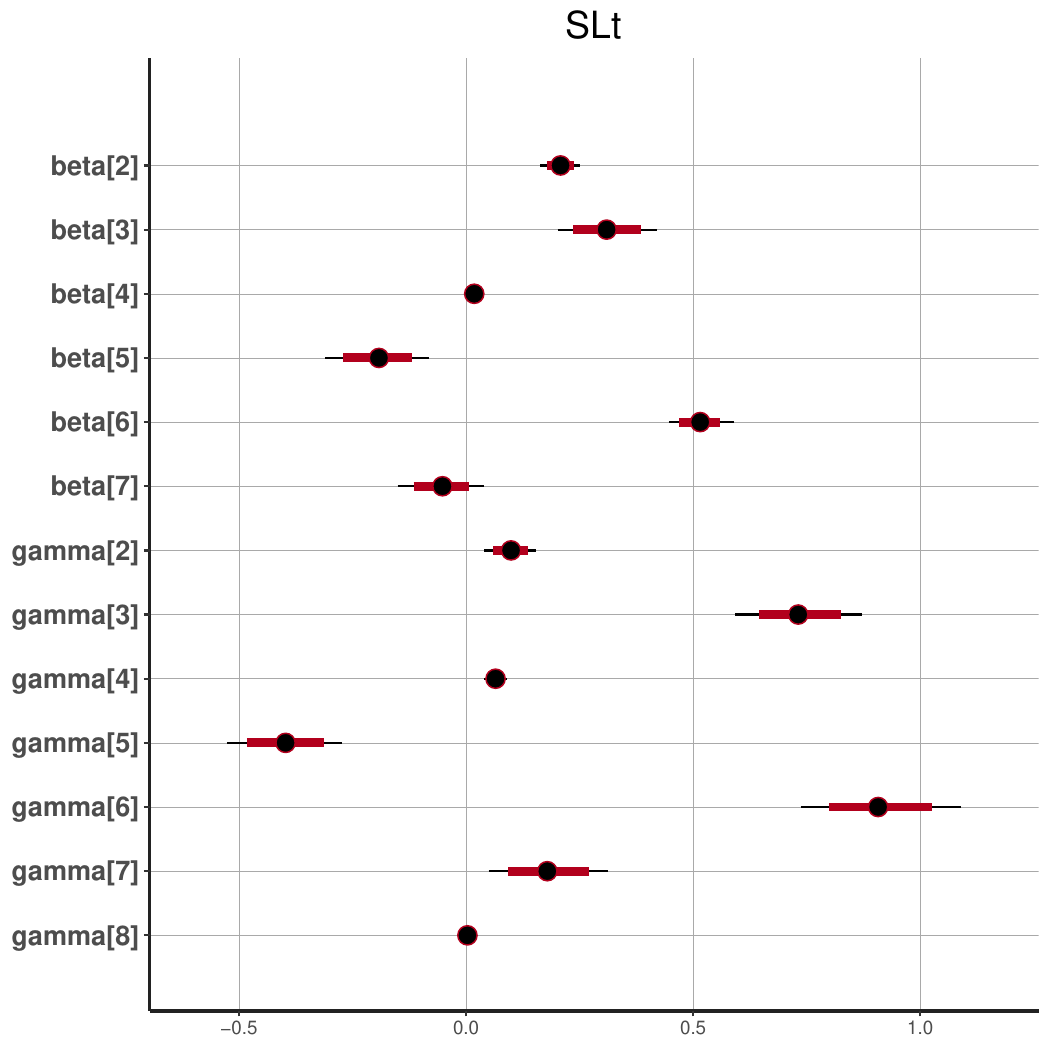} 
    \includegraphics[scale=0.24]{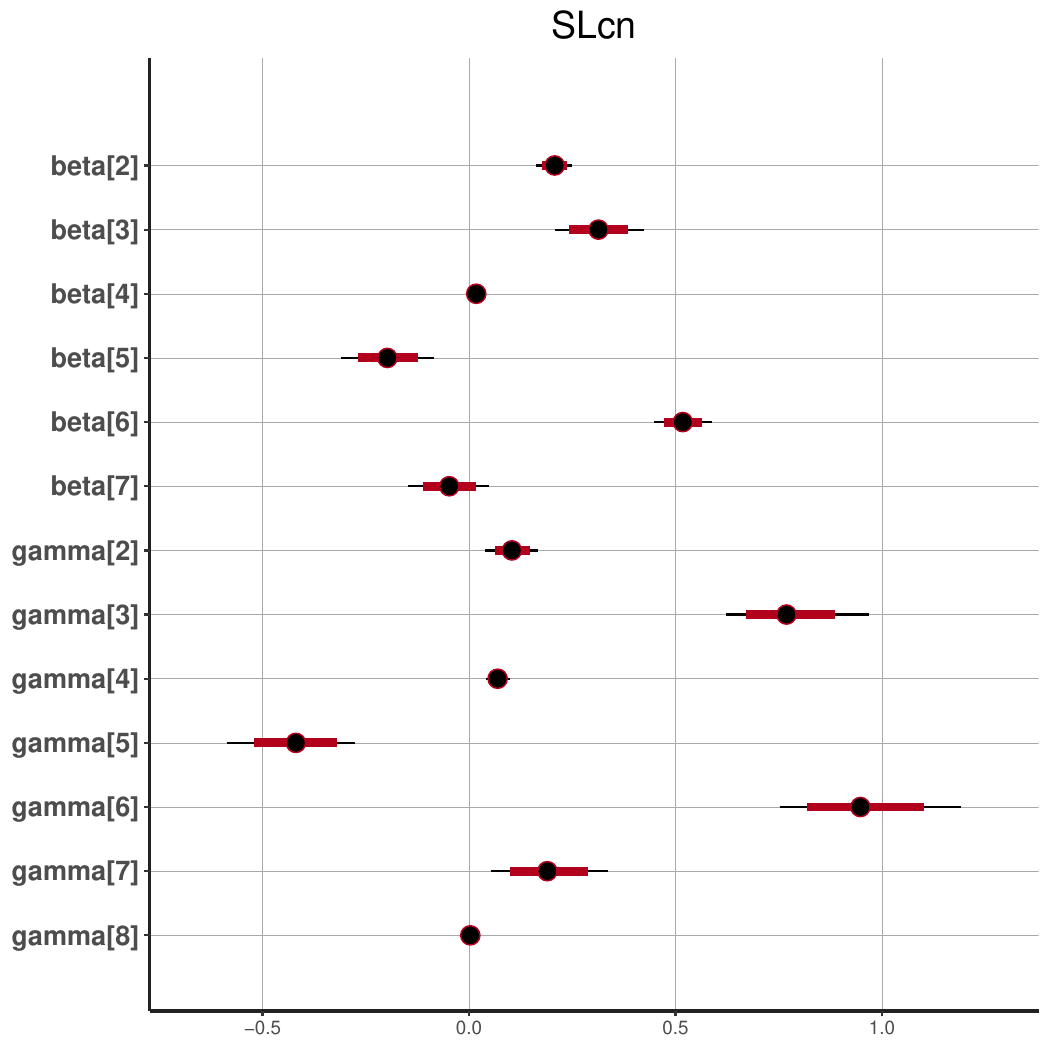} 
    \caption{Ambulatory expenditure data. Posterior density estimation with the median and {95\% credible interval (CI)} for $\bbeta$ and $\bgamma$ parameters of SL models.}
    \label{fig:expenditure1}
\end{figure}

Bayesian estimation results for the SLn, SLt, and SLcn models are summarized in Table~\ref{tab:pexpenditure} and visualized in Figure~\ref{fig:expenditure1}. Overall, the findings align with those of \citet{marchenko2012heckman} and \citet{lachosHeckman}. In the outcome equation, age and female are positively associated with ambulatory spending, while blhisp exerts a consistently negative effect. Education and insurance status show weak and statistically uncertain effects. Totchr has the largest positive effect, underscoring the role of chronic conditions. 

In the selection equation, age, female, and education increase the probability of observing positive expenditures, while blhisp reduces it. Totchr again has the strongest positive effect. Income exhibits only a negligible and statistically weak influence, lending support to the exclusion restriction. 
As detailed in Supplementary S.4, posterior density plots and correlation plots are provided for all parameters. These show that most covariates in both the outcome and selection equations exhibit very weak posterior correlations, indicating that multicollinearity is not a concern. This lack of strong association supports the stability of the posterior inference and enhances the interpretability of the individual regression effects.

Regarding the selection parameter $\rho$, the SLn model yields an HPD interval including zero $(-0.40,0.13)$, suggesting weak evidence of selection bias under normality. By contrast, both the SLt $(-0.51,-0.09)$ and SLcn $(-0.51,-0.07)$ models produce intervals excluding zero, indicating significant selection effects under non-Gaussian errors. The flexible models also provide insights into tail behavior and contamination. In the SLt model, the degrees of freedom parameter $\nu$ is estimated between 8.21 and 17.91 (95\% HPD), while in the SLcn model the contamination probabilities are estimated as $\nu_1=0.24$ and $\nu_2=0.39$, reflecting heavy-tailed behavior and the presence of outliers. Model comparison based on LOOIC, WAIC, and CPO (Table~\ref{tab:pexpenditure}) consistently favors the SLt model, highlighting the advantages of heavy-tailed error specifications in capturing the distributional features of ambulatory expenditures.

\begin{table}[!ht]
\centering
\caption{Ambulatory expenditure data. Posterior estimates of parameters and Bayesian model selection values.}
\label{tab:pexpenditure}
\setlength{\tabcolsep}{4pt}
\begin{tabular}{cccccccccccccc}
  \hline
  \hline
 \multicolumn{2}{c}{Criterion} & \multicolumn{4}{c}{SLn} & 
 \multicolumn{4}{c}{SLt} & \multicolumn{4}{c}{SLcn}  \\
   \hline
 \multicolumn{2}{c}{LOOIC} & \multicolumn{4}{c}{11706.64} & \multicolumn{4}{c}{11680.12} & \multicolumn{4}{c}{11680.74} \\
 \multicolumn{2}{c}{WAIC} & \multicolumn{4}{c}{11706.57} & \multicolumn{4}{c}{11680.03} & \multicolumn{4}{c}{11680.63} \\
 \multicolumn{2}{c}{CPO} & \multicolumn{4}{c}{-5853.278} & \multicolumn{4}{c}{-5840.01} & \multicolumn{4}{c}{-5840.32} \\
   \hline 
   \hline
 & & ME & SD & \multicolumn{2}{c}{HPD (95\%)} & ME & SD & \multicolumn{2}{c}{HPD (95\%)}& ME & SD & \multicolumn{2}{c}{HPD (95\%)} \\ 
\hline
&&&&&&&&&&&&&\\
Outcome model & & & & & & & & & & & & & \\
$Intercept $ & $\beta_1$ & 5.06 & 0.23 & 4.60 & 5.51 & 5.19 & 0.21 & 4.81 & 5.61 & 5.18 & 0.20 & 4.78 & 5.58 \\  
$Age $ & $\beta_2$ & 0.21 & 0.02 & 0.16 & 0.26 & 0.21 & 0.02 & 0.16 & 0.25 & 0.21 & 0.02 & 0.16 & 0.25 \\
$Female$ & $\beta_3$ & 0.35 & 0.06 & 0.23 & 0.46 & 0.31 & 0.06 & 0.20 & 0.42 & 0.31 & 0.06 & 0.21 & 0.42 \\ 
$Educ$ & $\beta_4$ & 0.02 & 0.01 & -0.00 & 0.04 & 0.02 & 0.01 & -0.00 & 0.04 & 0.02 & 0.01 & -0.00 & 0.04 \\ 
$Blhisp$ & $\beta_5$ & -0.21 & 0.06 & -0.32 & -0.11 & -0.19 & 0.06 & -0.31 & -0.08 & -0.20 & 0.06 & -0.31 & -0.08 \\ 
$Totchr$ & $\beta_6$ & 0.54 & 0.04 & 0.46 & 0.61 & 0.52 & 0.04 & 0.44 & 0.58 & 0.52 & 0.04 & 0.45 & 0.59 \\  
$Ins$ & $\beta_7$ &-0.03 & 0.05 & -0.13 & 0.07 & -0.05 & 0.05 & -0.15 & 0.04 & -0.05 & 0.05 & -0.14 & 0.06 \\ 
&&&&&&&&&&&&&\\
Selection model & & & & & & & & & & & & & \\
$Intercept$ & $\gamma_1$ & -0.68 & 0.19 & -1.05 & -0.30 & -0.75 & 0.21 & -1.15 & -0.34 & -0.79 & 0.23 & -1.25 & -0.36 \\ 
$Age$ & $\gamma_2$ & 0.09 & 0.03 & 0.03 & 0.14 & 0.10 & 0.03 & 0.04 & 0.15 & 0.11 & 0.03 & 0.04 & 0.17 \\ 
$Female$ & $\gamma_3$ & 0.66 & 0.06 & 0.55 & 0.78 & 0.73 & 0.07 & 0.59 & 0.86 & 0.78 & 0.09 & 0.61 & 0.94 \\ 
$Educ$ & $\gamma_4$ & 0.06 & 0.01 & 0.04 & 0.09 & 0.06 & 0.01 & 0.04 & 0.09 & 0.07 & 0.01 & 0.04 & 0.10 \\ 
$Blhisp$ & $\gamma_5$ & -0.37 & 0.06 & -0.49 & -0.25 & -0.40 & 0.07 & -0.52 & -0.27 & -0.42 & 0.08 & -0.57 & -0.27 \\ 
$Totchr$ & $\gamma_6$ &0.80 & 0.07 & 0.66 & 0.94 & 0.91 & 0.09 & 0.72 & 1.07 & 0.96 & 0.11 & 0.74 & 1.16 \\ 
$Ins$ & $\gamma_7$ & 0.17 & 0.06 & 0.05 & 0.29 & 0.18 & 0.07 & 0.05 & 0.31 & 0.19 & 0.07 & 0.06 & 0.34 \\ 
$Income$ & $\gamma_8$ & 0.00 & 0.00 & 0.00 & 0.01 & 0.00 & 0.00 & 0.00 & 0.01 & 0.00 & 0.00 & 0.00 & 0.01 \\ 
&&&&&&&&&&&&&\\
&$\sigma^2$ & 1.63 & 0.05 & 1.53 & 1.73 & 1.42 & 0.06 & 1.30 & 1.53 & 1.27 & 0.14 & 0.97 & 1.51 \\ 
&$\rho$ &  -0.15 & 0.14 & -0.40 & 0.13 & -0.31 & 0.11 & -0.51 & -0.09 & -0.29 & 0.11 & -0.51 & -0.07 \\  
&&&&&&&&&&&&&\\
& $\nu$  & & & & & 12.33 & 2.62 & 8.21 & 17.91 & & & & \\ 
& $\nu_1$  & & & & & & & & & 0.24 & 0.13 & 0.05 & 0.52 \\ 
& $\nu_2$  & & & & & & & & & 0.39 & 0.06 & 0.27 & 0.51 \\ 
 \hline
   \hline
\end{tabular}
\end{table}

\subsection{Mroz: Labor Supply Data}\label{application_Mroz}

\begin{table}[!ht]
\centering
\caption{Labor Supply Data. Posterior estimates of parameters and Bayesian model selection values}
\label{tab:pwage}
\setlength{\tabcolsep}{4pt}
\begin{tabular}{cccccccccccccc}
  \hline
  \hline
 \multicolumn{2}{c}{Criterion} & \multicolumn{4}{c}{SLn} & 
 \multicolumn{4}{c}{SLt} & \multicolumn{4}{c}{SLcn}  \\
   \hline
\multicolumn{2}{c}{LOOIC} & \multicolumn{4}{c}{1791.972} & \multicolumn{4}{c}{1703.452} & \multicolumn{4}{c}{1702.727} \\
\multicolumn{2}{c}{WAIC} & \multicolumn{4}{c}{1791.788} & \multicolumn{4}{c}{1703.38} & \multicolumn{4}{c}{1702.634} \\
\multicolumn{2}{c}{CPO} & \multicolumn{4}{c}{-895.9688} & \multicolumn{4}{c}{-851.6868} & \multicolumn{4}{c}{-851.3202} \\
   \hline 
   \hline
 & & ME & SD & \multicolumn{2}{c}{HPD (95\%)} & ME & SD & \multicolumn{2}{c}{HPD (95\%)}& ME & SD & \multicolumn{2}{c}{HPD (95\%)} \\ 
\hline
&&&&&&&&&&&&&\\
Outcome model&&&&&&&&&&&&&\\
$Intercept$&$\beta_1$ & 0.63 & 0.22 & 0.21 & 1.04 & 0.31 & 0.17 & -0.02 & 0.65 & 0.33 & 0.17 & 0.01 & 0.67 \\  
$Education$&$\beta_2$ & 0.07 & 0.02 & 0.04 & 0.10 & 0.09 & 0.01 & 0.06 & 0.11 & 0.09 & 0.01 & 0.06 & 0.11 \\ 
$City$&$\beta_3$ & 0.11 & 0.08 & -0.04 & 0.26 & 0.09 & 0.06 & -0.02 & 0.20 & 0.08 & 0.06 & -0.03 & 0.20 \\ 
&&&&&&&&&&&&&\\
Selection model&&&&&&&&&&&&&\\
$Intercept$&$\gamma_1$ & 3.80 & 0.76 & 2.32 & 5.30 & 5.89 & 1.02 & 3.83 & 7.85 & 6.28 & 1.21 & 3.90 & 8.70 \\ 
$Hwage$&$\gamma_2$ & -0.11 & 0.02 & -0.14 & -0.07 & -0.15 & 0.02 & -0.20 & -0.11 & -0.16 & 0.03 & -0.22 & -0.11 \\ 
$Youngkids$&$\gamma_3$ & -0.43 & 0.09 & -0.59 & -0.25 & -0.60 & 0.12 & -0.83 & -0.37 & -0.66 & 0.13 & -0.89 & -0.40 \\ 
$Tax$&$\gamma_4$ & -5.79 & 0.85 & -7.60 & -4.28 & -8.42 & 1.17 & -10.72 & -6.10 & -8.99 & 1.41 & -11.94 & -6.45 \\  
$Feduc$&$\gamma_5$ & -0.02 & 0.01 & -0.04 & 0.01 & -0.01 & 0.02 & -0.04 & 0.02 & -0.01 & 0.02 & -0.04 & 0.02 \\
$Educ$&$\gamma_6$ & 0.11 & 0.02 & 0.06 & 0.16 & 0.12 & 0.03 & 0.06 & 0.18 & 0.13 & 0.03 & 0.06 & 0.19 \\ 
$City$&$\gamma_7$ & -0.04 & 0.10 & -0.24 & 0.17 & -0.10 & 0.12 & -0.34 & 0.14 & -0.10 & 0.14 & -0.37 & 0.15 \\
&&&&&&&&&&&&&\\
 &$\sigma^2$ & 0.63 & 0.06 & 0.53 & 0.77 & 0.25 & 0.03 & 0.18 & 0.32 & 0.23 & 0.04 & 0.15 & 0.30 \\ 
 &$\rho$ & -0.75 & 0.06 & -0.86 & -0.63 & -0.70 & 0.07 & -0.83 & -0.56 & -0.70 & 0.07 & -0.82 & -0.56 \\ 
&&&&&&&&&&&&&\\
 &$\nu$ & & & & & 3.05 & 0.51 & 2.12 & 4.01 & & & & \\ 
 &$\nu_1$ & & & & & & & & & 0.25 & 0.07 & 0.13 & 0.38 \\ 
 &$\nu_2$ & & & & & & & & & 0.13 & 0.02 & 0.08 & 0.18 \\ 
  \hline
  \hline
\end{tabular}
\end{table}

\begin{figure}[!ht]
    \centering
    \includegraphics[scale=0.24]{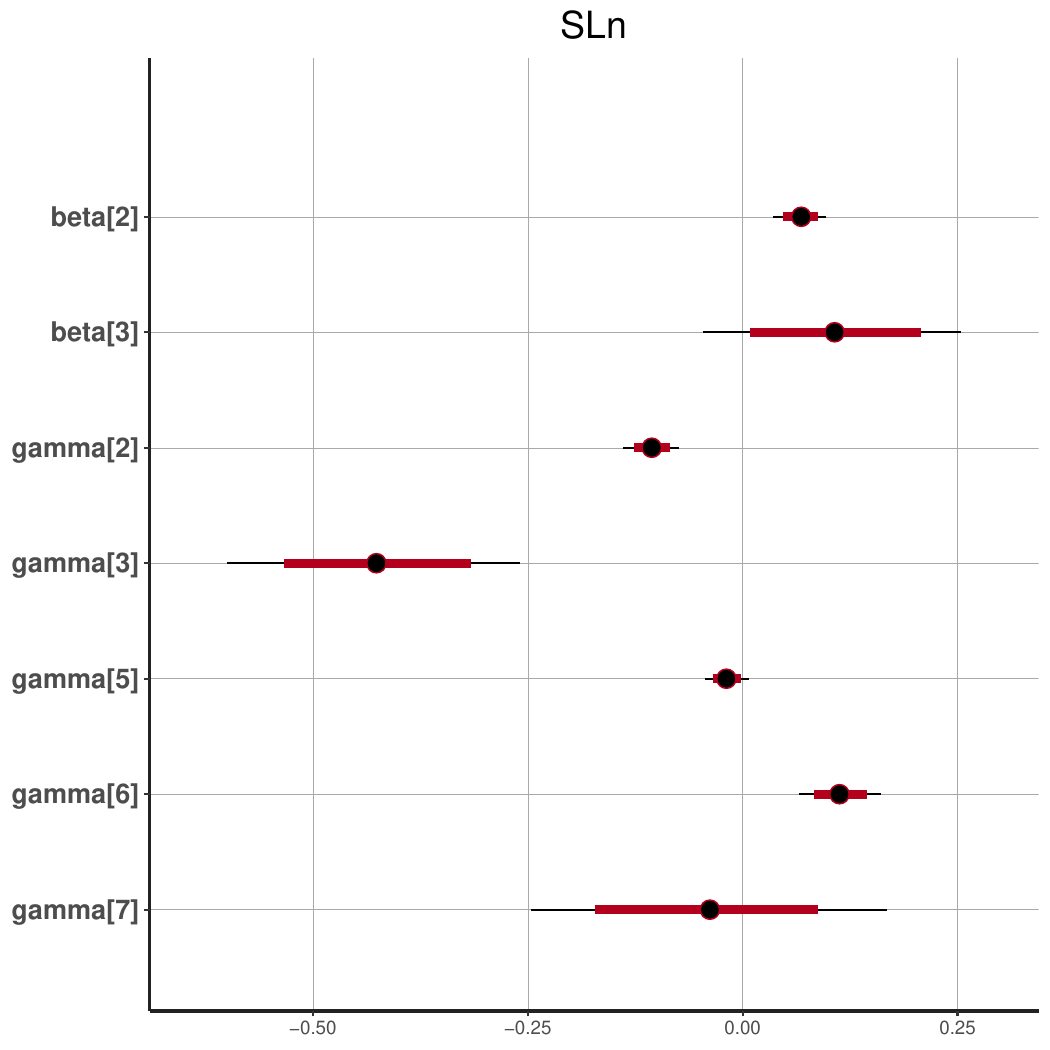} 
     \includegraphics[scale=0.24]{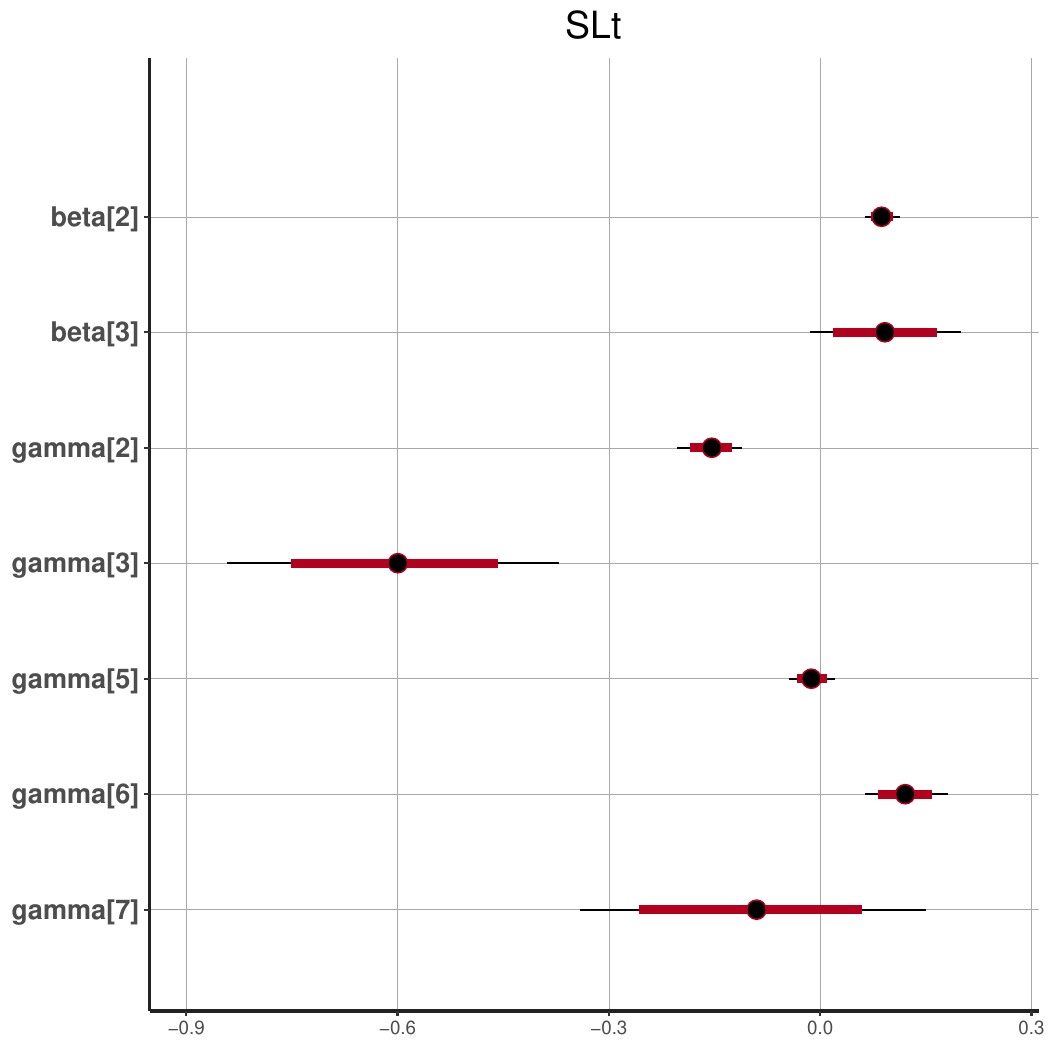} 
    \includegraphics[scale=0.24]{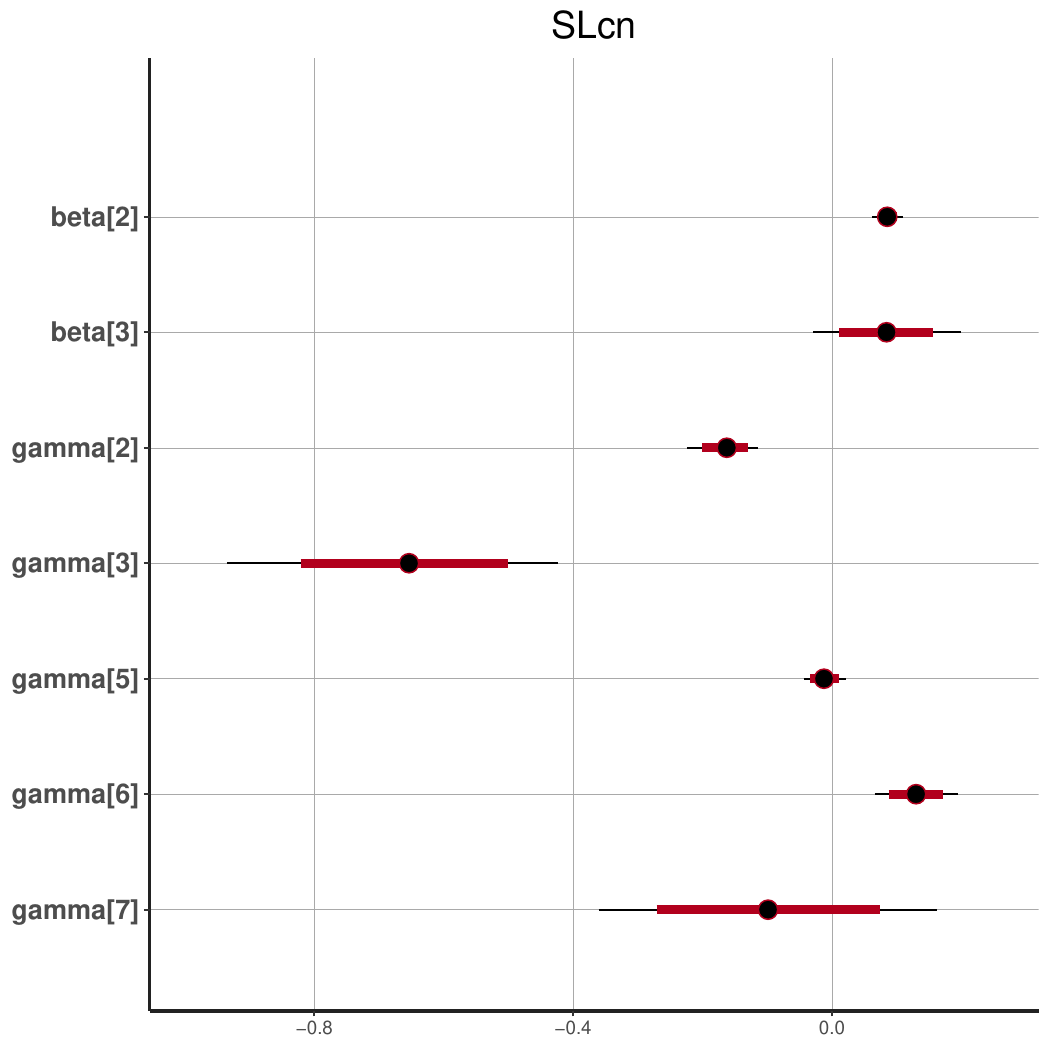} 
    \caption{Labor Supply Data. Posterior density estimation with the median and {95\% credible interval (CI)} for $\bbeta$ and $\bgamma$ parameters of SL models.}
    \label{fig:labor1}
\end{figure}

The second application focuses on missing econometric data, where we re-analyzed the dataset originally presented in \citet{mroz1999discrete} to estimate the wage offer function for married women.   
The dataset consists of observations on 753 married white women for 21 variables and can be found in the \textsf{R} package \texttt{AER} \citep{kleiber2020package}. The outcome of interest is the logarithm of female \texttt{wage}, which is missing for 325 individuals (43.2\% of the data) and observed for 428 individuals (56.8\%).  We follow \citet{ogundimu2016sample}, where the covariates in the outcome equation are education status and city, i.e., $\x = (1, \texttt{educ}, \texttt{city})$. The selection equation uses husband's wage, number of children 5 years old or younger, marginal tax rate of the wife, the wife's father's educational attainment, as well as education status and city, i.e., $\w = (\texttt{hwage}, \texttt{youngkids}, \texttt{tax}, \texttt{feduc}, \x)$.

The estimation results for the SLn, SLt, and SLcn models are presented in Table \ref{tab:pwage}, and Figure \ref{fig:labor1} displays the posterior estimates from the MCMC chains for each specification. For the parameter $\rho$, the 95\% HPD intervals are broadly similar across the three models. Notable differences appear in the intercept estimates when comparing the SLn model with the SLt and SLcn models in both the outcome and selection equations, as well as for the variable \textit{tax} in the selection equation.
Turning to the interpretation of the covariate effects on women’s labor-force participation, we find that the husband’s wage has a positive impact, the presence of young children has a negative impact, and education has a positive impact. In the SLt model, the additional parameter for the degrees of freedom is estimated at $\nu = 3.05$. In the SLcn model, the contamination parameters indicate $\nu_2 = 0.13$, reflecting low variance attributable to outliers, and $\nu_1 = 0.25$. Model comparison criteria in Table \ref{tab:pwage} suggest that LOOIC and WAIC favor the SLcn model, which also achieves the highest CPO value.
As detailed in Supplementary S.4, correlation plots of the posterior distributions show that in the selection equation, \textit{Hwage} and \textit{tax} are the only variables with a positive correlation. In the outcome equation, education and city exhibit virtually no correlation, indicating that multicollinearity is not a concern.

\subsection{Outlier Analysis for the SLcn model}\label{residual_analysis}
\begin{figure}[!ht]
\captionsetup{position=top}
\centering
    \begin{subfigure}[b]{0.45\textwidth} 
        \includegraphics[width=\textwidth]{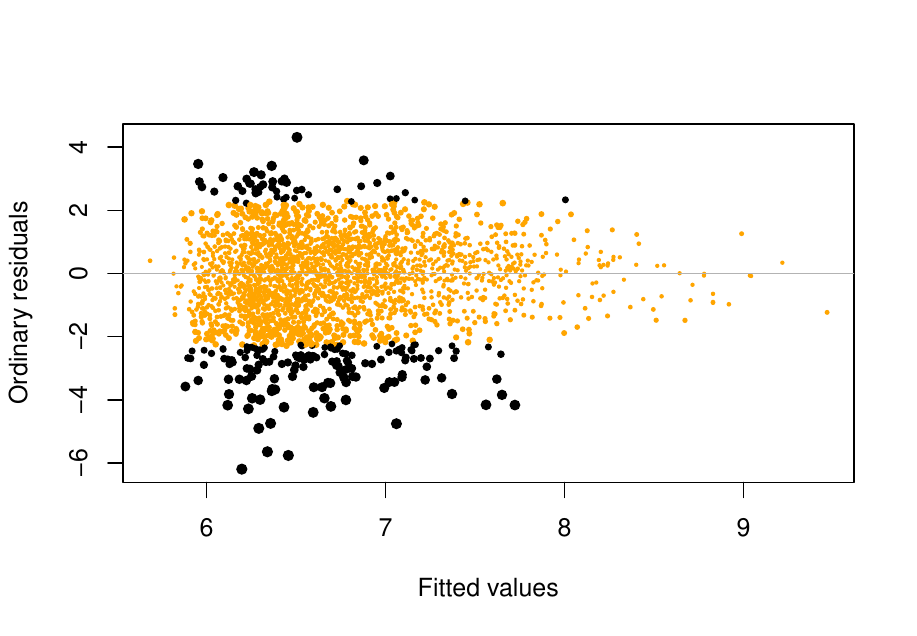}
    \caption{Ordinary residuals vs. Fitted values}
        \label{fig:A1residualplot}
    \end{subfigure}
    \begin{subfigure}[b]{0.45\textwidth}
        \centering
        \includegraphics[width=\textwidth]{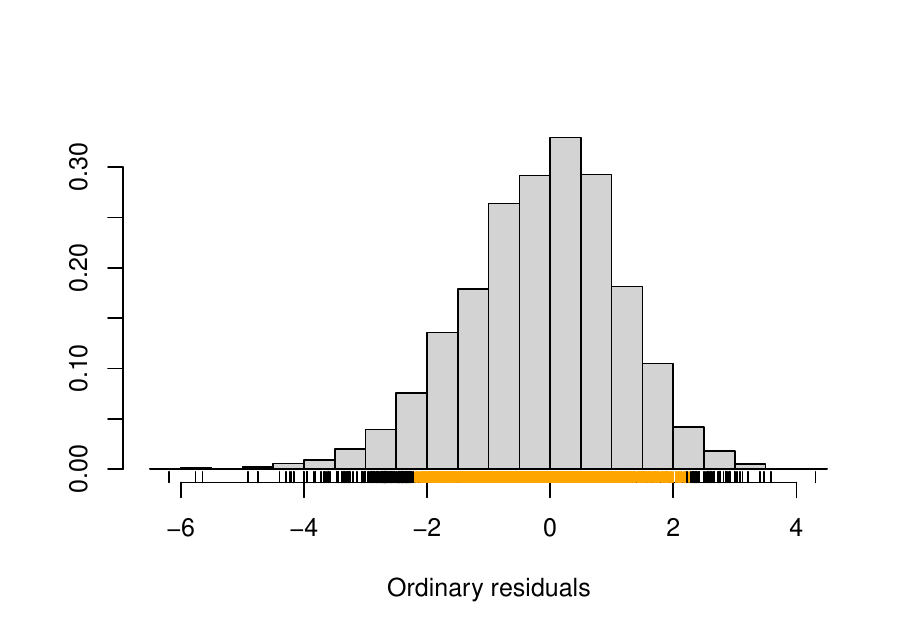}
        \caption{Histogram of ordinary residuals}
        \label{fig:A1residualhist}
    \end{subfigure}
    \begin{subfigure}[b]{0.45\textwidth}            
        \includegraphics[width=\textwidth]{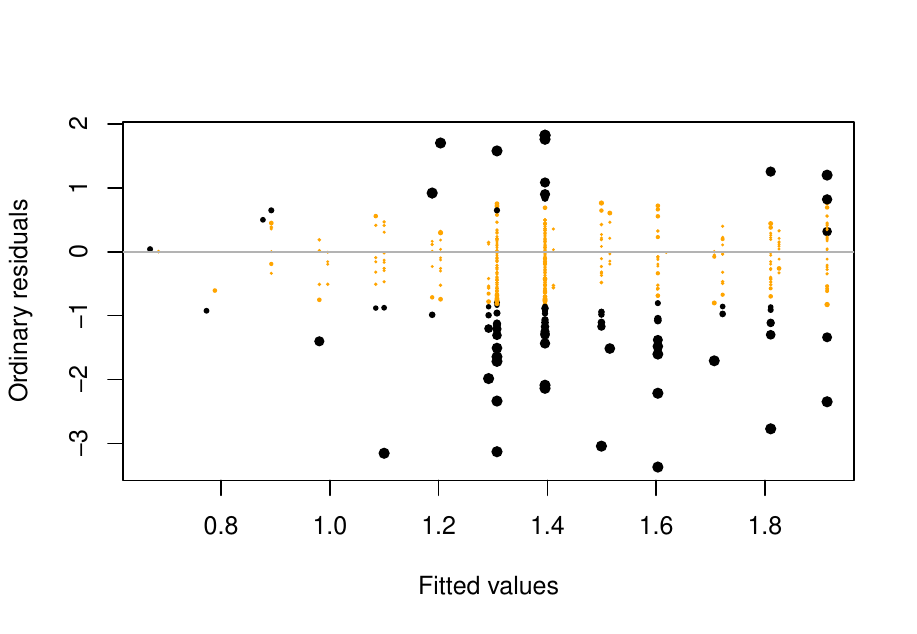}
        \caption{Ordinary residuals vs. Fitted values}
        \label{fig:E1residualplot}
    \end{subfigure}
    \begin{subfigure}[b]{0.45\textwidth}
        \centering
        \includegraphics[width=\textwidth]{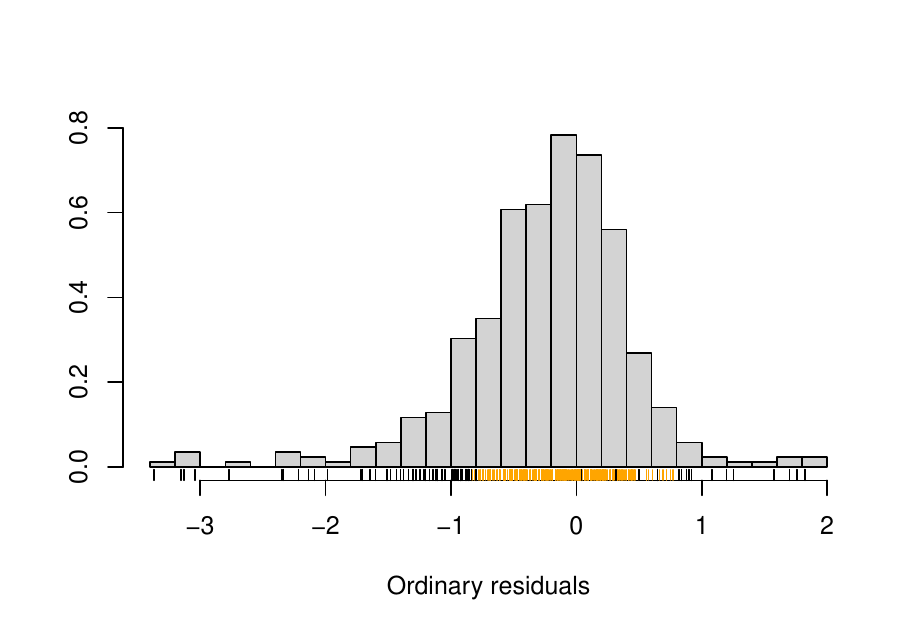}
        \caption{Histogram of ordinary residuals}
        \label{fig:E1residualhist}
    \end{subfigure}
\caption{Residual analysis for ambulatory expenditure (Upper Panel) and labor supply data (Lower Panel). The left column shows ordinary residuals versus fitted values, while the right column presents histograms of the residuals.}
\label{fig:residual_comparison}
\end{figure}
{Finally, Fig.~\ref{fig:residual_comparison} displays the plot of the classical residuals versus the fitted values for the SLcn model, along with a histogram of these residuals. The color of the bullets is determined by the automatic inlier/outlier detection rule discussed in Section~\ref{sec:Automatic mild outlier detection}: black represents outliers (with a probability greater than 0.5 of being a regular observation), while orange indicates inliers (with a probability lower than 0.5 of being a regular observation). These plots allow us to visualize the detected outliers, which are 196 (7.0\%) of the 2802 observed data for the ambulatory expenditure (Upper Panel) and 49 (11.4\%) of the 428 observed data for the labor supply data (Lower Panel). The residual plots suggest that inliers exhibit greater variability compared to outliers. Furthermore, the residual plots show a random dispersion of points forming a constant-width band around the horizontal line at zero, as expected from an appropriate model. }

\section{Conclusions}
\label{sec:6}

In this paper, we have proposed a fully Bayesian approach for analyzing Heckman selection models, specifically extending them to handle non-standard error distributions, such as the Student's-t and contaminated normal distributions. These distributions are more flexible and better suited for capturing heavy-tailed behaviors in real-world data, addressing a limitation of the traditional bivariate normal assumption commonly used in selection models. By employing {Stan's NUTS algorithm} for posterior simulation, we provide an efficient solution for parameter estimation, overcoming computational challenges often encountered in likelihood-based methods like the {EM} algorithm \citep{lachosHeckman}.

Through extensive simulation studies and real-world applications, we have demonstrated the efficacy of the Bayesian framework in improving inference accuracy and handling the complexities of sample selection bias. These models and methods are implemented in the \textsf{R} package \texttt{HeckmanStan}, available on GitHub at \url{https://github.com/heeju-lim/HeckmanStan}, which provides a practical tool for researchers and practitioners working with Heckman selection models in various fields. 

The proposed framework, while effective for handling selection bias in a single outcome, does not currently accommodate several and/or heterogeneous responses \citep{wang2024mixtures, lin2025finite}. Additionally, it does not address situations where both censoring and missingness occur simultaneously in the outcome data as discussed in \citet{wang2025flexible} and \citet{lin2025multivariate} .  Future work could extend the current model to support multivariate (several) and/or heterogeneous responses and develop a unified approach for jointly handling censoring and missing data. Another promising direction would be to investigate objective Bayesian approaches for key parameters such as the degrees of freedom $\nu$ in the SLt model and the contamination probability $\nu_1$ in the SLcn model, which could reduce subjectivity in prior specification and further enhance the robustness and consistency of the conclusions \citep{fonseca2008objective,ordonez2024penalized}.

Overall, our work highlights the flexibility and robustness of Bayesian methods for addressing heavy-tailed sample selection bias. We anticipate that it will be straightforward to extend these results to model skewness and heavy tails simultaneously. If we assume a bivariate skew-t \citep{arellano2010multivariate} or a bivariate skew-contaminated-normal distribution \citep{Lachos_Ghosh_Arellano_2009} for the outcome and selection equations, we can develop a new class of sample selection model, extending the work by \citet{ogundimu2016sample}.

\bigskip

\noindent \textbf{Acknowledgments}\\
The authors sincerely thank the Editor, Associate Editor, and anonymous reviewers for their valuable feedback and constructive criticism, which have significantly enhanced the presentation and quality of this paper. Victor H. Lachos acknowledges the partial financial support from the Office of the Vice President for Research (REP) and UConn - CLAS’s Summer Research Funding Initiative 2023. The second author is grateful to Coordenação de Aperfeiçoamento de Pessoal de Nível Superior - Brazil (CAPES) for financial support.

\bibliography{sn-new-bibliography}

\end{document}